\documentclass[preprint,10pt,letterpaper]{aastex}

\usepackage{multirow}

\def\Msun{M_\sun}
\def\Rsun{R_\sun}
\def\Lsun{L_\sun}

\slugcomment{Accepted for publication in \textit{The Astrophysical Journal}. \copyright\ 2013. The American Astronomical Society. All rights reserved.}

\begin{document}

\title{Disk-Related Bursts and Fades in Young Stars}

\author{Krzysztof Findeisen, Lynne Hillenbrand}
\affil{Cahill Center for Astronomy and Astrophysics, California Institute of Technology, MC 249-17, Pasadena, CA 91125}
\email{krzys@astro.caltech.edu, lah@astro.caltech.edu}

% Begin PTF photometric pipeline builders' list
\author{Eran Ofek}
\affil{Benoziyo Center for Astrophysics, Department of Particle Physics and Astrophysics, Weizmann Institute of Science, Rehovot 76100, Israel}
\author{David Levitan, Branimir Sesar}
\affil{Cahill Center for Astronomy and Astrophysics, California Institute of Technology, MC 249-17, Pasadena, CA 91125}
\author{Russ Laher, Jason Surace}
\affil{Spitzer Science Center, California Institute of Technology,  MC 314-6, Pasadena, CA 91125}

\begin{abstract}
We present first results from a new, multiyear, time domain survey of young stars in the North America Nebula complex using the Palomar Transient Factory. Our survey is providing an unprecedented view of aperiodic variability in young stars on timescales of days to years. 
The analyzed sample covers $R_\mathrm{PTF} \approx 13.5$-18 and spans a range of mid-infrared color, with larger-amplitude optical variables (exceeding 0.4~mag root-mean-squared) more likely to have mid-infrared evidence for circumstellar material. 
This paper characterizes infrared excess stars with distinct bursts above or fades below a baseline of lower-level variability, identifying 41 examples. The light curves exhibit a remarkable diversity of amplitudes, timescales, and morphologies, with a continuum of behaviors that can not be classified into distinct groups.
Among the bursters, we identify three particularly promising sources that may represent theoretically predicted short-timescale accretion instabilities. 
Finally, we find that fading behavior is approximately twice as common as bursting behavior on timescales of days to years, although the bursting and fading duty cycle for individual objects often varies from year to year. 
%Most bursters and all faders have a hot infrared excess consistent with an inner dust disk. 
\end{abstract}


\section{Introduction}

Variability in pre-main sequence stars can further our insight into physical processes associated with the formation and early evolution of both stars and planets. 
The observed flux variations are diagnostic of dynamic or radiative transfer effects that can occur on time scales ranging from hours to decades, or possibly longer. Driving phenomena include physical processes associated with envelope-to-disk infall \citep{2010ApJ...719.1896V}, disk-to-star accretion \citep{1996ApJ...458L..79W,1998ApJ...497..342M}, differential rotation of a three-dimensional disk, rotation of the star, magnetic field interaction between the star and the disk \citep{RUK2004a,2012arXiv1209.1161R}, accretion-driven wind and outflow \citep{2012ApJ...758..100B}, and planet-disk interaction. 
Different amplitudes and time scales can be associated with each of the theoretically postulated physical phenomena. In addition, the observed behavior of any individual system can be modified by orientation with respect to the line of sight. 
The wide range of plausible aperiodic behavior originates for the most part in the circumstellar environment. Variability of circumstellar origin is superposed on an underlying periodic modulation that is expected due to rotation of surface inhomogeneities, analogous to enhanced sunspots, across the projected stellar disk, as well as any short time scale chromospheric flaring or other aperiodic activity. 

Historically, while many empirical studies of  pre-main sequence star variability have involved searches for periodic behavior, those addressing aperiodic variability usually focused on explanations involving stochastically time variable disk-to-star accretion, circumstellar extinction, or both. 
Examples of the former include the extreme ($>$2-6 mag) ``outburst events'' as exemplified by EX Lup and FU Ori objects \citep{FUOri_define}. These types of sources are interpreted as undergoing episodes of rapid mass accumulation due to an instability in the inner disk. In the context of stellar mass assembly history, the duration and frequency of such outbursts is important to establish empirically since these events are thought, based on theory, to play a determining role in setting the final mass of the star. Accretion outbursts may also determine a star's appearance to us on the so-called ``birthline'' in the canonical HR diagram of stellar evolution \citep{1997ApJ...475..770H,2009ApJ...702L..27B}, from which stellar masses and ages are usually derived without considering the effects of accretion history. For similar reasons, it is also important to understand the variation at less extreme levels in the disk-to-star mass accretion rates. The ``irregular variables'' identified by, e.g., \citet{HBM2002} at low amplitudes ($<$0.1-0.3 mag) and short time scales are thought to indicate non-steady accretion.

Examples of the latter, extinction-related, variability include UX Ori stars, which undergo distinct and somewhat long-lived extinction events, as well as the broader category of stars identified by color-color and color-magnitude trends consistent with shorter-timescale, random variation along reddening vectors by, e.g., \citet{CHS2001,CHS2002}. 
More recently, so-called ``dipper'' events \citep[e.g.,][]{SigOri,YSOvar_first} are attributed to repeated sub-day or several-day circumstellar extinction enhancements. Such repeating but aperiodic flux dips or eclipse-like events have been qualitatively explained by, e.g., \citet{FM2010} and \citet{FMR2012} using rotating non-axisymmetric disk models or by \citet{dust_turner} with a vertical disk turbulence model. Periodic versions of the dipper class are known as AA~Tau stars \citep[e.g.,][]{AATau_first}.

While variability is a common property of young pre-main-sequence stars and has been viewed as a key observational characteristic of newly formed stars since their discovery \citep[e.g.,][]{firstTTS}, the full breadth of variable phenomena has not been explored in quantitative detail. The expected periodic variability, indicative of stellar rotation, has been well studied \citep[e.g.,][]{2000ASPC..219..216G, HEM2007, IHA2008, RSM2006, CB2007}. The harder to interpret aperiodic variability, our focus, is cataloged but relatively unexplored in the literature. Many fundamental properties of aperiodic variability in young stars are still poorly quantified. As aperiodic variables constitute more than half to two-thirds of variable stars in star-forming regions, characterizing and understanding them is essential to completing our understanding of young star/disk physics. In addition to the photometric study reported here, spectroscopic monitoring may be required, such as reported by \citet{2011A&A...526A..97C} or \citet{spec_series}.

To further progress, we have carried out a multiyear optical wavelength monitoring program aimed at determining the distributions of variability time scales and amplitudes among members of a young stellar population. Specifically, we have surveyed several square degrees of the North America and Pelican Nebulae region \citep{hsfr_nan} using the wide field of view and reliable time coverage of the Palomar Transient Factory\footnote{\url{http://www.astro.caltech.edu/ptf/}} \citep[PTF;][]{PTF, PTF_sciencecase} in operation at the Palomar 48'' Samuel Oschin Telescope. A notable niche of these PTF data is the long duration of the time series at roughly nightly cadence. 
Future contributions will assess the overall variability statistics, categories of variability, and characteristic time scales and amplitudes of optical variability in the North America and Pelican Nebula region. Here, we present our PTF survey strategy and our methodology for identifying variable stars. We then investigate
two specific types of variability phenomena exhibited among an infrared-excess selected sample of objects.

The present paper focuses on observable optical variability among the $\sim 2100$ known and suspected members of the North America Nebula complex cataloged by \citet{RGS2011} based on mid-infrared selection techniques. Of these, 84\% are within our monitored field. Among the wide range of behaviors exhibited by variable stars, we consider the evidence for and typical properties of bursting or fading behavior, possibly mixed with other forms of variability. In the case of bursting stars, while accretion-related instabilities having timescales of a few tens of days have been predicted by a number of theoretical studies \citep[e.g.,][]{AK1990,RUK2004b}, no evidence for accretion bursts produced by such instabilities has been published \citep{Bouvier_PPV}, although accretion bursts on both shorter \citep{shortBurst} and longer \citep{EXLup_review} timescales has been observed.
We assess the frequency of these intermediate timescale instabilities. 
For fading stars, while the existence of extinction-related variability is well-established, results vary among authors as to the frequency of young stars exhibiting such behavior, as well as the typical timescales. We also address in this study the ratio of bursting to fading light curves for a typical T Tauri star population.

The remainder of this paper is organized as follows. In Section~\ref{data}, we present our photometric data and our detection thresholds for variability. Section~\ref{demo} discusses how we defined the burster and fader populations and their key properties. In Sections~\ref{bursters} and \ref{dippers}, we discuss the bursters and faders in more detail, with an emphasis on how the largest sample yet identified of such objects can constrain their underlying physics. In Section~\ref{oddballs}, we describe several noteworthy objects in more detail. Section~\ref{discussion} summarizes our results, describes limitations of our analysis, and suggests pathways for future work.

\section{Photometric Data}\label{data}

\subsection{Instrument and Survey}

\begin{figure}
\includegraphics[height=0.9\textheight]{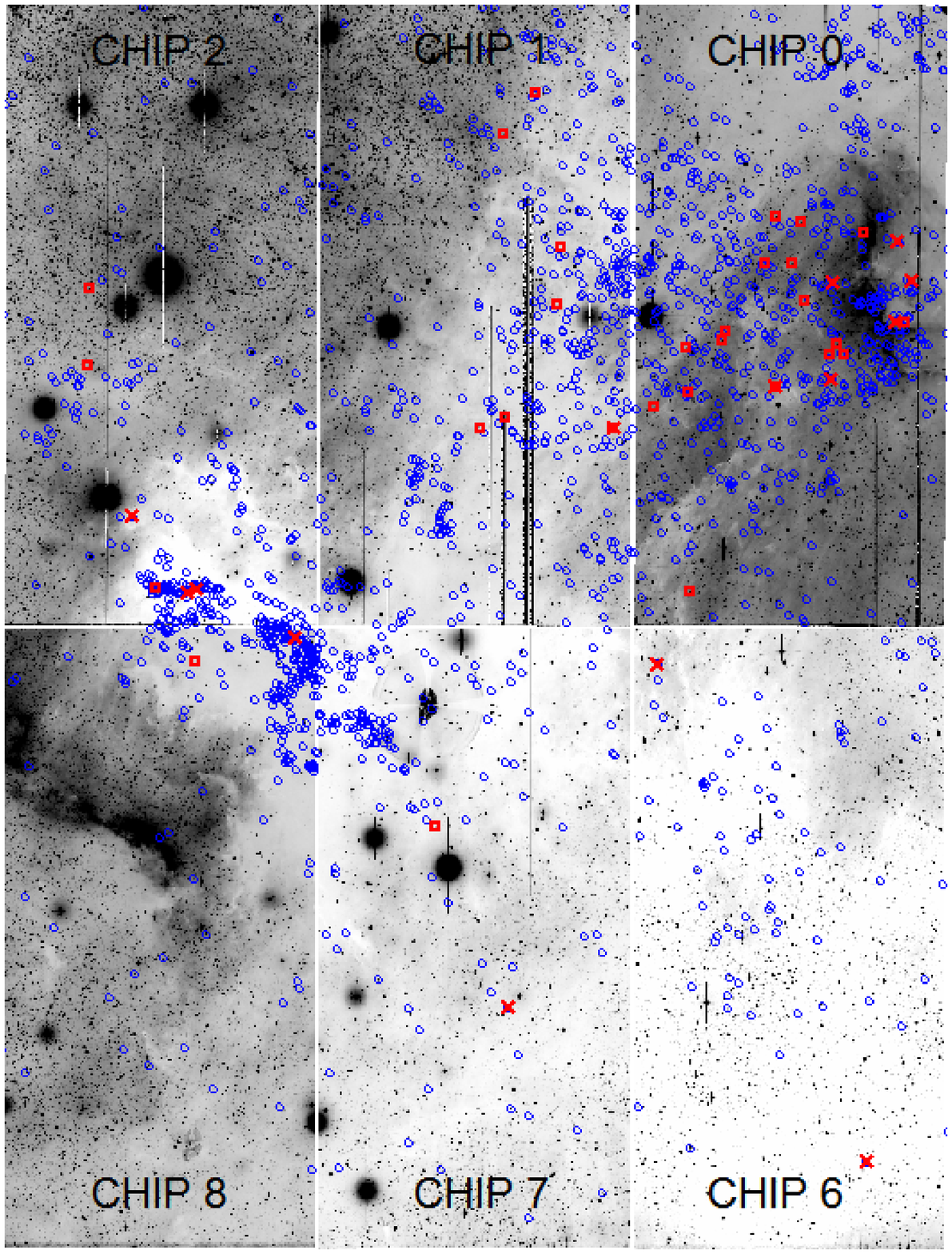}
\caption{The North America Nebula complex, as observed by PTF in a single epoch from 2009. Only six of the 11 PTF chips are shown; the remainder, to the left of this field, were off the nebula and probed the Galactic field population. The North America Nebula proper (NGC~7000) is on the left side of the frame, while the Pelican Nebula (IC~5070) is on the upper right, with the L935 dark cloud between them. The blue circles mark the positions of candidate members selected using infrared excess by \citet{RGS2011}. We also highlight stars with apparent bursting activity from Table~\ref{candidates} with red X's, and stars with apparent fading activity with red squares.
}\label{overview}
\end{figure}

The PTF survey camera is a mosaic of 11 chips, covering a total area of 7.26 square degrees with 1\arcsec\ pixels. This is a wide enough field to observe almost the entire North America Nebula complex in a single exposure, as shown in Figure~\ref{overview}.
PTF typically observes in the Mould-$R$ or $g^\prime$ bands. Because our targets are intrinsically red and further reddened by extinction, we conducted our entire survey in $R$ band, where a typical 60-second exposure reaches stars as faint as 20th magnitude. Throughout the survey, we took at least two exposures per night, separated by one hour.

\begin{figure}
\includegraphics[width=0.47\textwidth]{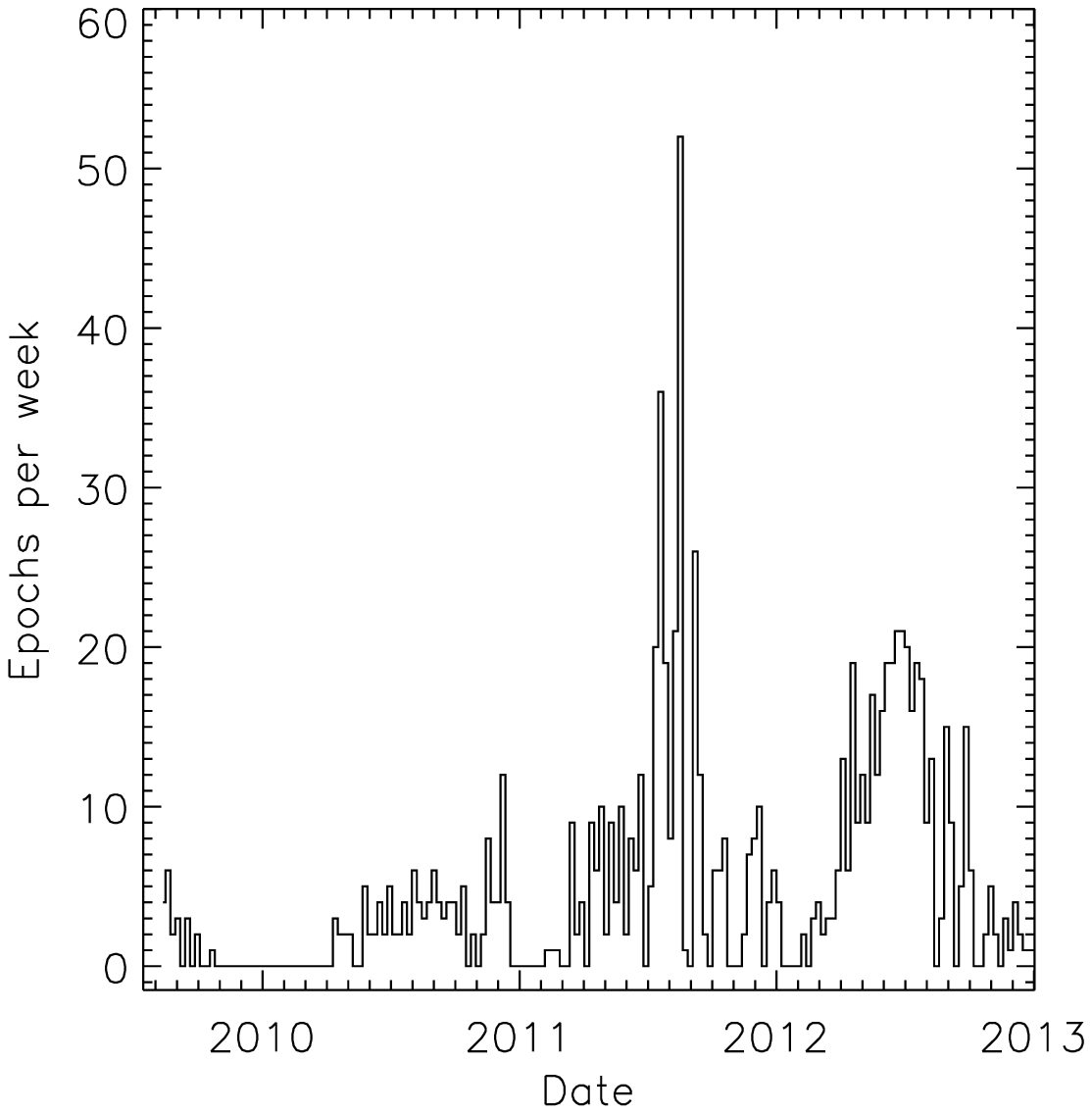}
\includegraphics[width=0.47\textwidth]{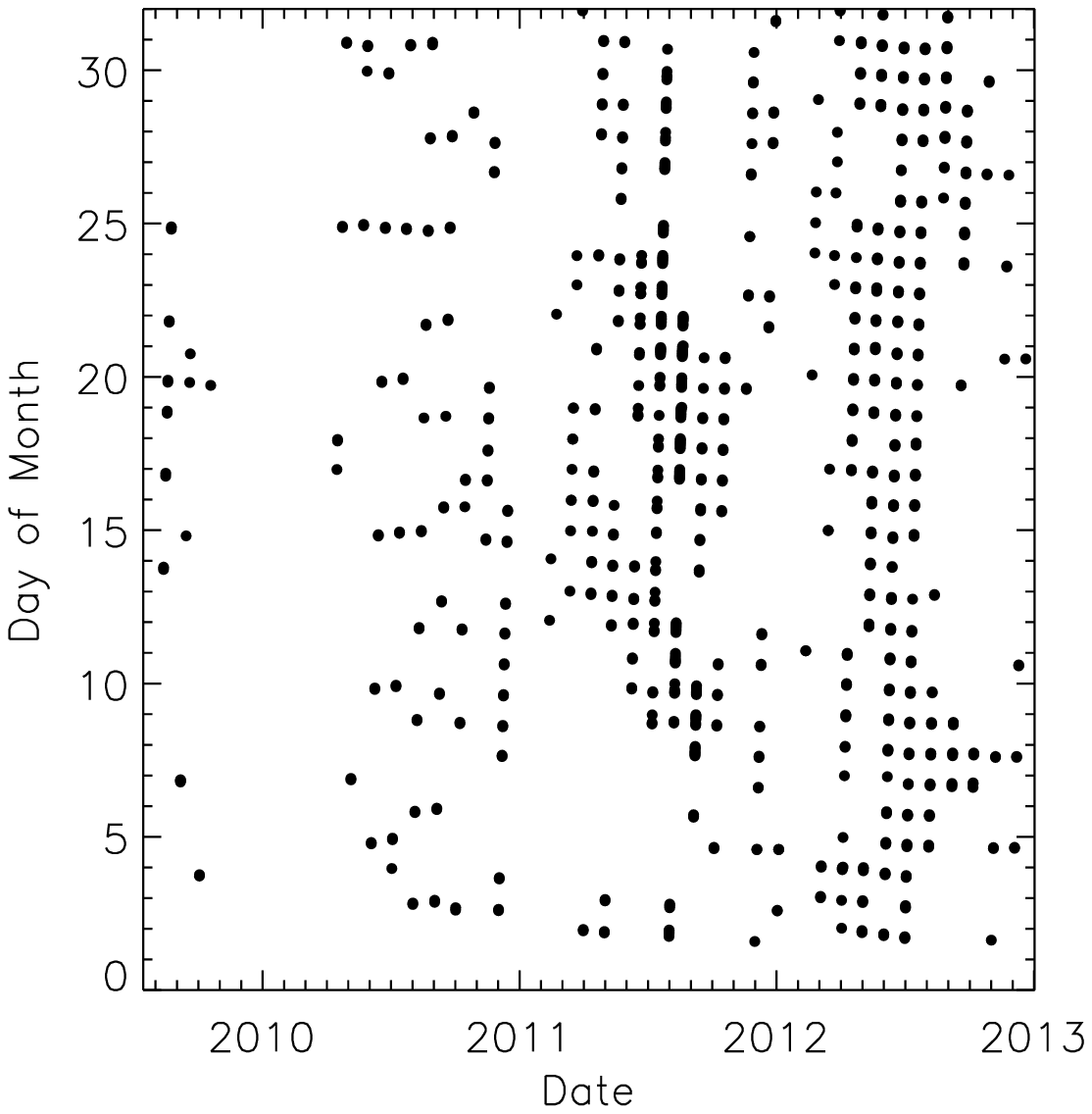}
\caption{The complex cadence used in the PTF North America Nebula survey. Date labels denote the beginning of each year. Left: histogram of the number of observations taken in each week of the survey. Right: individual observation times of all survey epochs, with each month dispersed along the vertical axis for clarity. The period of all-night, high-cadence monitoring appears as a set of elongated points in mid-2011. The cadence in 2011-2012 was close to nightly, while observations in 2009 and 2010 were more sporadic.}\label{cadencefig}
\end{figure}

Our survey cadence was complex as a result of changing operational factors. Our observations started in August 2009, continuing with observations every third night until October, when Palomar was shut down due to ash from fires. When we started our 2010 season in April, the cadence was lowered to every fifth night. From August to October 2010, we were able to observe every night, while the remainder of the season was hampered by poor weather. For our 2011 and 2012 seasons, from March 2011 to January 2012 and from March 2012 to December 2012, we were able to observe every night, but only during bright time, as the PTF project had started observing exclusively with the $g^\prime$ filter during dark time. In addition, during July and August 2011, we obtained hourly exposures all night, in both bright and dark time. We illustrate our observing pattern in Figure~\ref{cadencefig}.

Our survey represents one of the most uninterrupted, multiyear optical variability surveys of a star-forming region, featuring 884 epochs across 352 nights between 2009 and 2012. 
Our largest data gaps are the 1-3 month gaps in the winter, when the region is not visible from Palomar, and the two-week gaps during dark time in most months when the $R$ filter is not available. Aside from these regular gaps, we have two years of uninterrupted nightly coverage, except for occasional weather gaps, and another year and a half of lower-cadence data for probing long-term variability.

\subsection{Reduction Pipeline}

All the PTF data for our field was processed by the PTF Photometric Pipeline. Images were debiased, flatfielded, and astrometrically calibrated, with source catalogs generated by SExtractor (Laher et al., in prep). An absolute photometric calibration good to a systematic limit of $\sim2\%$ was generated using SDSS sources observed throughout the night \citep{IPAC}. Relative photometric calibration further refined the photometry, particularly on nonphotometric nights (Levitan et al., in prep; for algorithm details see \citet{relative1} and \citet{relative2}).
The PTF Photometric Pipeline photometry is typically repeatable to 0.5-1\% for bright (15th mag) nonvariable sources in sparse fields on photometric nights. Photometry for typical sources in our field is less reliable, of the order of 2-3\%, because nebula emission and source crowding introduce additional errors.

\label{flaglist}
The pipeline flagged photometric points as bad detections if the sources were automatically identified as part of airplane, satellite, or cosmic-ray tracks; if they fell on a high dark current, unusually noisy, or poorly illuminated area of the chip; if they fell on a chip edge; if they contained dead pixels; if they were affected by bleeding from bright stars; if they contained saturated pixels; or if they had neighbors biasing their photometry. We removed any sources from our sample that were flagged in more than half the epochs, as discussed in the next section, and we removed all flagged epochs from our light curves before plotting or analyzing them.

\subsection{Identifying the Variables}\label{variableselect}

To determine which sources were variable during the observation period, we grouped all PTF detections with $14 \le R \le 20$~mag into half-magnitude bins on a chip-by-chip basis. The width of the bins (0.5 mag) was chosen so that the brightest and least populated bin (14-14.5~mag) had roughly 100 sources on most chips. We then computed the median RMS of all the stars in each bin, and fit the medians by an equation of the form:
\begin{equation}\label{rmsfit}
RMS = \sqrt{a^2 + (b \times 10^{-0.4p(mag-14)})^2}
\end{equation}
This equation is partly motivated as the sum of a systematic term and a flux-dependent term; the exponent of the flux-dependence $p$ was allowed to vary because the natural choice, $p = \frac{1}{2}$ (i.e., noise that scales as the square root of the flux, as expected from photon noise), was too shallow. In practice we found $p \sim \frac{2}{3}$ for most chips. We list the fit parameters in Table~\ref{rmsthresholds}.

The curve found by fitting Equation~\ref{rmsfit} describes the locus of nonvariable stars on a given chip. We defined the boundary between variable and nonvariable stars to be 1.75 times the median RMS. This cutoff was determined empirically, rather than analytically, to avoid making assumptions about the noise properties of the data. We set the cutoff by visually inspecting light curves with both $R \sim 14$ and $R \sim 16$; at RMS values lower than 1.75 times the threshold the light curves were indistinguishable from noise, while at higher values the light curves were clearly structured on short timescales.

\begin{figure}
\includegraphics[width=0.47\textwidth]{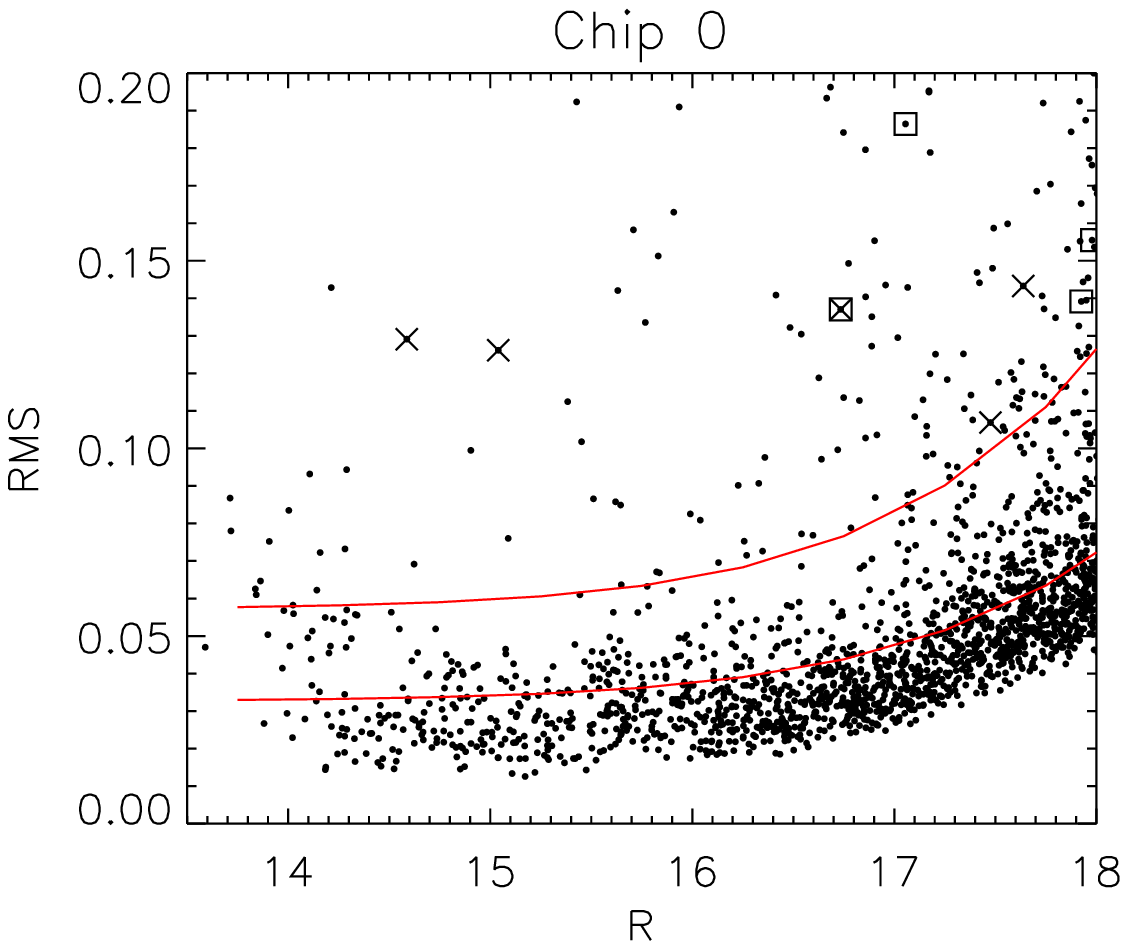}
\includegraphics[width=0.47\textwidth]{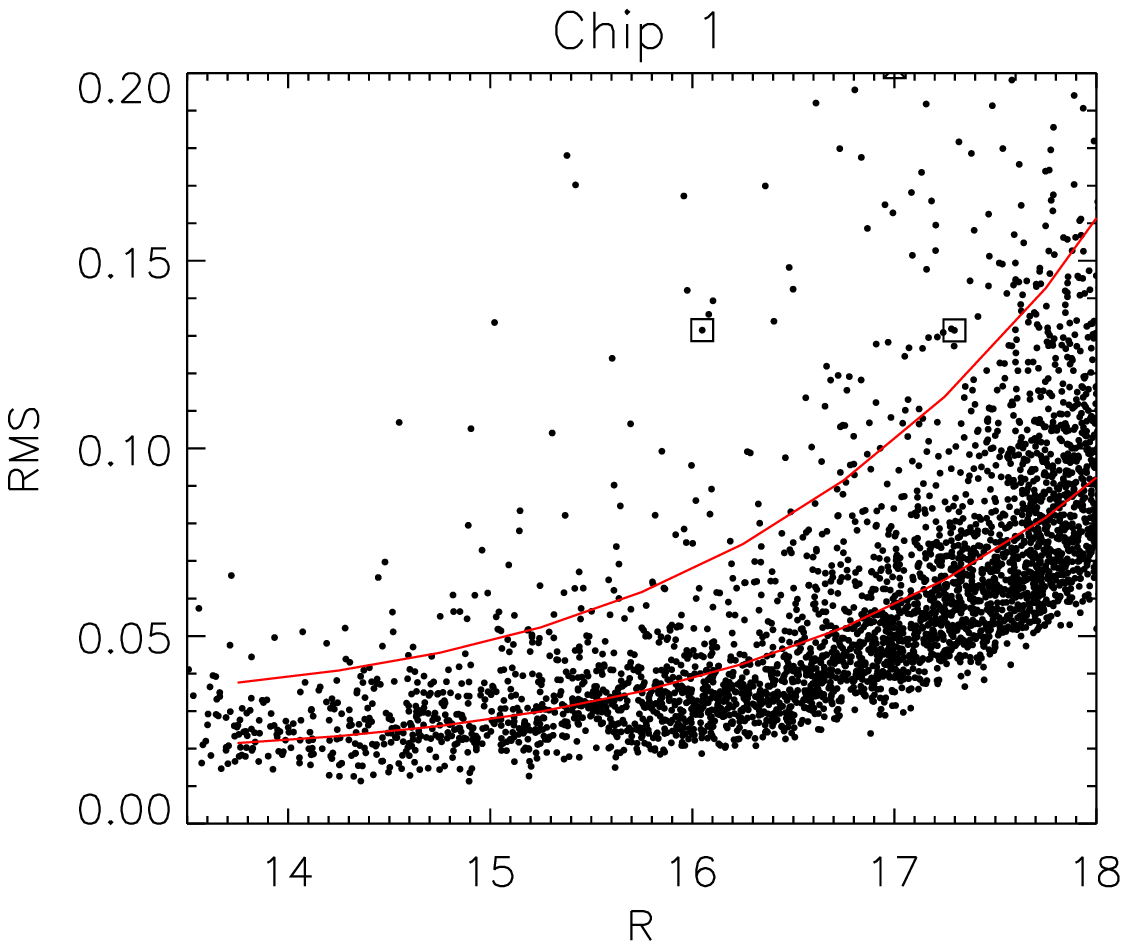}
\includegraphics[width=0.47\textwidth]{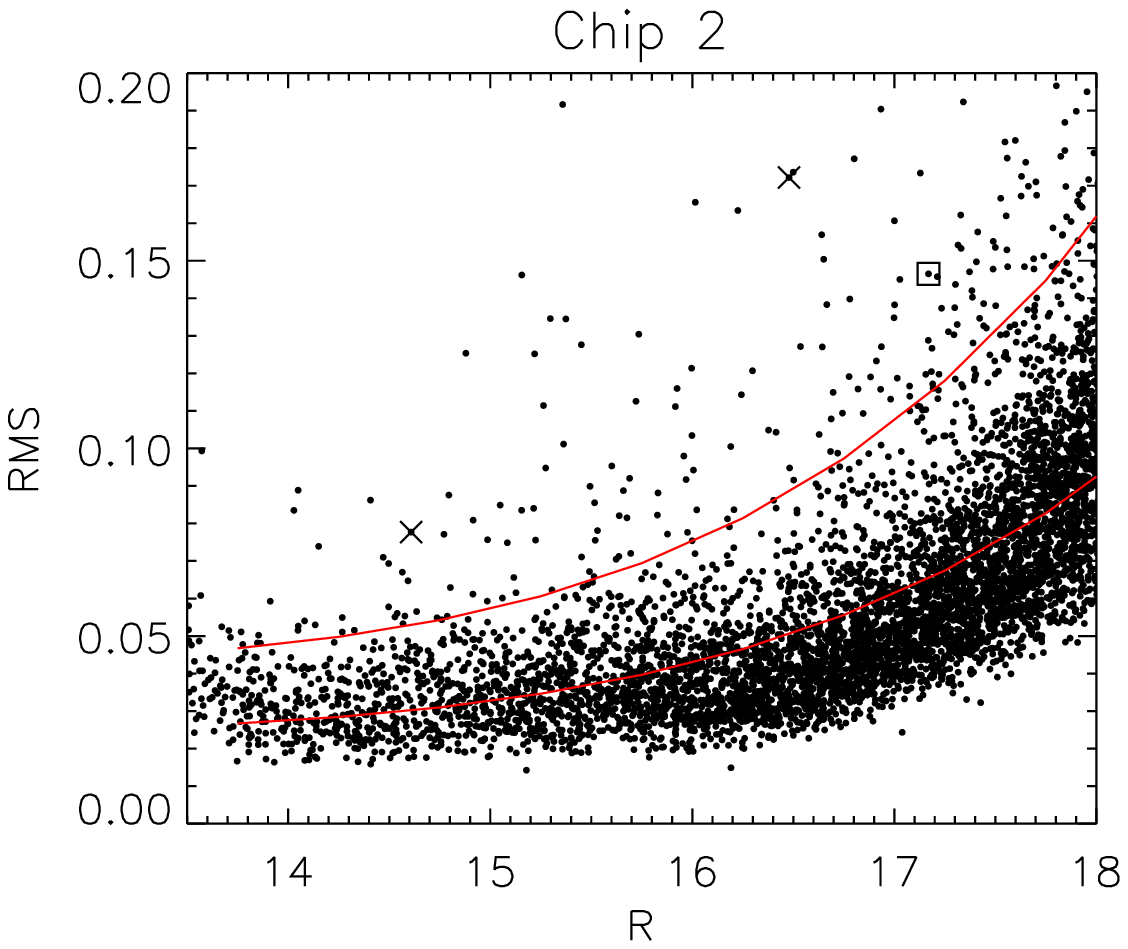}
\includegraphics[width=0.47\textwidth]{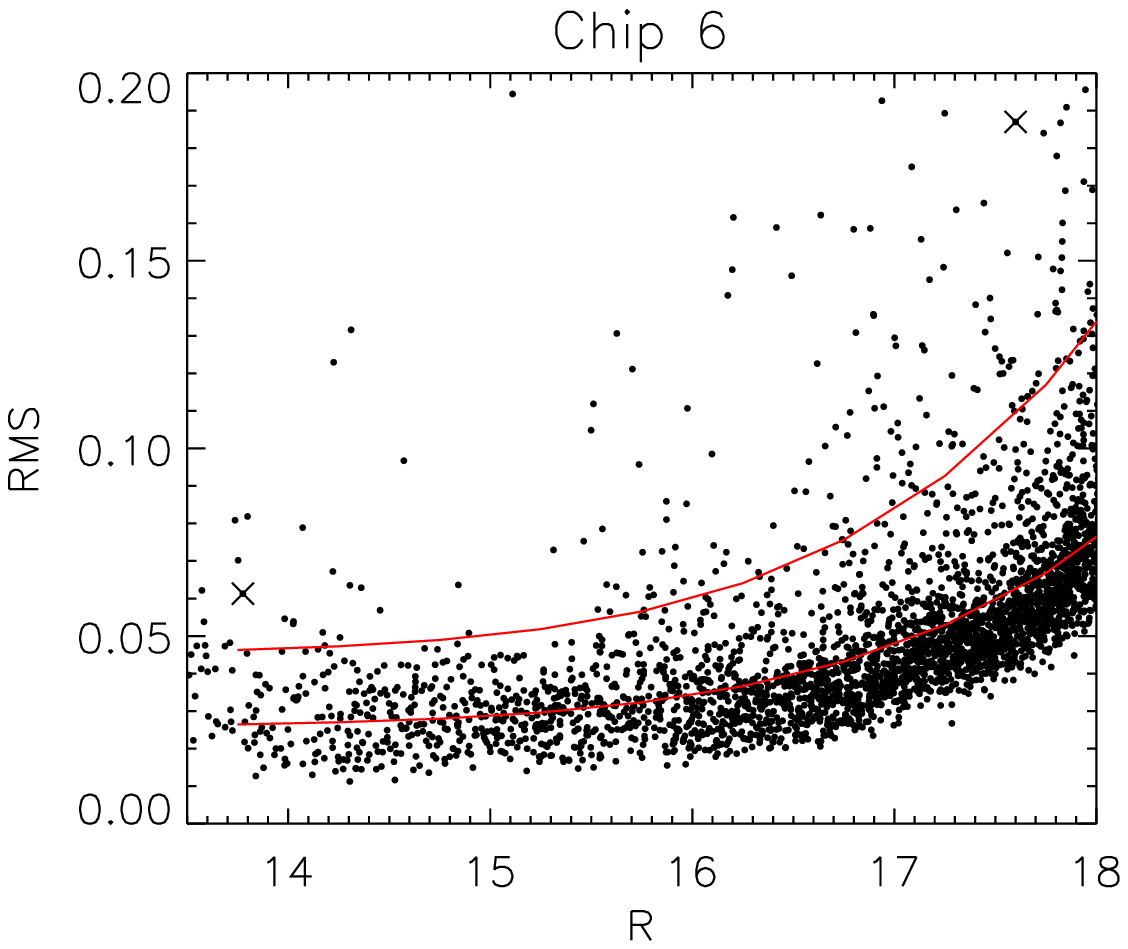}
\includegraphics[width=0.47\textwidth]{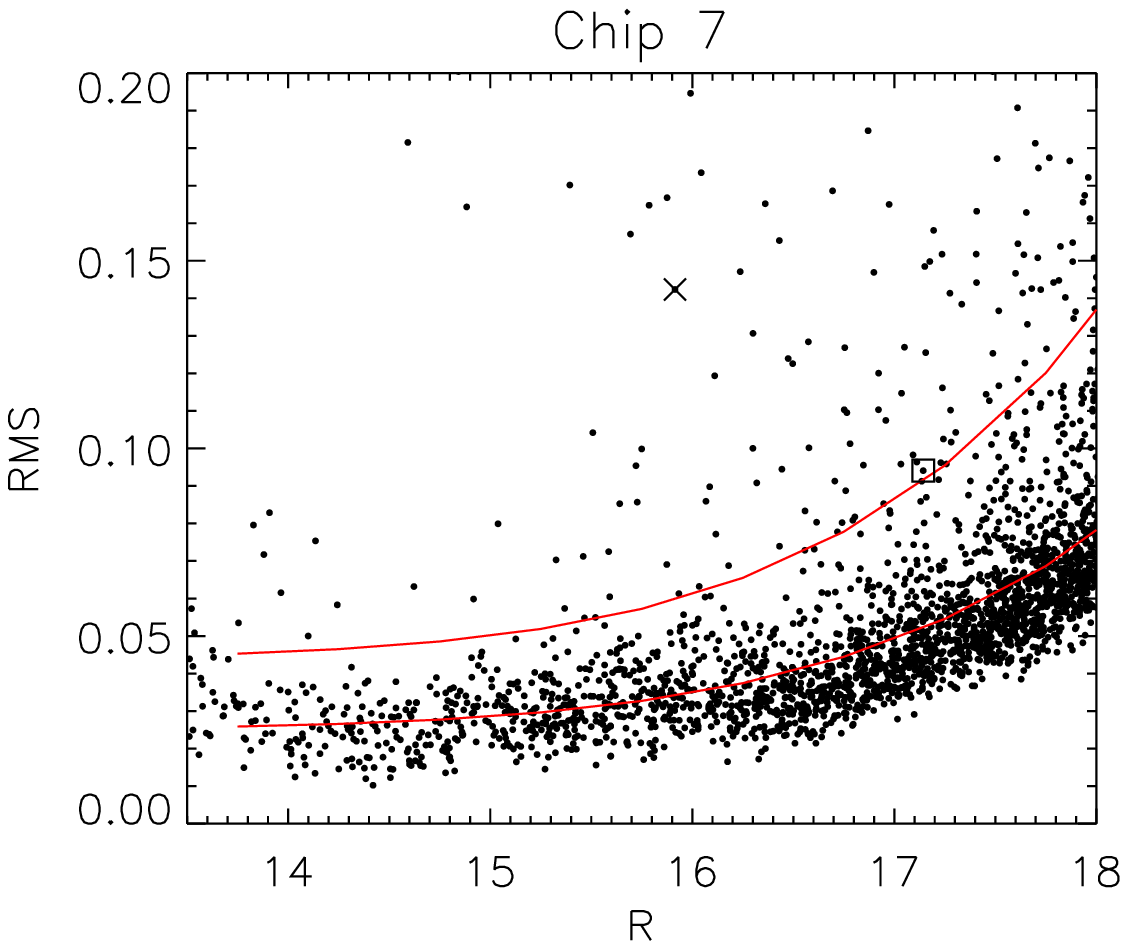}
\qquad\ \ \ \includegraphics[width=0.47\textwidth]{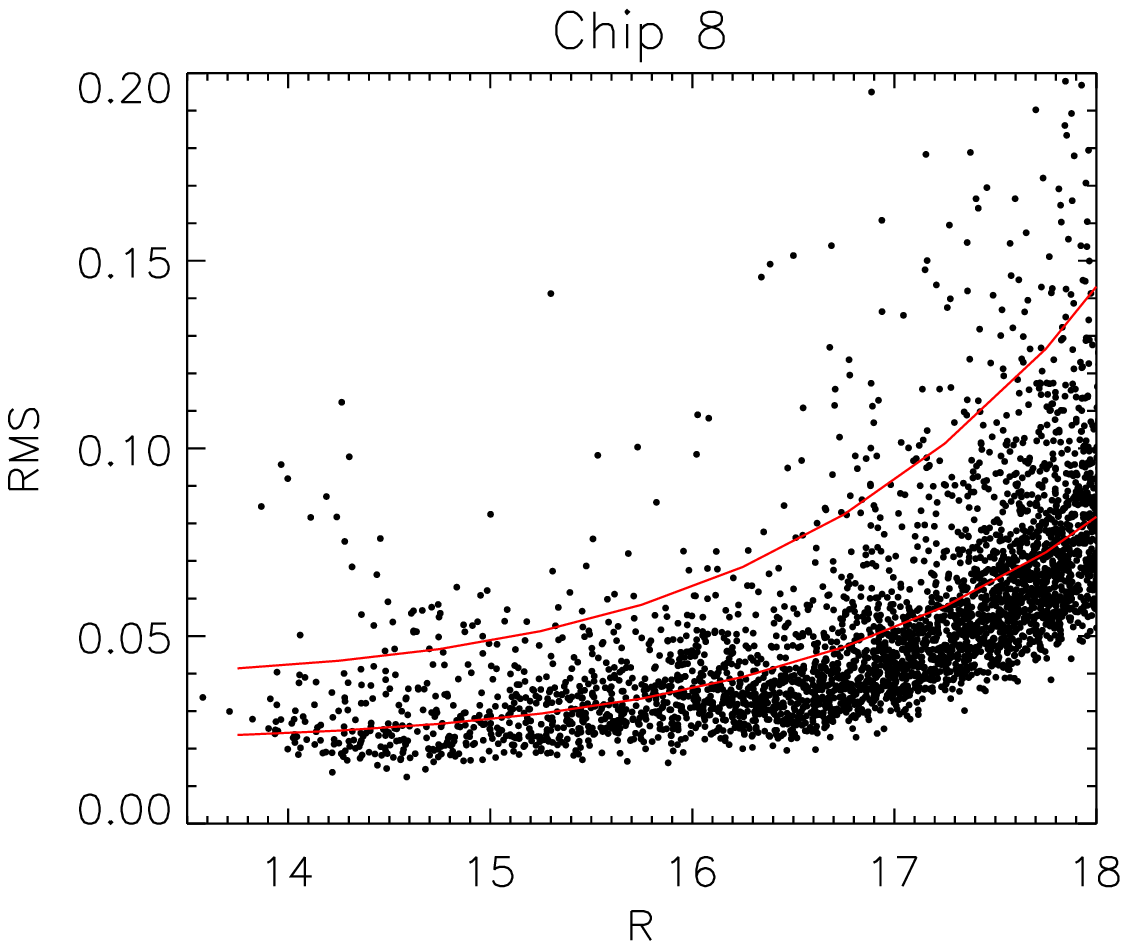}
\caption{RMS scatter vs. median magnitude for all sources with flags (listed in Section~\ref{flaglist}) in fewer than half the epochs. The fit to the median RMS as a function of magnitude is plotted as the lower red curve, while our variability detection threshold (1.75 times the median) is plotted as the upper red curve. As in Figure~\ref{overview}, X's mark candidate bursting stars while squares mark candidate fading stars. 27 high-amplitude variables are beyond the upper edge of the Chip~0 plot, 10 each above the upper edge of the Chip~1 and Chip~2 plots, and 1-5 off the upper edge of each of the others.}\label{rmsvsmag}
\end{figure}

We show in Figure~\ref{rmsvsmag} plots of RMS vs. magnitude for the six chips that covered the star-forming complex, along with the median fit and the variability detection boundary for each chip. For 14th magnitude stars, we are sensitive to variability with an RMS amplitude of a few percent, while below 16th magnitude, we can probe only 10\% flux variations. In Table~\ref{rmsthresholds}, we list the number of PTF sources and the number and fraction identified as photometrically variable using the methods outlined above. 
Nearly 3000 variables projected on the dark cloud and the associated nebulae are identified.  Their RMS amplitudes range from 0.03 to 1.1~mag. 

\section{Bursting and Fading Among Infrared Excess Sources}\label{demo}

\subsection{Sample Selection}

Because the North America Nebula complex is located in the plane of the Galaxy, a significant number of our high quality light curves are those of foreground or background field stars. In our first reconnaissance of the variability properties of the region, we therefore concentrated 
on variable stars among 
a list of candidate North America Nebula members identified by \citet{RGS2011}. 
Specifically, \citeauthor{RGS2011} used infrared colors, primarily Spitzer IRAC $3.6\textrm{~$\mu$m} - \textrm{MIPS  } 24\textrm{~$\mu$m}$, to identify stars surrounded by circumstellar dust. 
Additional 
considerations included location in various color-magnitude diagrams that help distinguish young stars from contaminating dusty sources such as extragalactic AGN and galactic late-type giants. 
Each source was assigned a spectral energy distribution class based on the slope of a linear fit to all available photometry between 2 and 24~$\mu$m. Class~I sources have rising slopes and are interpreted as objects with not only circumstellar disks, but likely more spherically distributed envelopes as well. Flat-spectrum sources have roughly constant $\lambda F_\lambda$ over the 2-24~$\mu$m range and have a similar interpretation. Class~II sources are consistent with traditional disk SEDs. Class~III sources have the steepest slopes; most have no excess in the IRAC bands but were selected based on an excess at 24~$\mu$m. 
Only 6 of the Class~III sources in \citet{RGS2011} were not selected using either IRAC or MIPS excess criteria. 
\citeauthor{RGS2011} note that, since their primary selection is based on infrared data, they are incomplete with respect to Class~III sources.

\label{completeness}
Of the 2082 candidates from \citet{RGS2011}, 601 had a counterpart in the PTF source catalog. 
As we show in Table~\ref{ptfcomplete}, the recovery rate by PTF depended strongly on the type of IR excess. Only 5\% of the relatively red Class~I sources in the PTF field had detections, while fully 93\% of the relatively blue Class~III sources were detected by PTF. The strong correlation between (infrared) source color and recovery rate, in the sense that redder sources are recovered less often, suggests that most of the sources we did not recover in PTF were missed because they were below our optical detection limits. However, we also know, from image inspection, that the PTF pipeline had difficulty identifying and extracting sources from crowded or nebulous regions. 
If we assume that all the Class~III sources must be bright enough to detect in the optical if they are visible in the Spitzer bands even with a small infrared excess, then source extraction problems should dominate the 7\% missing Class~III sources. Presumably, roughly 7\% of the rest of the sample also fell in regions where the PTF pipeline could not reliably identify sources. 
We note that, while the overall incompleteness does not affect our main science goals, the bias away from Class~III sources in the parent sample and the bias away from Class~I sources from cross-matching to PTF do limit our ability to examine how variability properties change with the degree of infrared excess.

From our sample of 601 infrared excess selected candidate members with PTF counterparts, we restricted our attention to the 253 sources brighter than a median $R_\mathrm{PTF} = 18$. The detailed breakdown by SED type is given in Table~\ref{ptfcomplete}. We found from experience that the photometric quality for sources fainter than $R_\mathrm{PTF} \sim 18$ was such that, while we could determine whether a source was variable, we could not consistently assess the structure of the variability. 
Considering only sources whose light curves had bad photometry flags (see Section~\ref{flaglist} for a list) in fewer than half the epochs further reduced the sample to 186 stars, which are shown in Figure~\ref{cmd_cand}. The figure shows no trend with $R_\mathrm{PTF}$ except for more sources at fainter magnitudes, suggesting our magnitude limits avoid any substantial systematics. High-amplitude sources (RMS $\gtrsim$ 0.3-0.4~mag) tend to be associated with strong infrared excess, while low amplitudes are found in both strong- and weak-excess sources.

From this sample of 186, we studied in more detail the 117 that showed significant variability in PTF, as defined in Section~\ref{variableselect}. Both cuts are presented in more detail in Table~\ref{ptfcomplete}. 
These infrared excess selected variables include most of the high amplitude variables in the field, as shown in Figure~\ref{rmshisto}; most of the low amplitude variables in the field lack an infrared excess and are not part of our sample. 
The 117 infrared-excess variable sources, along with the other variables in the field, exhibit a wide range of time scales and amplitudes in their light curves. We sought to categorize the lightcurves and hereafter we focus on those that can be identified as bursting or fading.

\label{definitions}
When selecting sources for inclusion on the list of bursters or faders, we defined a burst in a light curve as a period of elevated fluxes above a (local) floor of relatively constant brightness. We did not place any explicit restriction on the length of the candidate burst. However, we tended to require elevated fluxes in multiple consecutive epochs to be certain that a brighter measurement was not a measurement error, and we required that the period of elevated fluxes be short enough that we could recognize the remainder of the light curve as a well-defined ``quiescent'' state. We defined fades analogously: a period of lowered fluxes, with the caveats that we believed the lower fluxes represented real variability and that the lower fluxes were distinct from the normal variability of the star. Both definitions were necessarily subjective, and we review possible selection effects in Section~\ref{biases}.

We visually inspected all 117 light curves for bursting or fading activity. For comparison, we also inspected 100 randomly chosen variable PTF sources that did \emph{not} have an infrared excess, mixing them with the sample of 117 so that we did not know whether any particular light curve was from the target sample or the control group. We designated a star as a burster or a fader if it had at least one bursting or fading event during the monitoring period.

\subsection{Burster and Fader Statistics}

We identified 14 stars with candidate bursts and 29 stars with candidate fades, with two stars showing both bursting and fading behavior. The sources are listed in Table~\ref{lcstats}, with their photometric behavior summarized in Table~\ref{candidates}. Light curves of all 41 stars are available online from the PTF website\footnote{\url{http://www.astro.caltech.edu/ptf/}}. The sources are also highlighted in Figures~\ref{overview}, \ref{rmsvsmag}, and \ref{cmd_cand}.

For comparison, in the control group of 100 sources with no infrared excess, we saw only two stars that appeared to have one burst each, and no faders other than eclipsing binaries. 
The burst detected in one of the stars turned out to be a transient scattered light artifact we had failed to spot at the time of the original analysis. The other may also have been identified as a burster because of a systematic error in the data or in our visual inspection, or it may represent real astrophysical variability in the field. In the former case we expect $\sim 2$ of the bursters in our target sample to be mislabeled, while in the latter we expect $\sim 1$ false positives.

The stars listed in Tables~\ref{lcstats} and \ref{candidates}, some of which are highlighted in Figures~\ref{flaregallery} and \ref{dipgallery}, show a wide variety of behaviors. We see variability from a few tenths of a magnitude to several magnitudes. The bursts or fades last anywhere from around a day, the shortest timescale resolvable in most of our data, to hundreds of days. Events may repeat as frequently as once a week, or can appear only once in the three-year monitoring period. Nearly all the bursters and faders are aperiodic, with the exception of two faders that are discussed further in Section~\ref{dipperproperties}.

\subsection{Spectroscopic Characterization}

We pursued optical spectroscopy of both the variable star selected sample (this paper) and the infrared-excess selected sample of \citet{RGS2011} using the MMT, Keck Observatory, Palomar Observatory, and Kitt Peak National Observatory.

We observed 164 variable infrared-excess sources in the North America Nebula using the DEIMOS multi-object spectrograph \citep{deimos} at Keck on 2012 July 18-19, using the 600 line/mm grating. The sample included 19 bursters or faders. PTF was monitoring the field during both nights that spectra were taken, allowing us to determine the photometric state represented by each spectrum for all stars except those varying significantly in less than a day.

The spectra were reduced using a modified version of the DEEP2 pipeline \citep{deep2_text,deep2_code}, provided to us by Evan Kirby. The spectra were bias-corrected, dome-flatfielded, and lamp-calibrated, but were not flux-calibrated. We corrected for sky and nebula emission by subtracting the off-source spectrum visible within each slit. The final spectra covered approximately the 4400-9500~\AA\ range at 5~\AA\ resolution, although the range covered by the spectrum of any particular star could shift by $\sim 500$~\AA\ in either direction depending on the position of the star's slit on the instrument mask. 
Many cosmic rays were left uncorrected by the pipeline, so when making the figures in this paper, we cleaned the cosmic rays by hand for clarity. 

194 sources selected by either variability or infrared excess were observed using the Hectospec multi-object spectrograph \citep{Hectospec} on the MMT on 2012 November 3, December 4, and December 6, using the 270 lines/mm grating. The sample included 22 bursters or faders. The data were pipeline processed at the Harvard-Smithsonian Center for Astrophysics \citep{Hecto_pipeline}. The final spectra cover 3700-9100~\AA\ at 6~\AA\ resolution. PTF observed the region on November 3 and December 6, but to interpret the December 4 spectra we had to interpolate between photometry from December 3 and December 6.

One of the authors (L.~H.) had previously obtained low-resolution optical spectra of sources in the North America Nebula complex with the HYDRA multi-object spectrograph \citep{hydra} on the 3.5m WIYN telescope at Kitt Peak, using the 316 line/mm grating, on six nights between 1998 June 2 and 1998 July 21. L.~H. also took spectra using the (now decommissioned) Norris multi-object spectrograph \citep{norris} on the 5m Hale Telescope at Palomar, using the 600 line/mm grating, on 1998 August 14-15, 1999 July 21-23, and 1999 September 2-5. We took spectra of 27 bursters or faders during these runs. The HYDRA and Norris spectra do not have concurrent photometry.

The HYDRA and Norris observations were reduced for us using custom routines written in IDL. The routines applied corrections for bias, dome flats, cosmic rays, scattered light, and wavelength calibration. The spectra were not flux-calibrated. Sky and nebula emission were corrected by taking a shorter sky exposure offset 6-10\arcsec\ from the target position, and subtracting the counts in the sky exposure from the corresponding source spectrum, after scaling to the difference in observing times. In several configurations the sky emission was scaled by an additional 10-20\% to account for changes in the sky transmission. The HYDRA spectra covered 5000-10000~\AA\ at $R \sim 1500$, while the Norris spectra covered 6100-8750~\AA\ at $R \sim 2000$.

\section{The Burster Phenomenon}\label{bursters}

\subsection{Population Properties}

Upward excursions in young star light curves traditionally have been assumed to come from one of two mechanisms. Long-lasting, several-magnitude events in young stars with circumstellar accretion disks (e.g., EX~Lup, FU~Ori) are interpreted as dramatic increases in the accretion rate from the disk to the star. Events lasting a few hours or less and rising by at most a few tenths of a magnitude, particularly in disk-free stars, have been assumed to be associated with magnetic flares like those seen on the young field star UV~Cet or on the Sun. White light flares on the Sun last tens of minutes, while those on M dwarfs last up to several hours \citep[e.g.,][]{2010ApJ...714L..98K, 2011A&A...530A..84K}. As these timescales are set by the cooling times of dense chromospheric material at the base of the coronal loop, it is unlikely that magnetic flares can produce optical emission with much longer durations than observed. 

Of the 14 stars that show bursting behavior, only two, [OSP2002]~BRC~31~8 and FHO~1, have bursts lasting 1-2 hours, short enough to be plausible flares. The remainder must be driven by temporary increases in accretion, drops in extinction, or some other phenomenon. The bursters show a wide variety of behaviors. None are strictly periodic, although [OSP2002]~BRC~31~8 and FHO~29 do show enhanced photometric activity at roughly 300-day intervals. Some bursters, like FHO~26, repeat every few weeks. Others, like LkH$\alpha$~185 or FHO~4, show bursts only once a year or even more rarely. While [OSP2002]~BRC~31~8 and FHO~1 have very short bursts, too brief to resolve outside our highest-cadence monitoring in mid-2011, FHO~17 featured a burst lasting over 100 days, and FHO~18 showed bursts with a range of lengths from a few days to two weeks.

Despite their variety, the bursters do not fall naturally into distinct subclasses, forming instead a continuum of behaviors. We show in Figure~\ref{flarewidths} the joint distributions of burst amplitudes, burst widths, and burst separations for all 14 bursters. To avoid systematics associated with separating a burst or fade from the surrounding, sometimes complex, variability, and to avoid complications from varying sampling from event to event, the timescales and amplitudes in Figure~\ref{flarewidths} were estimated by eye and should be taken as illustrative values only. 
There is no pattern visible in the plot aside from a rough trend where longer bursts tend to be separated by longer intervals. 
The absence of distinct groups of bursters suggests that the diversity of sources can be explained by continuously varying the parameters of a single common scenario, rather than by invoking different mechanisms or different configurations for short- and long-timescale bursters. 

\begin{figure}
\includegraphics[width=0.47\textwidth]{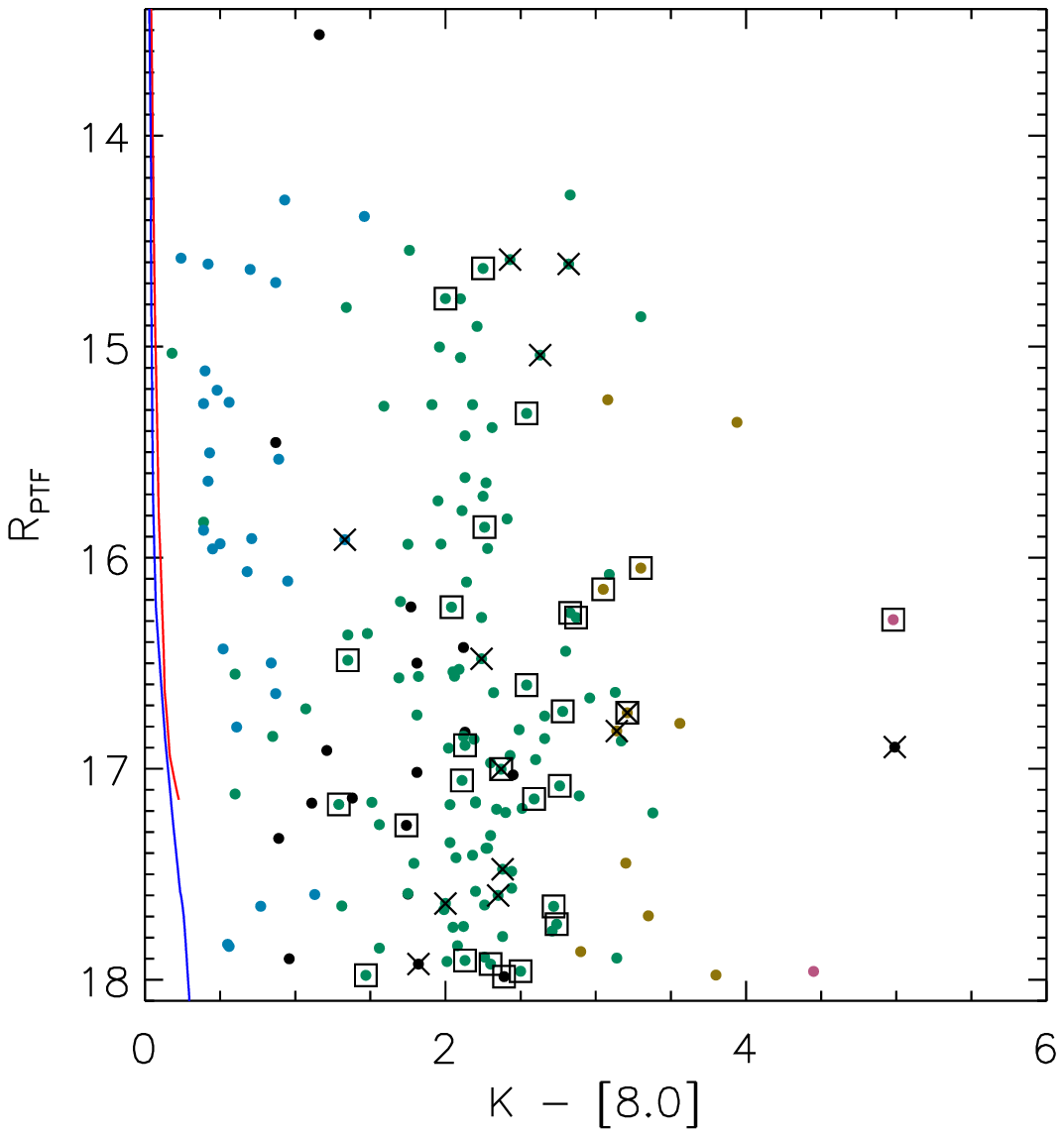}
\includegraphics[width=0.47\textwidth]{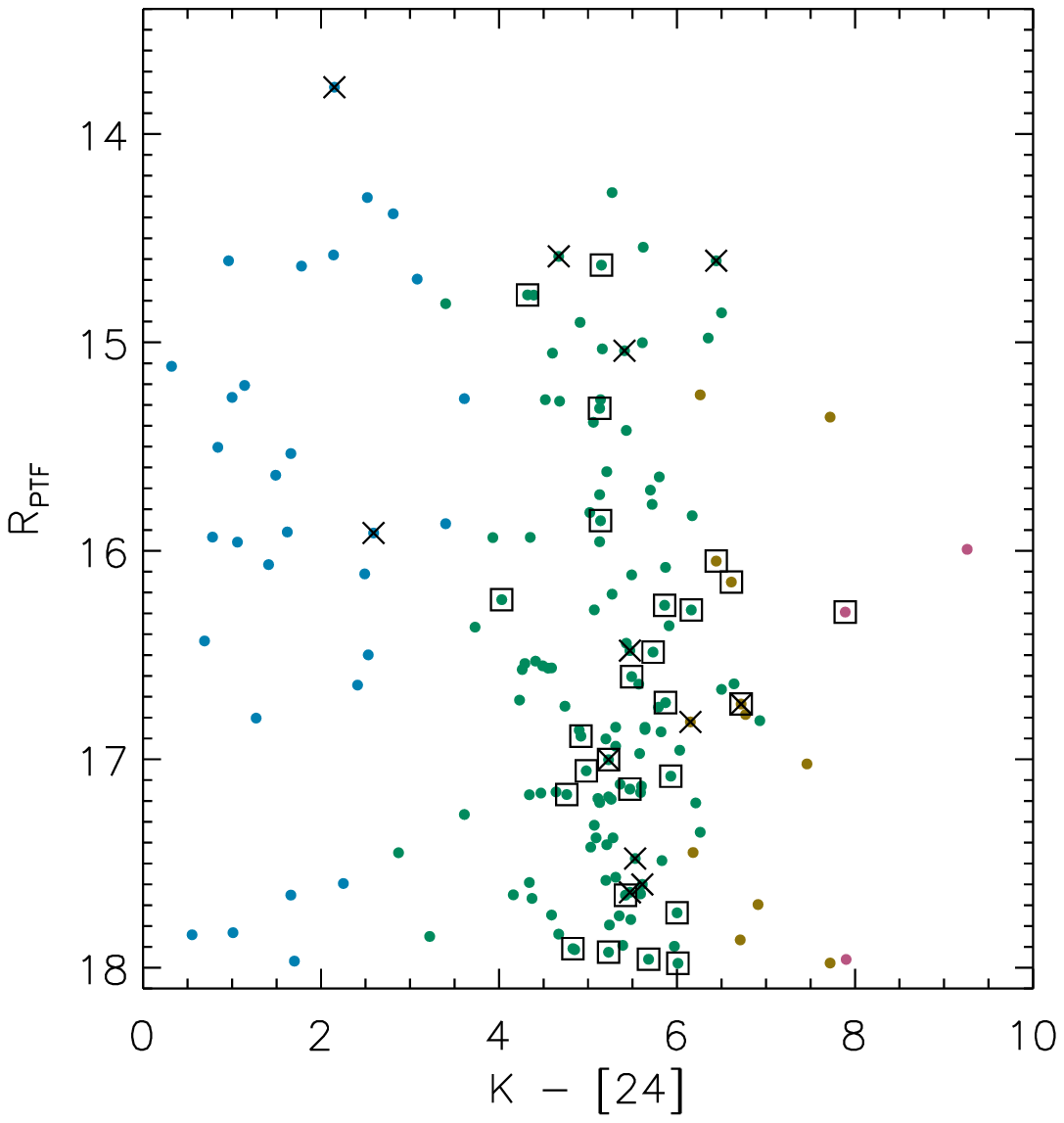}
\includegraphics[width=0.47\textwidth]{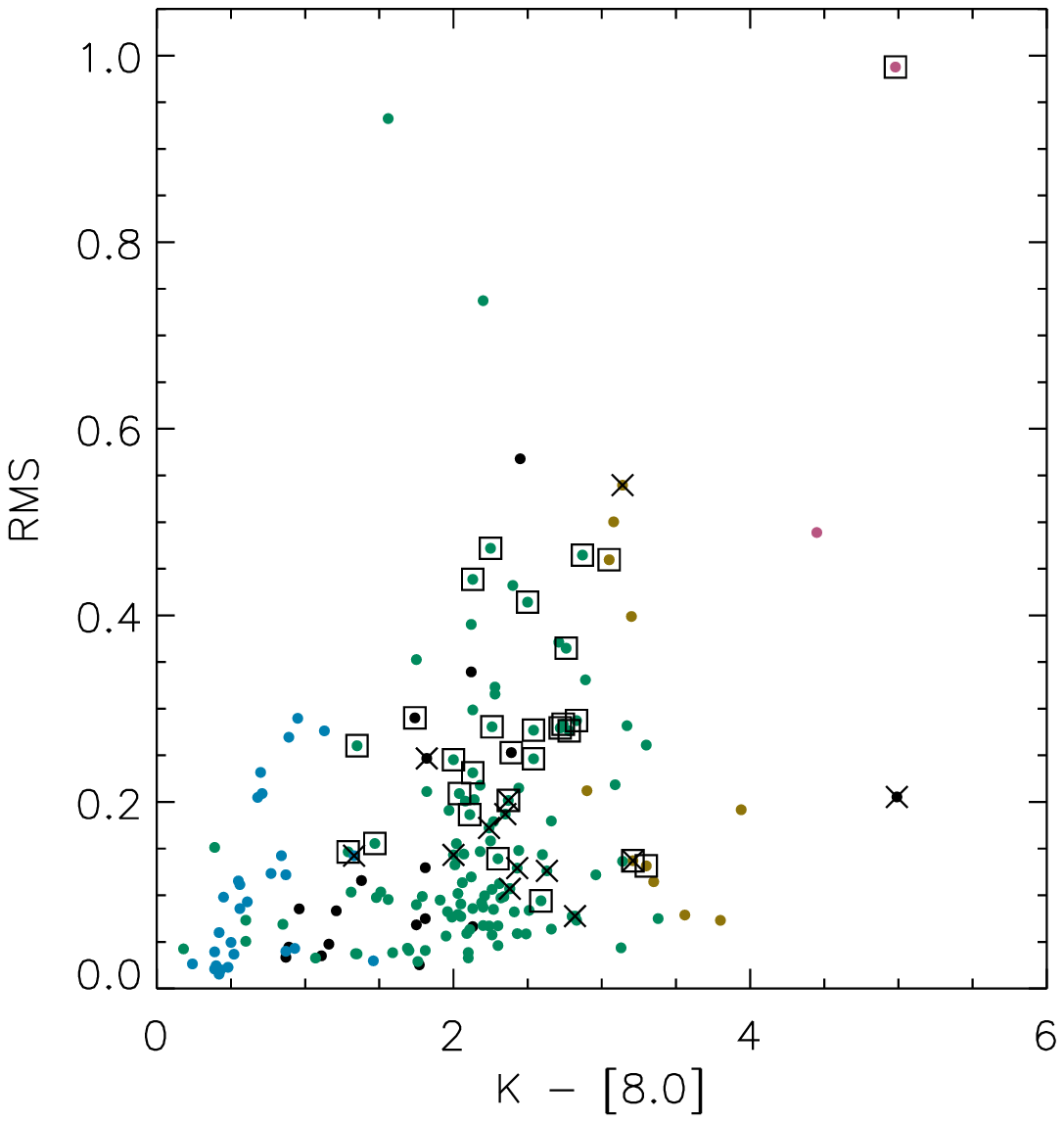}
\qquad\ \ \ \includegraphics[width=0.47\textwidth]{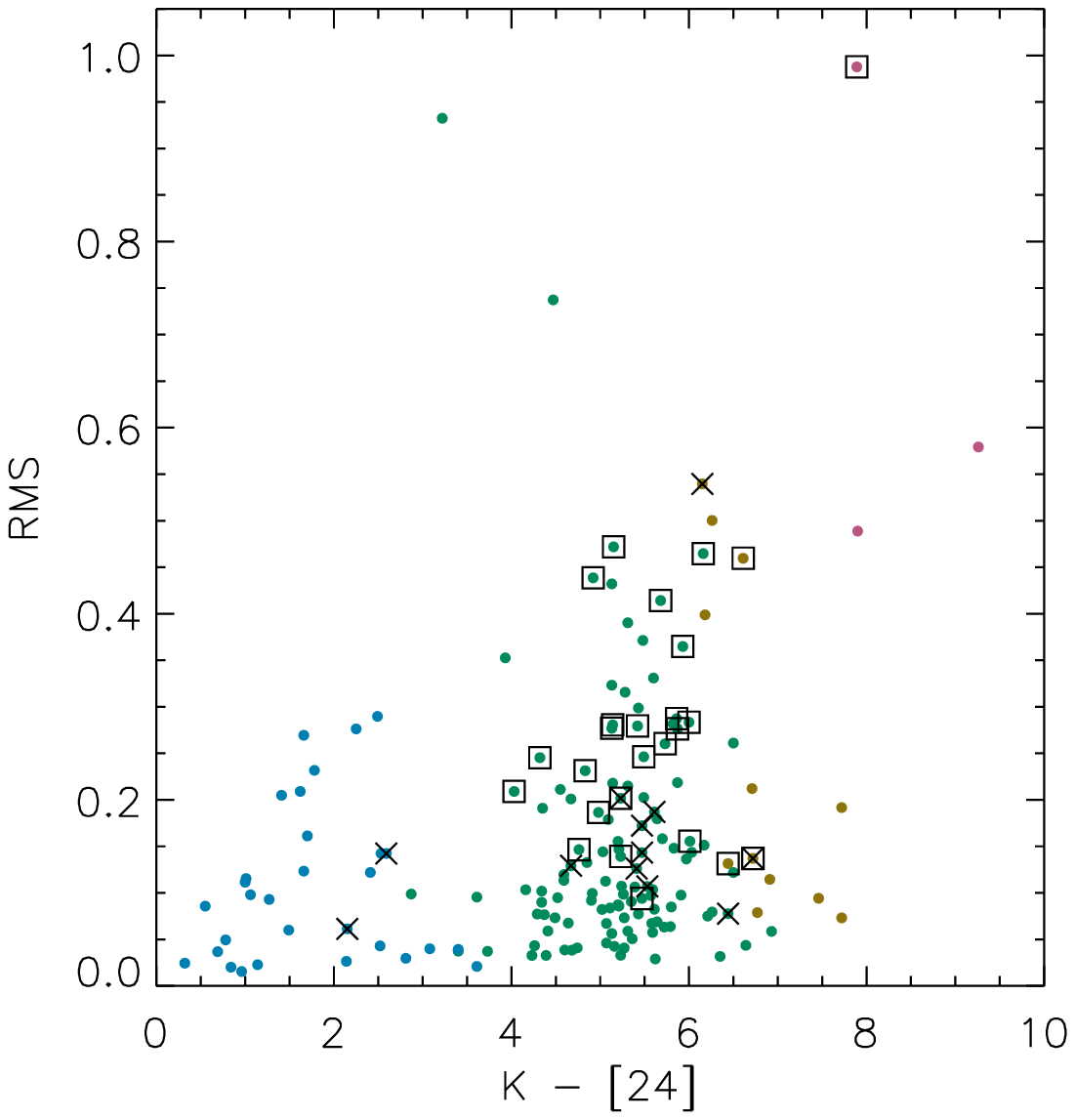}
\caption{PTF magnitude and IR color distributions for those PTF sources that have an infrared excess from \citet{RGS2011} and whose light curves have flags (listed in Section~\ref{flaglist}) in fewer than half the epochs. The color of the dots indicates the degree of infrared excess: blue dots are class~III sources, green ones class~II, yellow ones have a flat IR spectrum, while magenta sources are class~I sources. The black sources are those that were not detected in the Spitzer~24$\mu$m band, and so did not have an IR~excess class listed in \citet{RGS2011}. Not all sources appear on both plots, as some had missing 8$\mu$m or 24$\mu$m photometry. The curves in the upper left panel show synthetic photometry of \citet{SiessModels} isochrones for ages of 2~Myr (red) and 100~Myr (blue), at a distance of 600 pc, indicating the expected colors of stars with no infrared excess at all. 
As in Figures~\ref{overview} and \ref{rmsvsmag}, X's mark candidate bursting stars while squares mark candidate fading stars.}\label{cmd_cand}
\end{figure}

\begin{figure}
\includegraphics[height=0.47\textwidth]{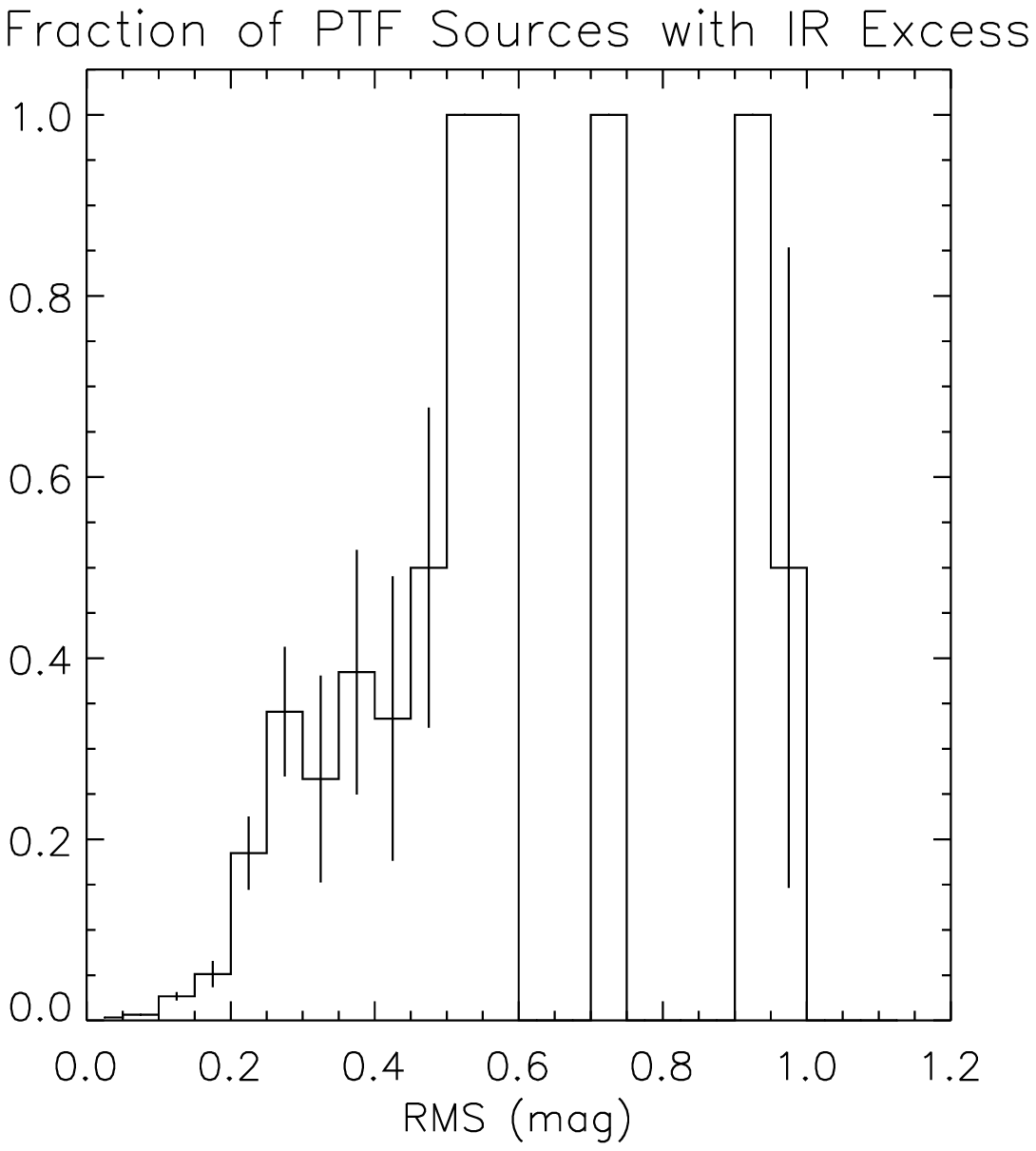}
\includegraphics[height=0.47\textwidth]{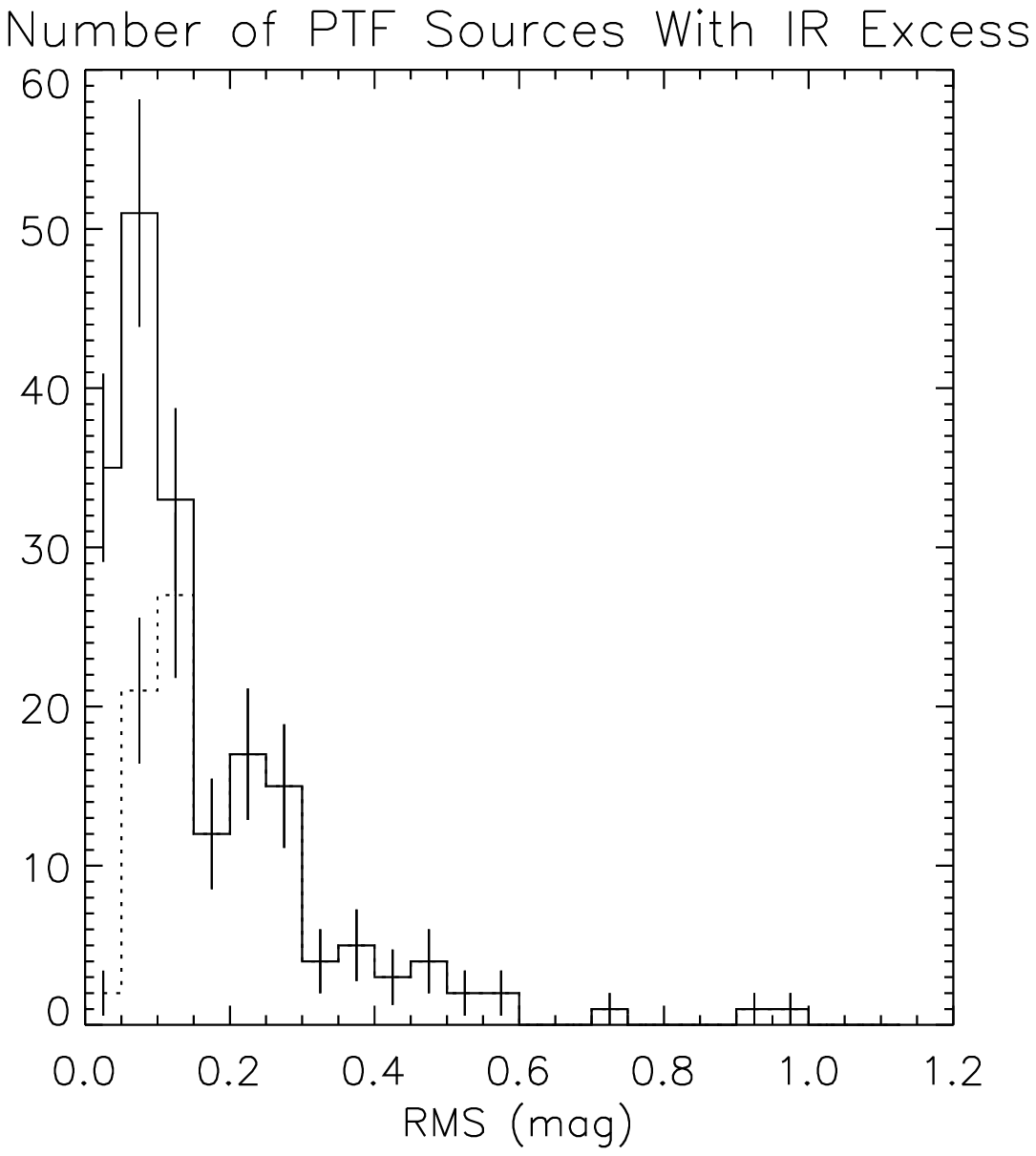}
\caption{RMS distribution of the sample, and its correlation with the presence of Spitzer infrared excess. 
Left: the fraction of sources with an infrared excess out of all PTF sources with $13.5 \le R \le 18$ and flags (listed in Section~\ref{flaglist}) in fewer than half the epochs. 
Right: the solid line denotes the distribution of RMS amplitudes for the 186 infrared excess sources from \citet{RGS2011} whose light curves have flags in fewer than half the epochs. The dashed line represents the subset of 117 that lie above the variability thresholds in Figure~\ref{rmsvsmag}.}\label{rmshisto}
\end{figure}

\begin{figure}
\includegraphics[width=0.47\textwidth]{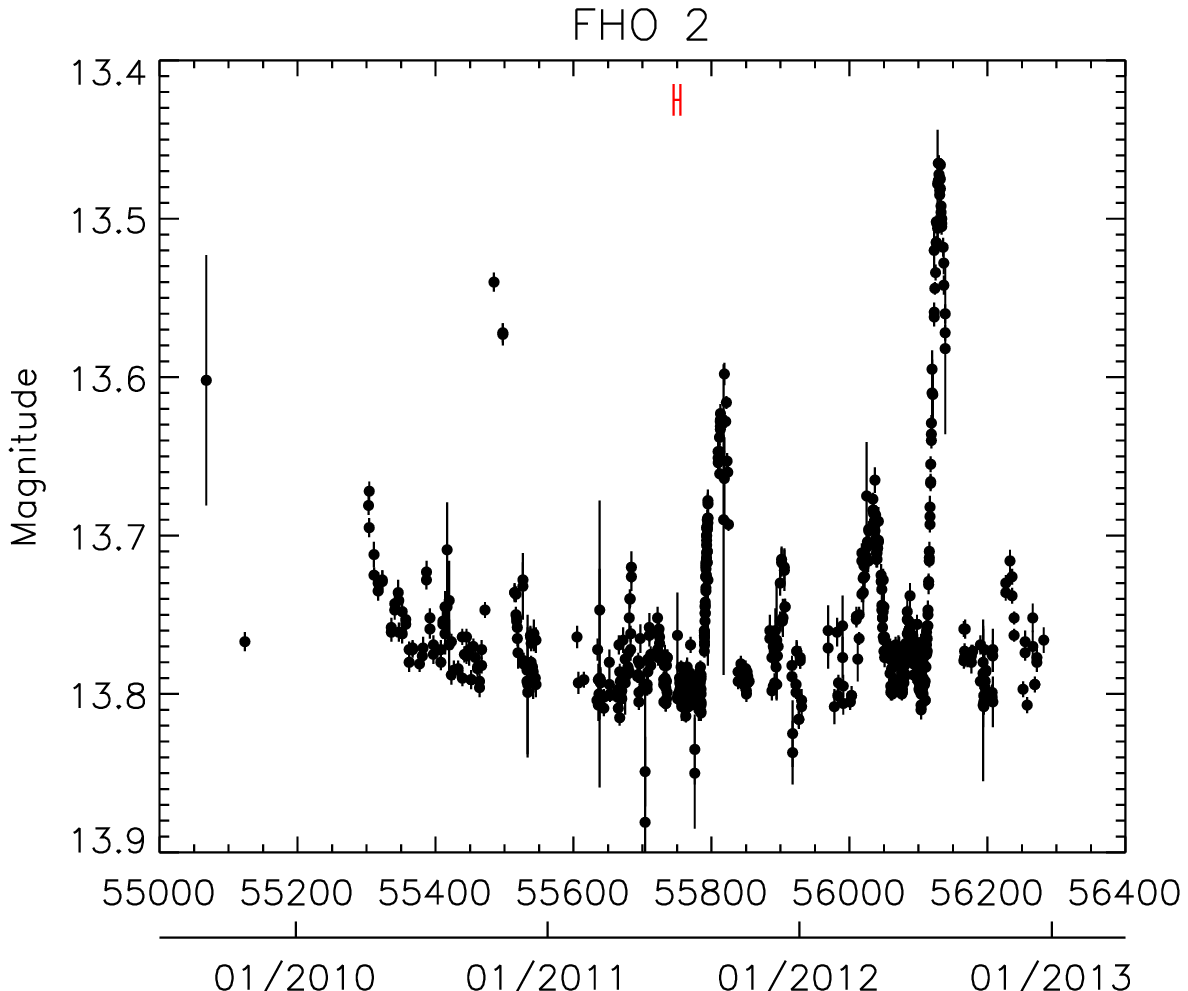}
\vspace{8pt}
\includegraphics[width=0.47\textwidth]{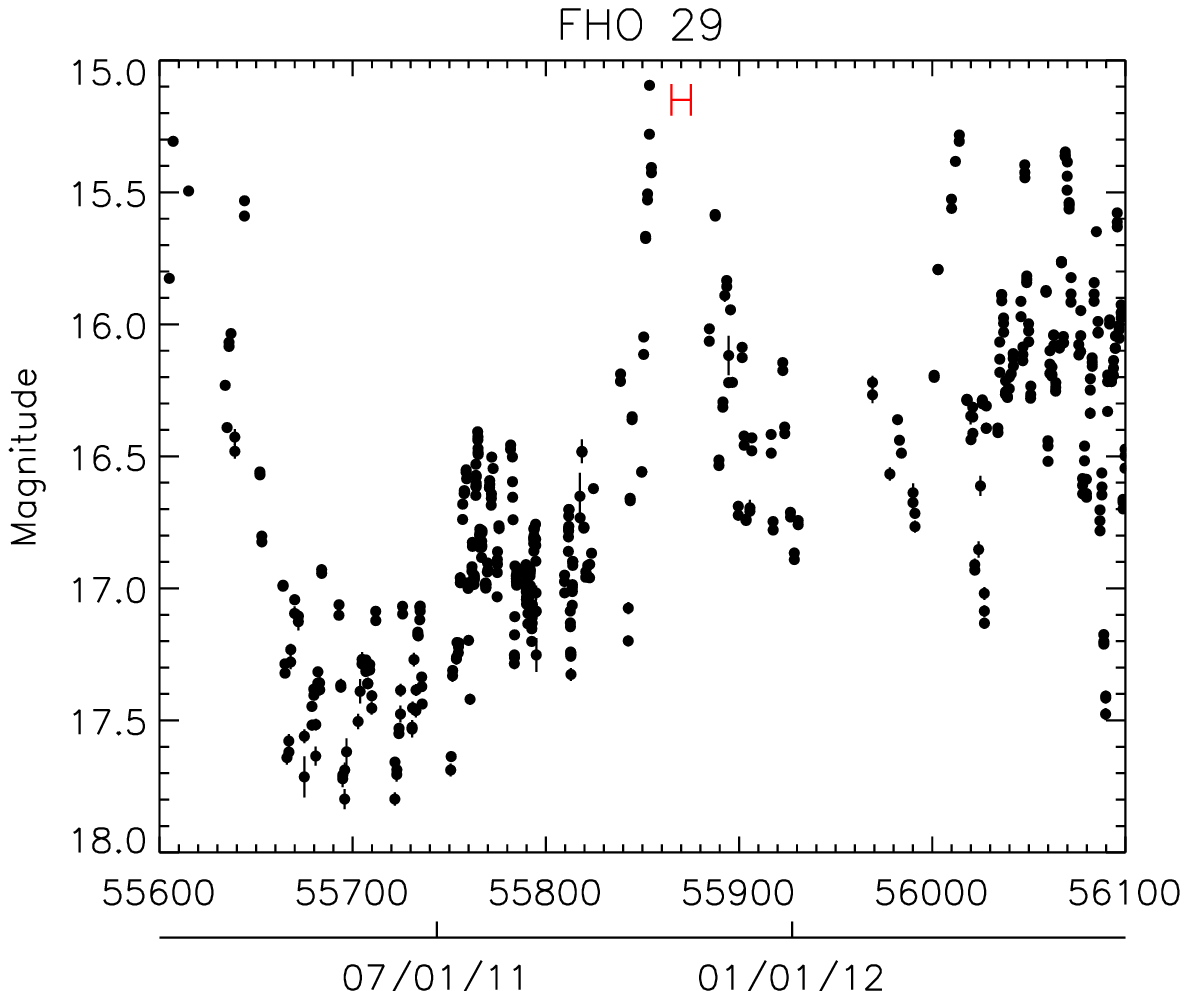}
\includegraphics[width=0.47\textwidth]{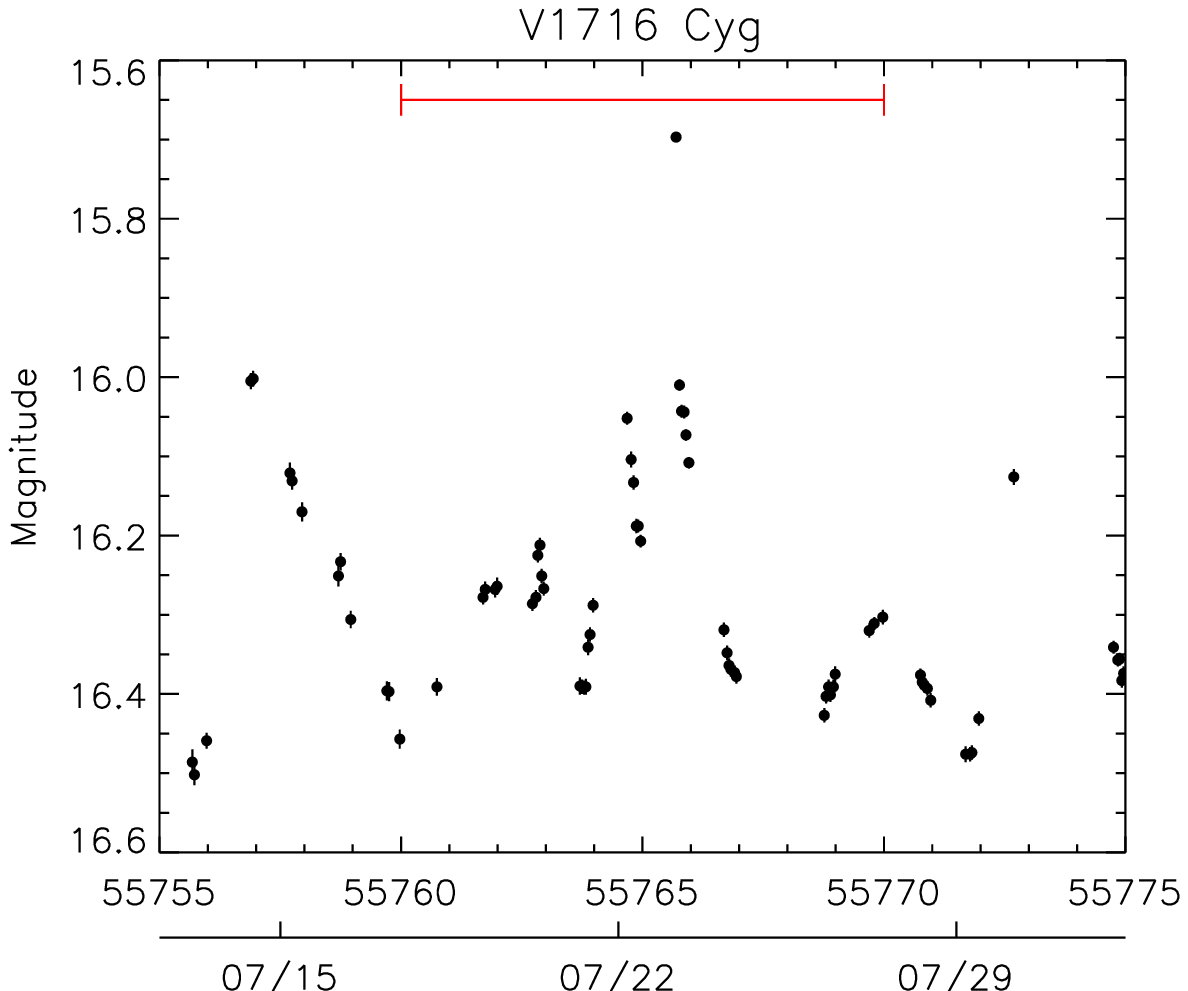}
\vspace{8pt}
\includegraphics[width=0.47\textwidth]{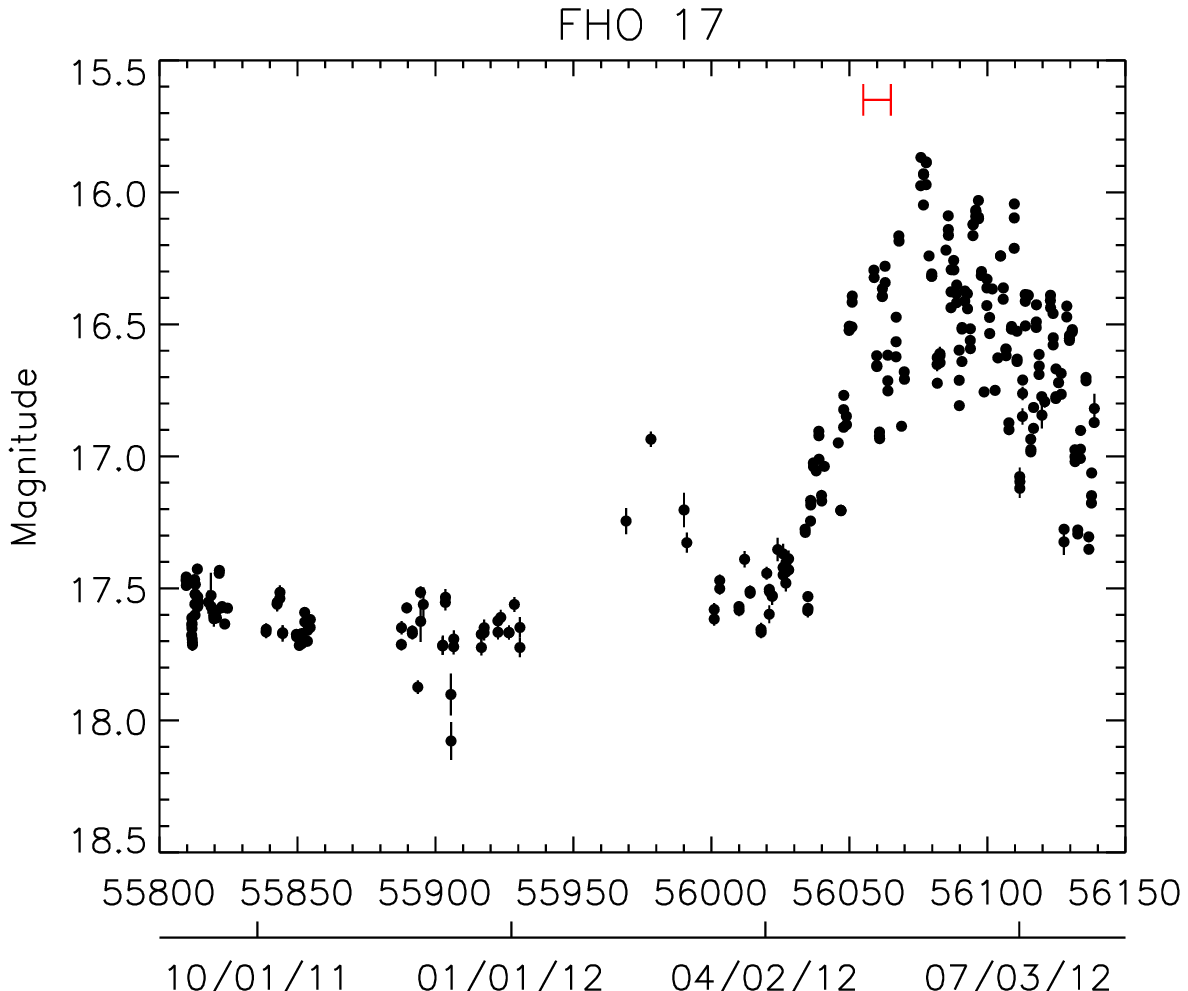}
\includegraphics[width=0.47\textwidth]{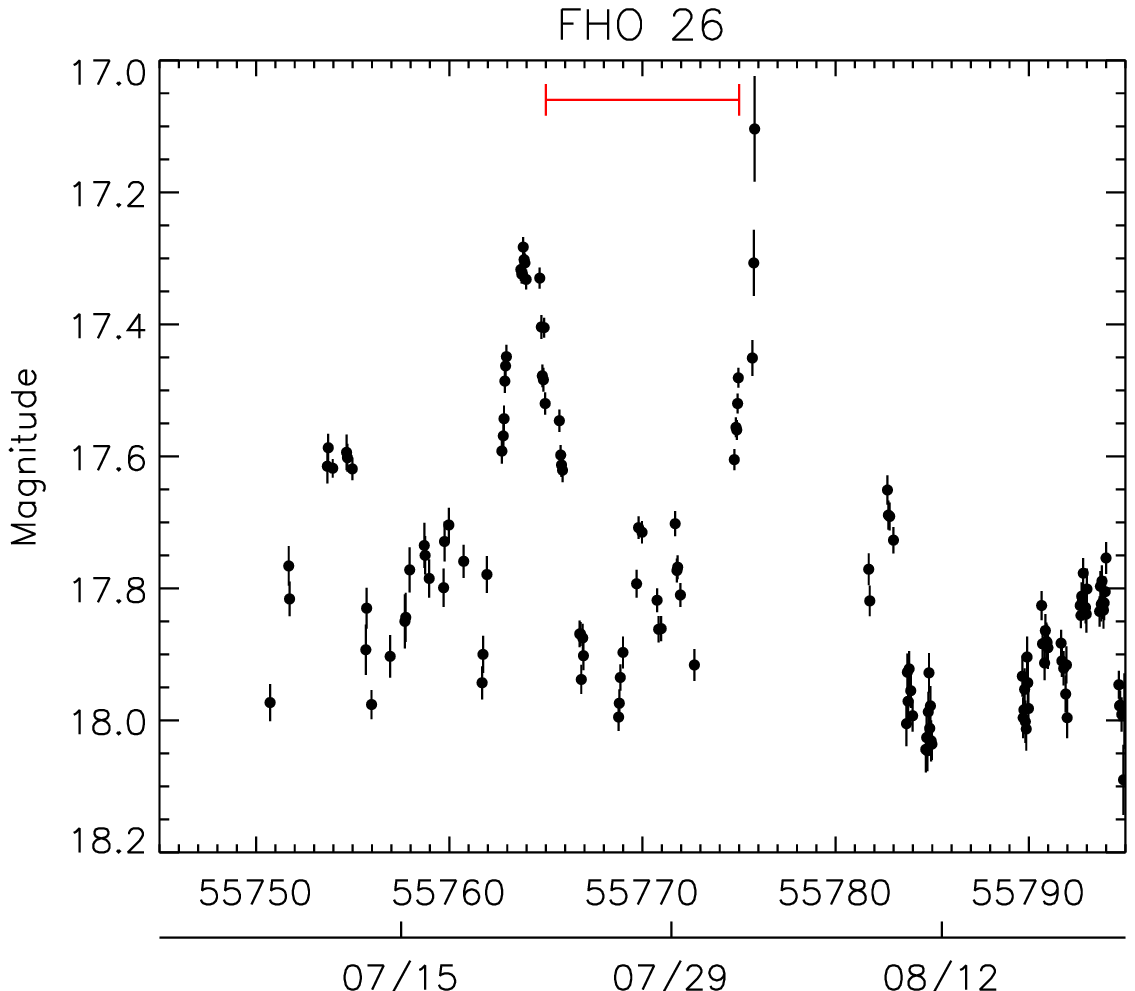}
\qquad\ \ \ \includegraphics[width=0.47\textwidth]{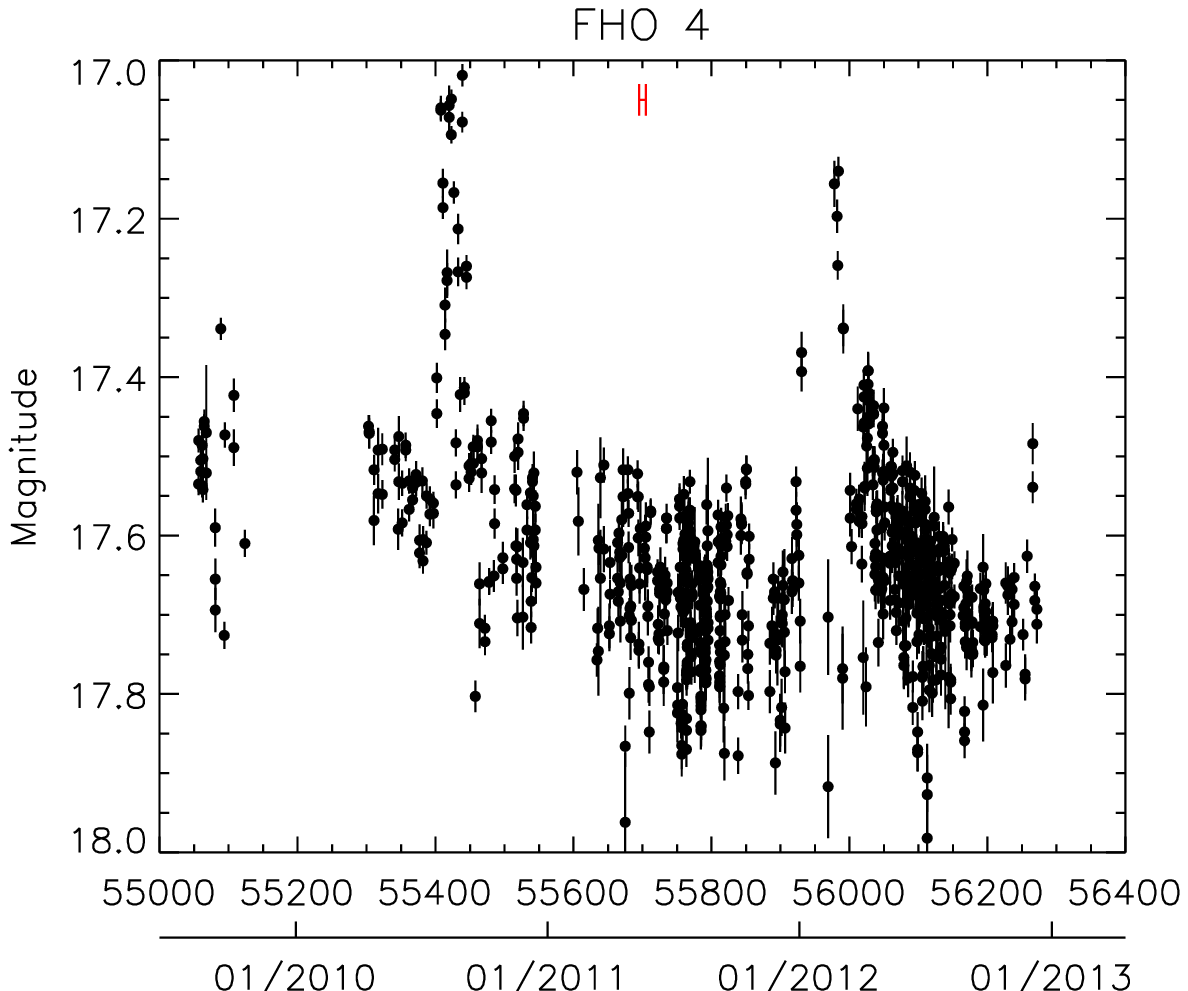}
\caption{Examples of the diverse behavior seen in our lightcurves for bursting stars. Variability amplitudes range from a few tenths of a magnitude to 2 magnitudes (top row). Detected bursts can last from less than two days to over a hundred (middle row), and can be separated by anywhere from 10~days to a year (bottom row). For scale, the horizontal bar near the top of each panel shows a 10~day interval. No points having any of the flags listed in Section~\ref{flaglist} are plotted.}\label{flaregallery}
\end{figure}

\begin{figure}
\includegraphics[width=0.47\textwidth]{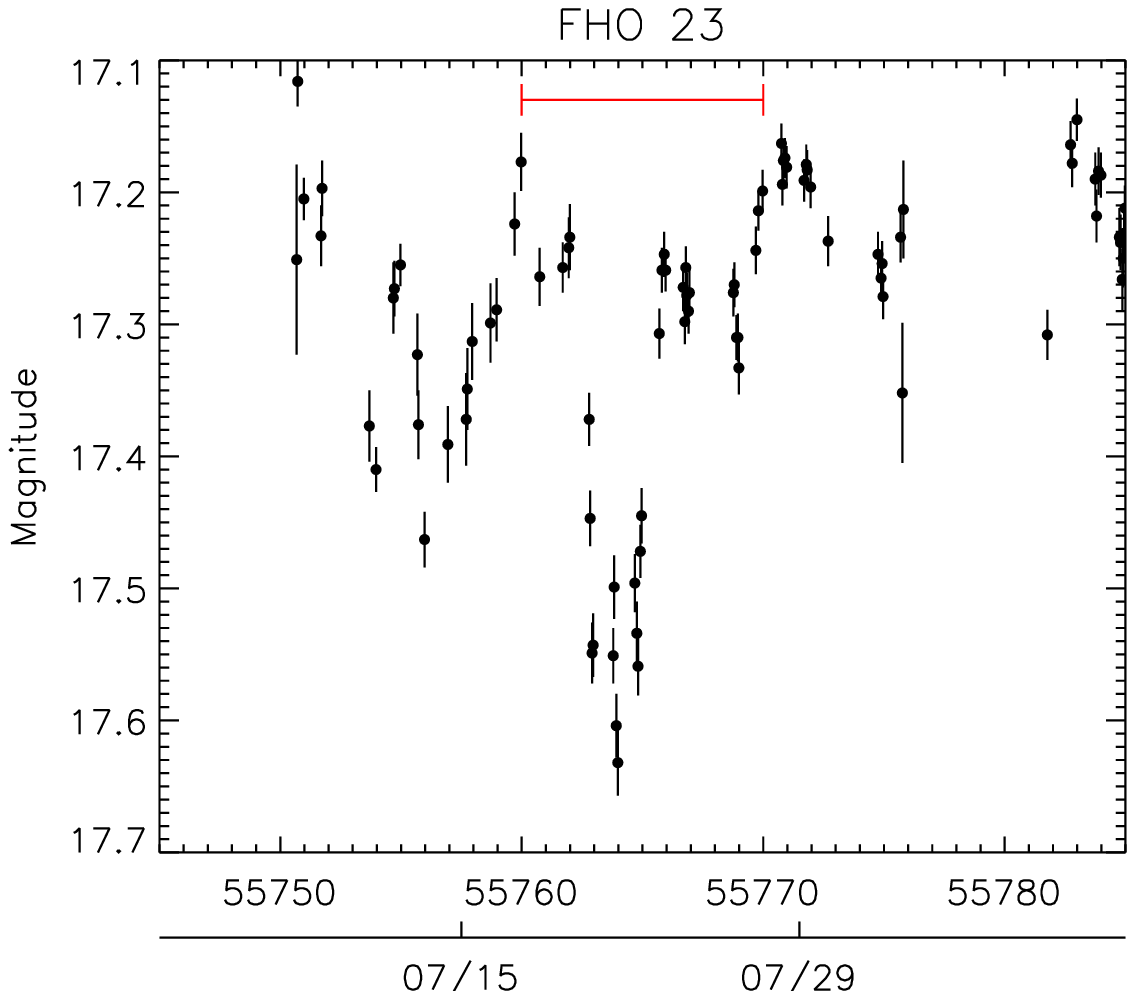}
\vspace{8pt}
\includegraphics[width=0.47\textwidth]{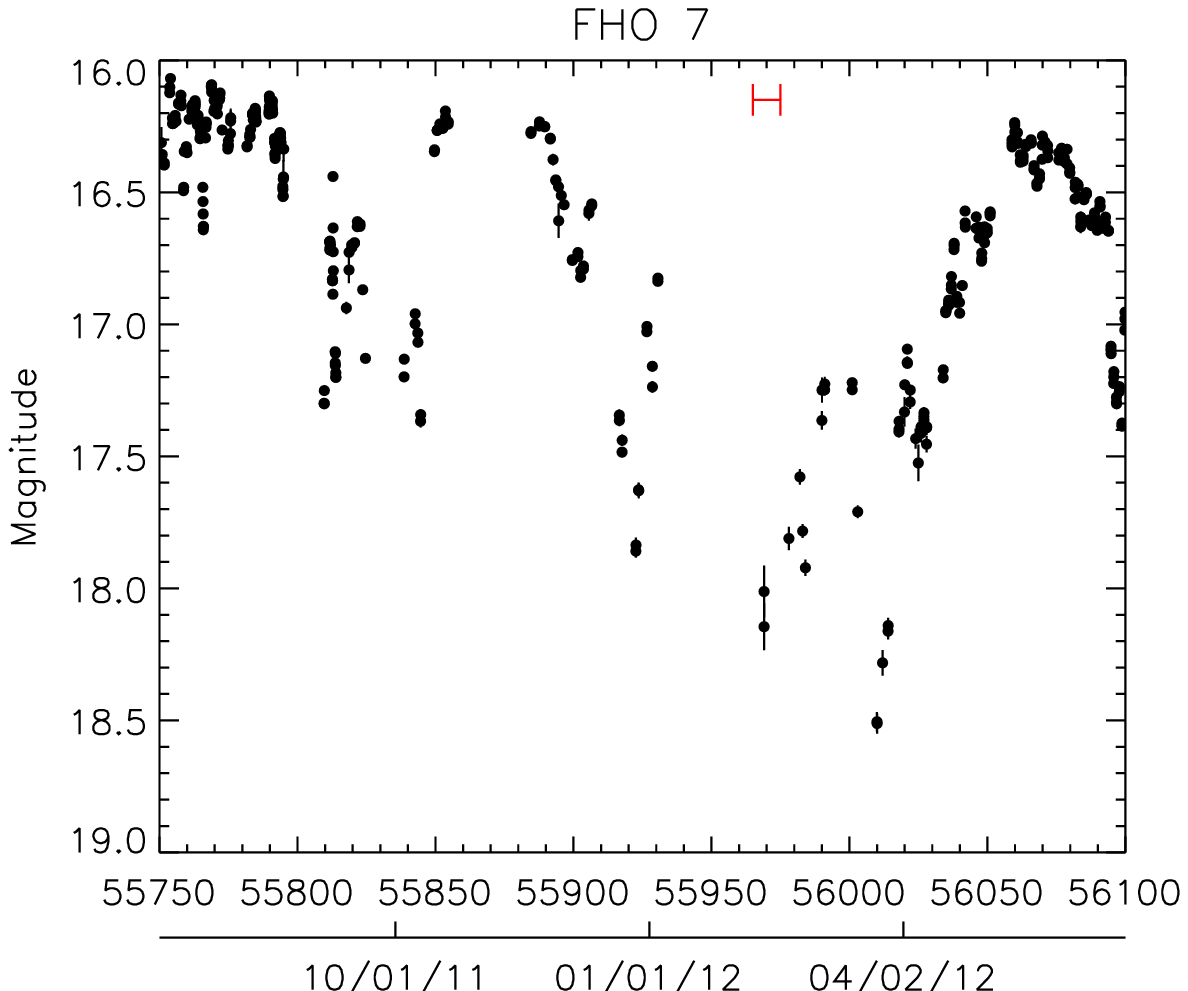}
\includegraphics[width=0.47\textwidth]{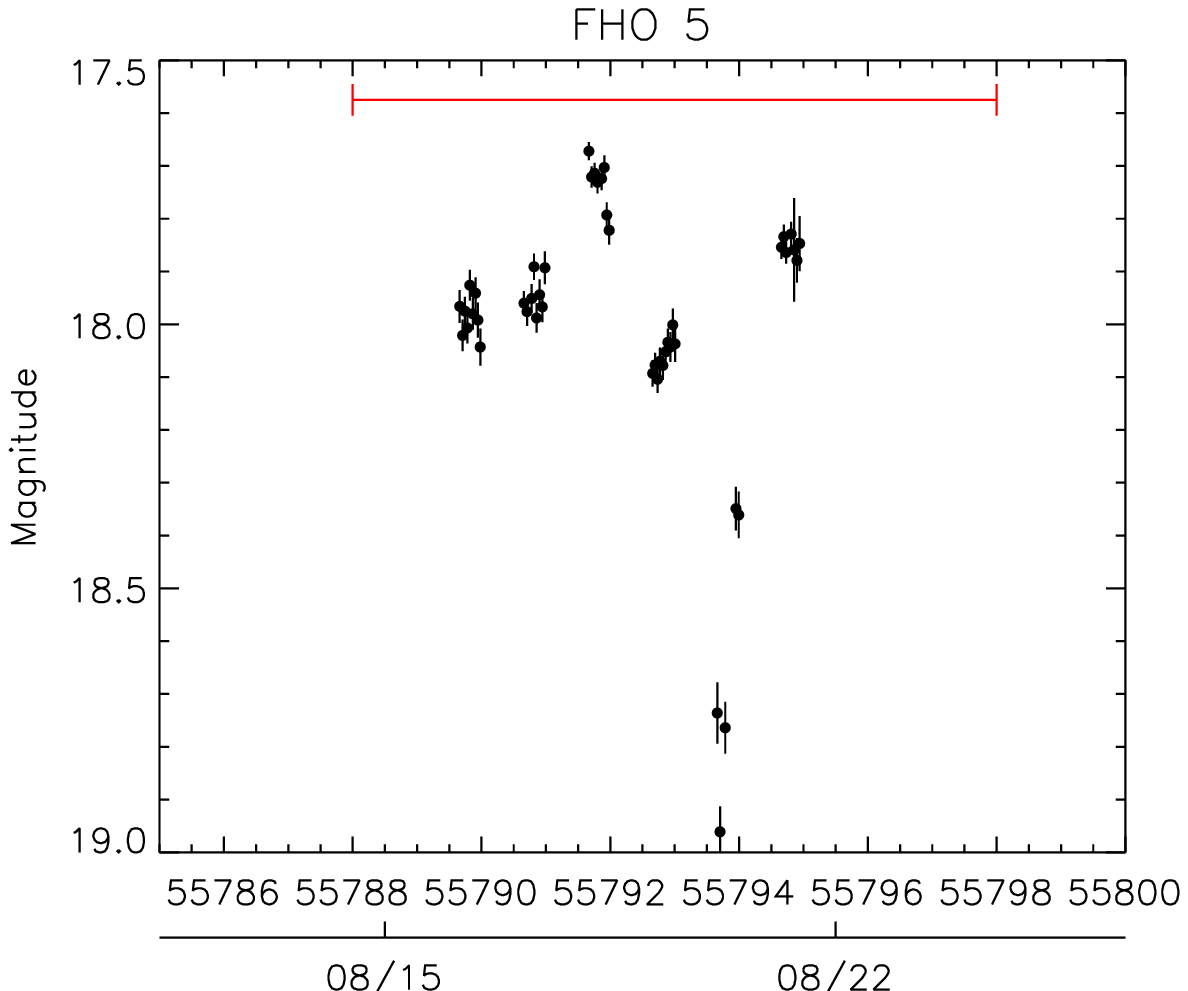}
\vspace{8pt}
\includegraphics[width=0.47\textwidth]{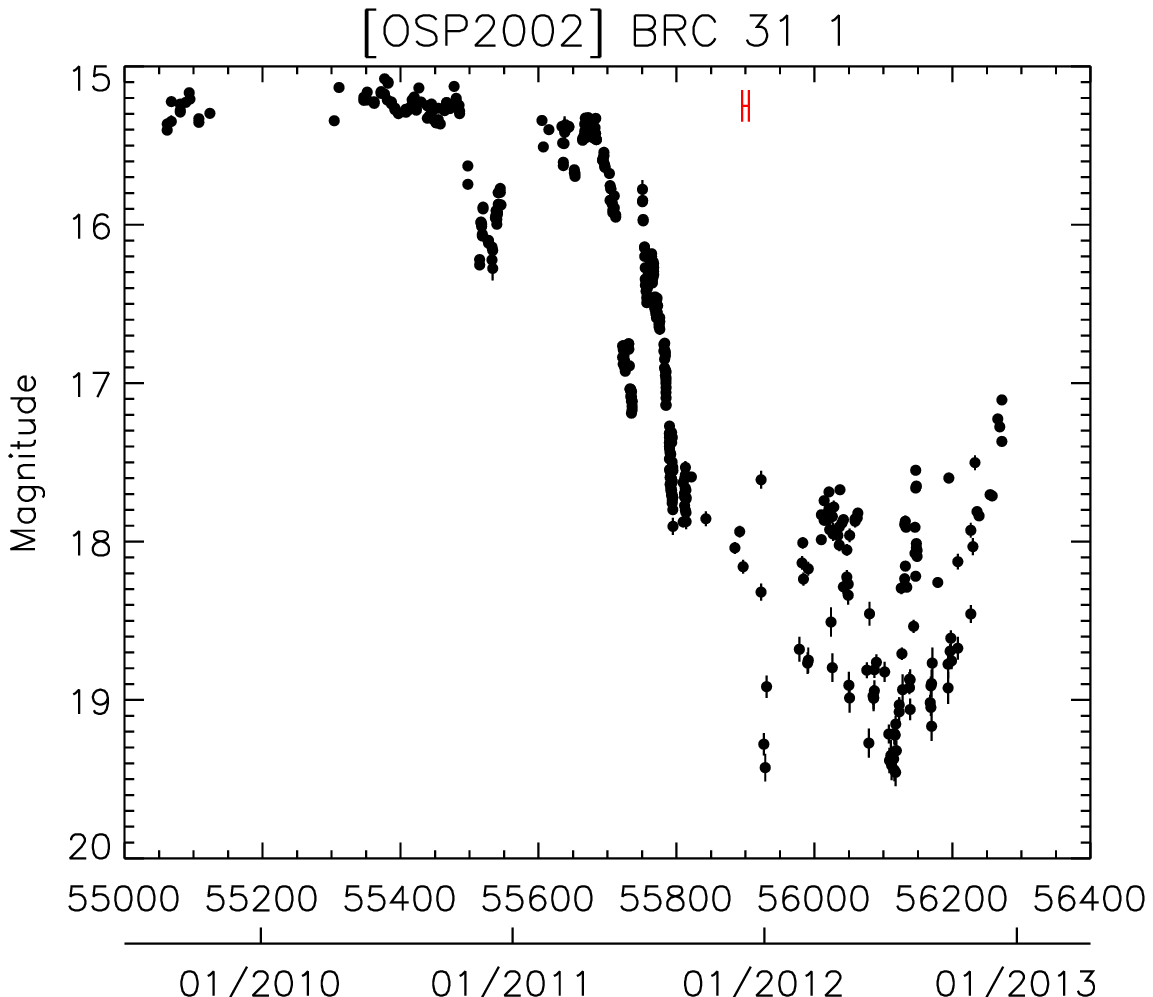}
\includegraphics[width=0.47\textwidth]{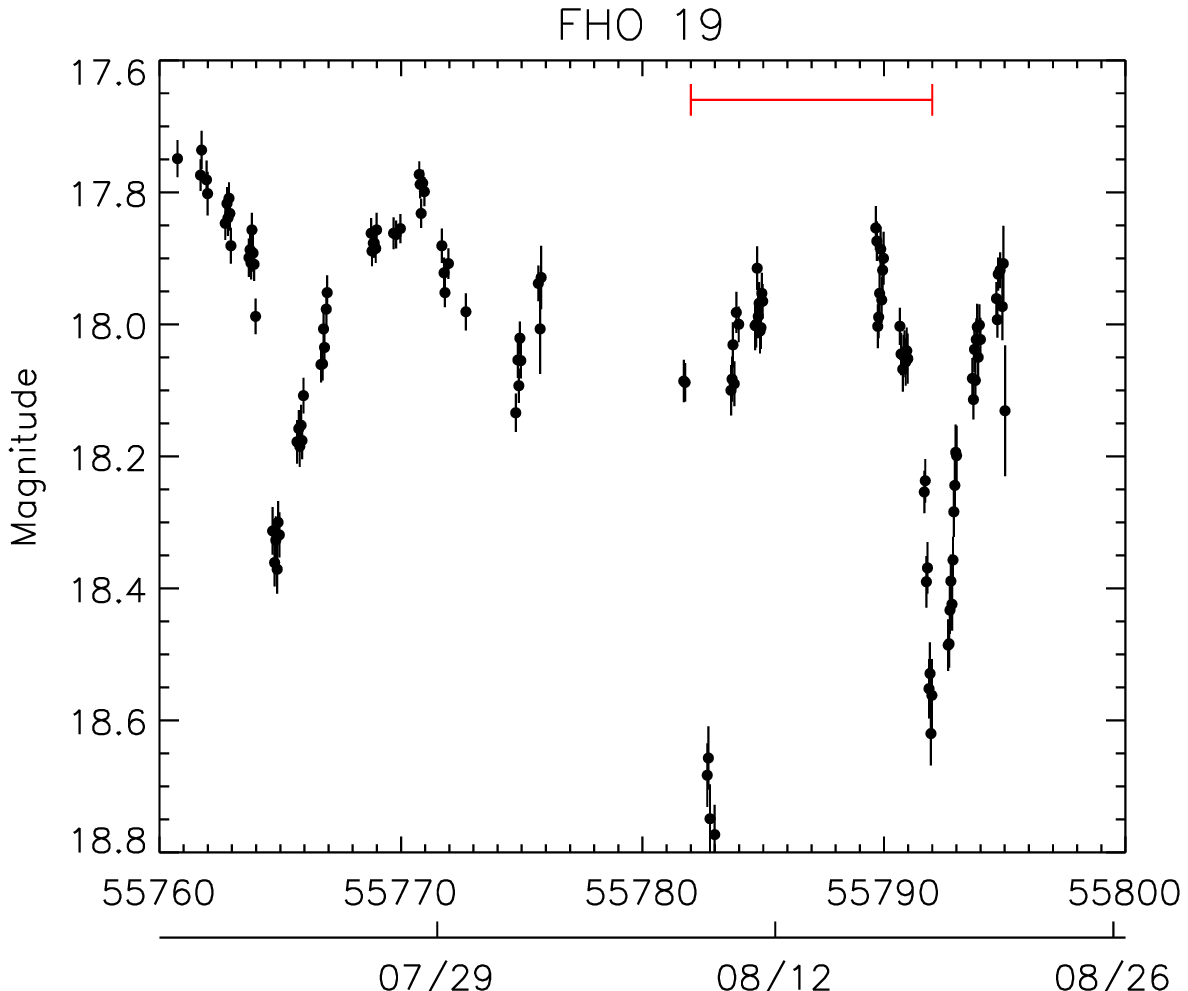}
\qquad\ \ \ \includegraphics[width=0.47\textwidth]{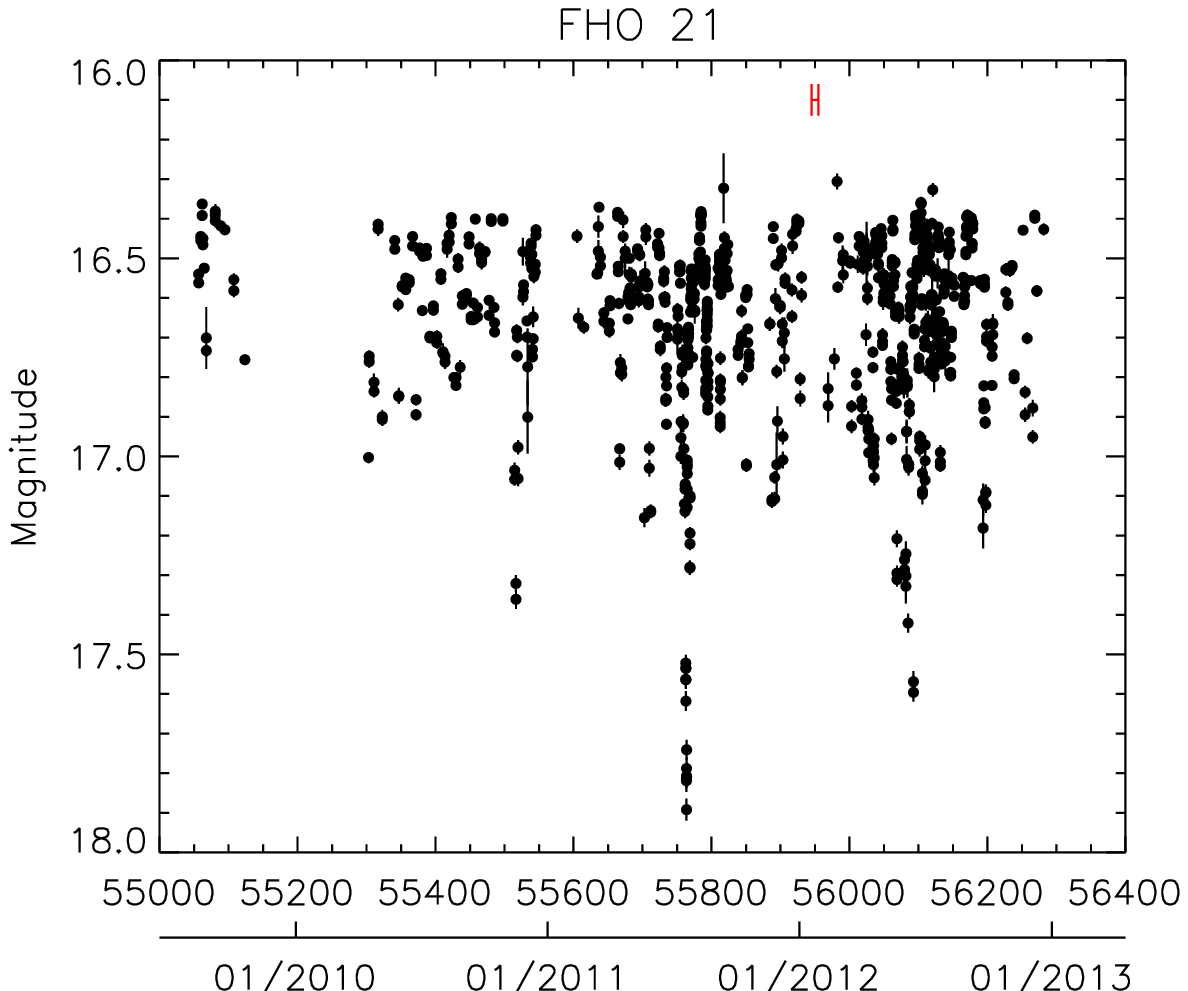}
\caption{Examples of the diverse behavior seen in our lightcurves for faders. Variability amplitudes range from a few tenths of a magnitude to nearly 2 magnitudes (top row). Fades can last anywhere from one day to over a year (middle row), and can be separated by anywhere from 9~days to over a year (bottom row). For scale, the horizontal bar near the top of each panel shows a 10~day interval. No points having any of the flags listed in Section~\ref{flaglist} are plotted.}\label{dipgallery}
\end{figure}

\begin{figure}
\includegraphics[width=0.47\textwidth]{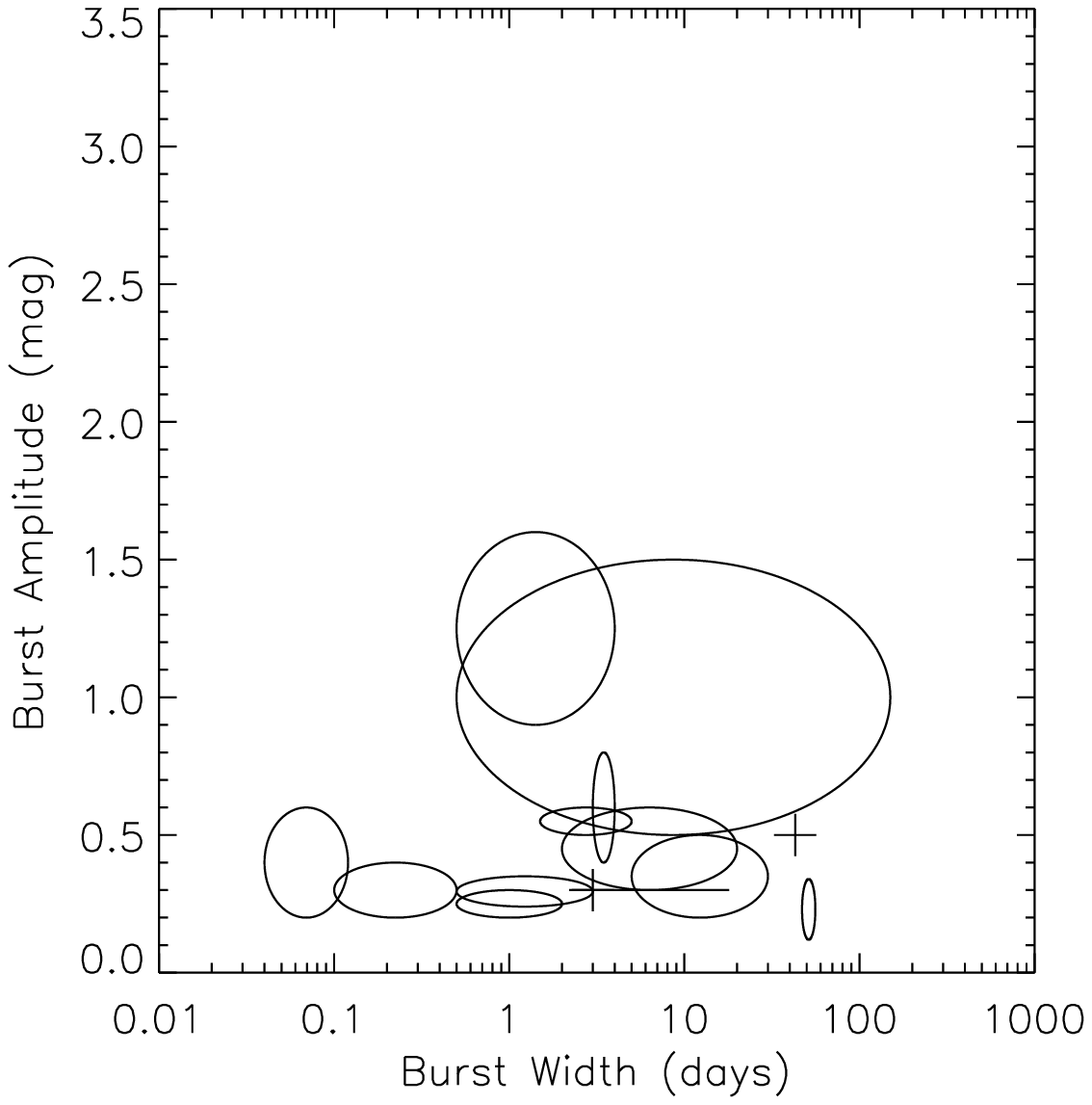}
\includegraphics[width=0.47\textwidth]{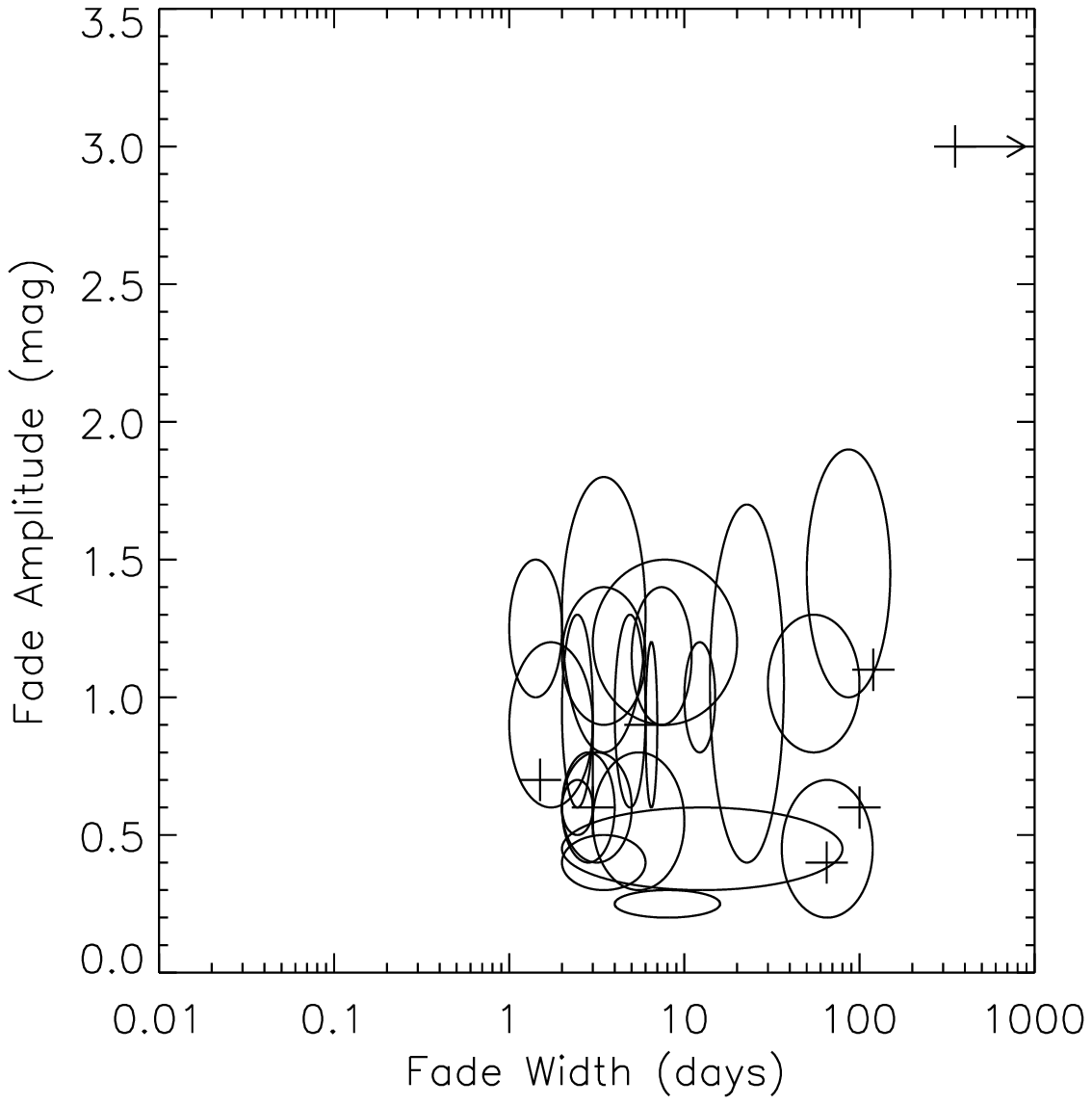}
\includegraphics[width=0.47\textwidth]{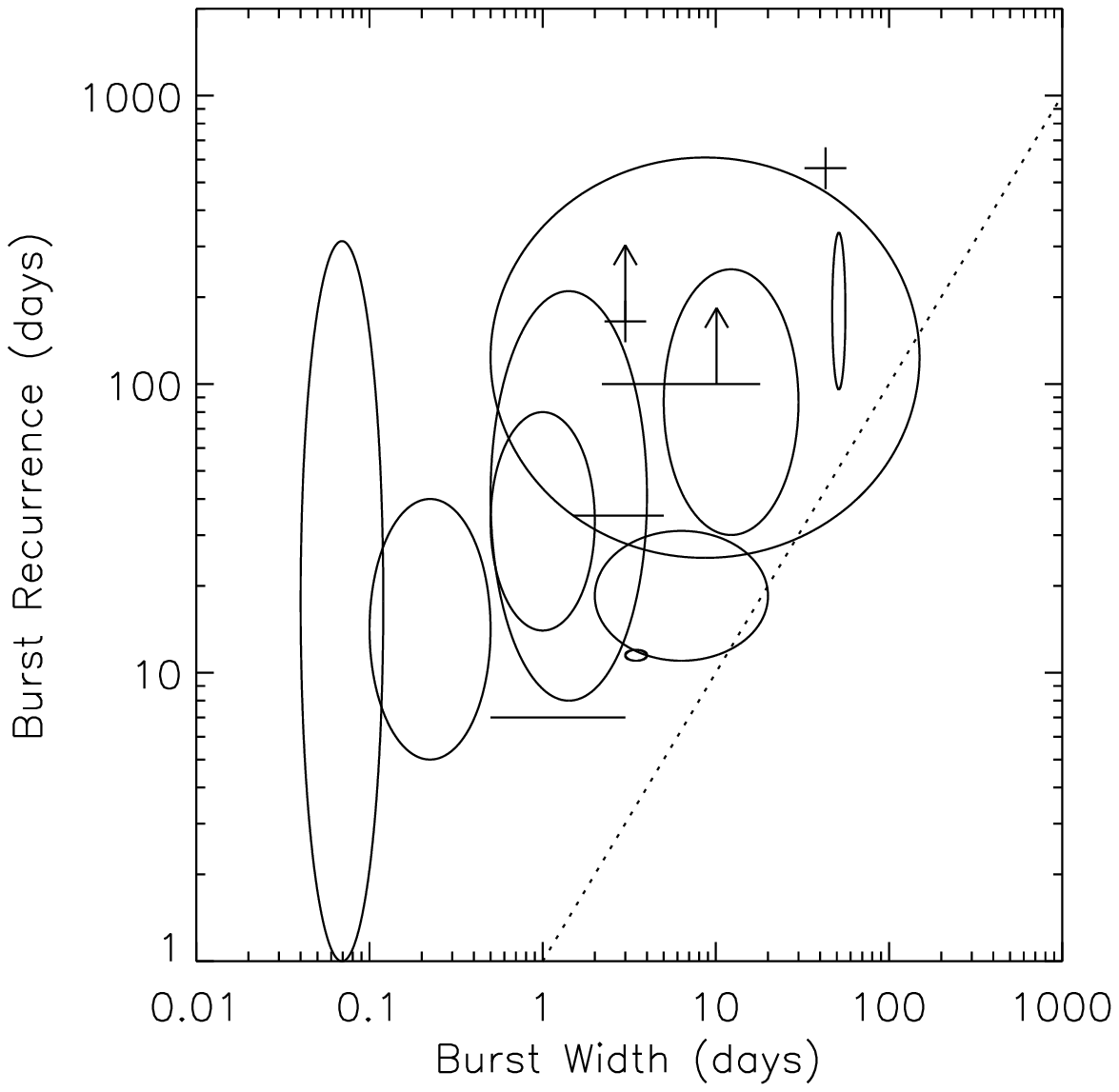}
\qquad\ \ \ \includegraphics[width=0.47\textwidth]{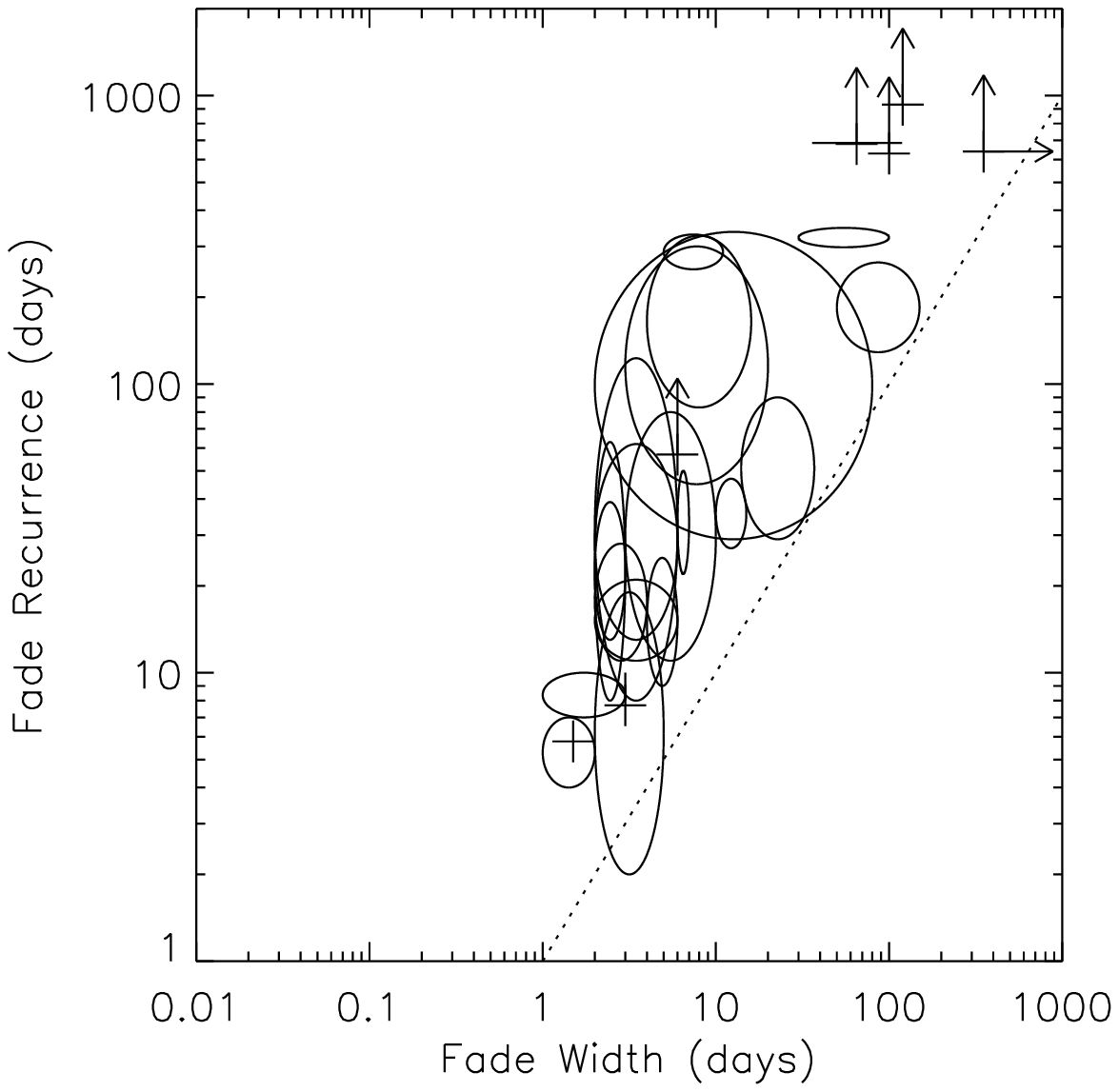}
\caption{Plots of the characteristic amplitudes and timescales for bursters (left) and faders (right), illustrating the wide variety of observed events.
The plus signs indicate sources that have either well-defined fixed values for their amplitudes and timescales (such as the two periodic AA~Tau analogs in the lower left corner of the fader panels), or that have only a single measurement (such as the single-event sources in the upper right of any panel). The ellipses represent sources that have bursts or fades of varying amplitudes, varying widths, or varying separations within a single light curve. The area below the dotted line on the lower two figures, where events would need to overlap each other, is not allowed, though some ellipses appear there because this analysis does not consider correlations between width and separation.}\label{flarewidths}\label{dipwidths}
\end{figure}

If either enhanced accretion or reduced extinction are responsible for bursting events, then stars with large infrared excess, and therefore more circumstellar material, may be more likely to show bursting behavior than stars with small infrared excess. Using the Kendall's~$\tau$ statistic \citep{kendalltau}, we found no evidence for a correlation between the Spitzer IRAC/MIPS $[3.6]-[24]$ color (56\% confidence) and the presence of bursting among our sample of 117 IR-excess sources, but we did find a marginally significant correlation (2.4\% confidence) with Spitzer IRAC $[3.6]-[8.0]$ color, in the sense that redder sources are more likely to show bursting behavior. We note that of the 14 bursters, only two, FHO~2 and FHO~24, are Class~III sources. The rest have $K - [8] > 1.8$ and $K - [24] > 5$ (see Figure~\ref{cmd_cand} for comparison to the rest of the sample). 
We note that since the K-band and Spitzer data are not coeval, the reported colors may be distorted by variability between the epochs of observation. However, mid-infrared variability is typically a few tenths of a magnitude or less \citep{irvar1, YSOvar_first, Flaherty_aperiodic}, and so should not dramatically affect a star's position in diagrams such as Figure~\ref{cmd_cand}.

While the weak correlation with $[3.6]-[8.0]$ color suggests that bursters are associated with stronger circumstellar disks, and therefore with the possibility of enhanced accretion or reduced circumstellar extinction, the absence of a similar correlation with $[3.6]-[24]$ color weakens this result. As noted in Section~\ref{completeness}, however, only a limited range of infrared color is well-represented in this sample. It is also possible that any correlation is being diluted by radiative transfer effects, geometry, or other factors that determine whether any particular star shows bursting behavior. 
We discuss how additional data could allow more conclusive tests in Section~\ref{future}.

\subsection{Constraints on Short-Term Accretion Outbursts}\label{shortoutbursts}

Magnetic or viscous instabilities acting at the boundary between the stellar magnetosphere and the circumstellar disk are expected to produce short bursts of accretion on timescales of weeks to months for certain regimes of disk properties \citep[e.g.,][]{AK1990, GW1999, RUK2004b, RUK2005}. However, variability from such outbursts has never been observed \citep{Bouvier_PPV}. The consistent cadence and long time coverage of our PTF survey have allowed the most sensitive search to date for such accretion events.

\begin{figure}
\includegraphics[width=0.47\textwidth]{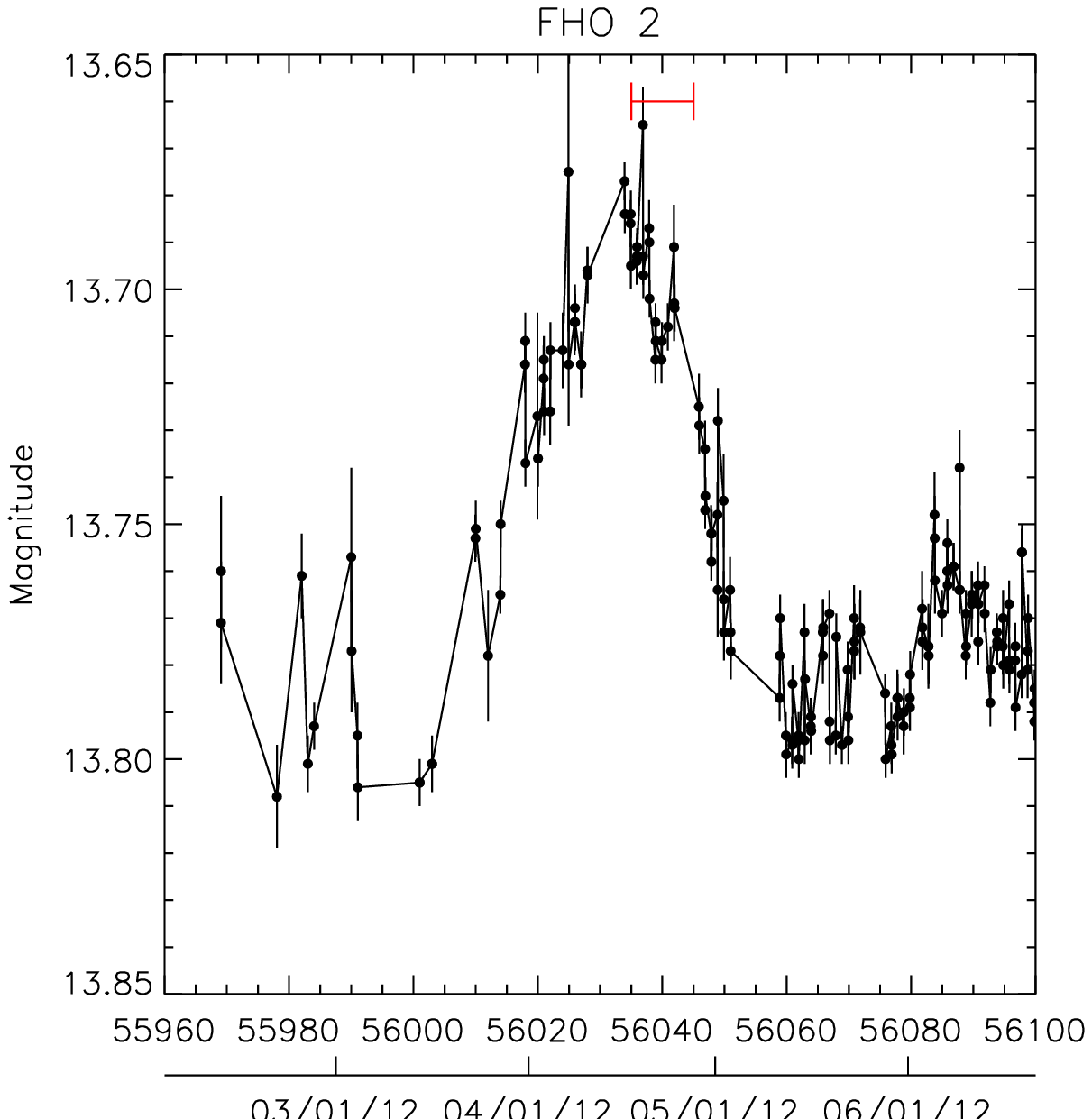}
\vspace{8pt}
\includegraphics[width=0.47\textwidth]{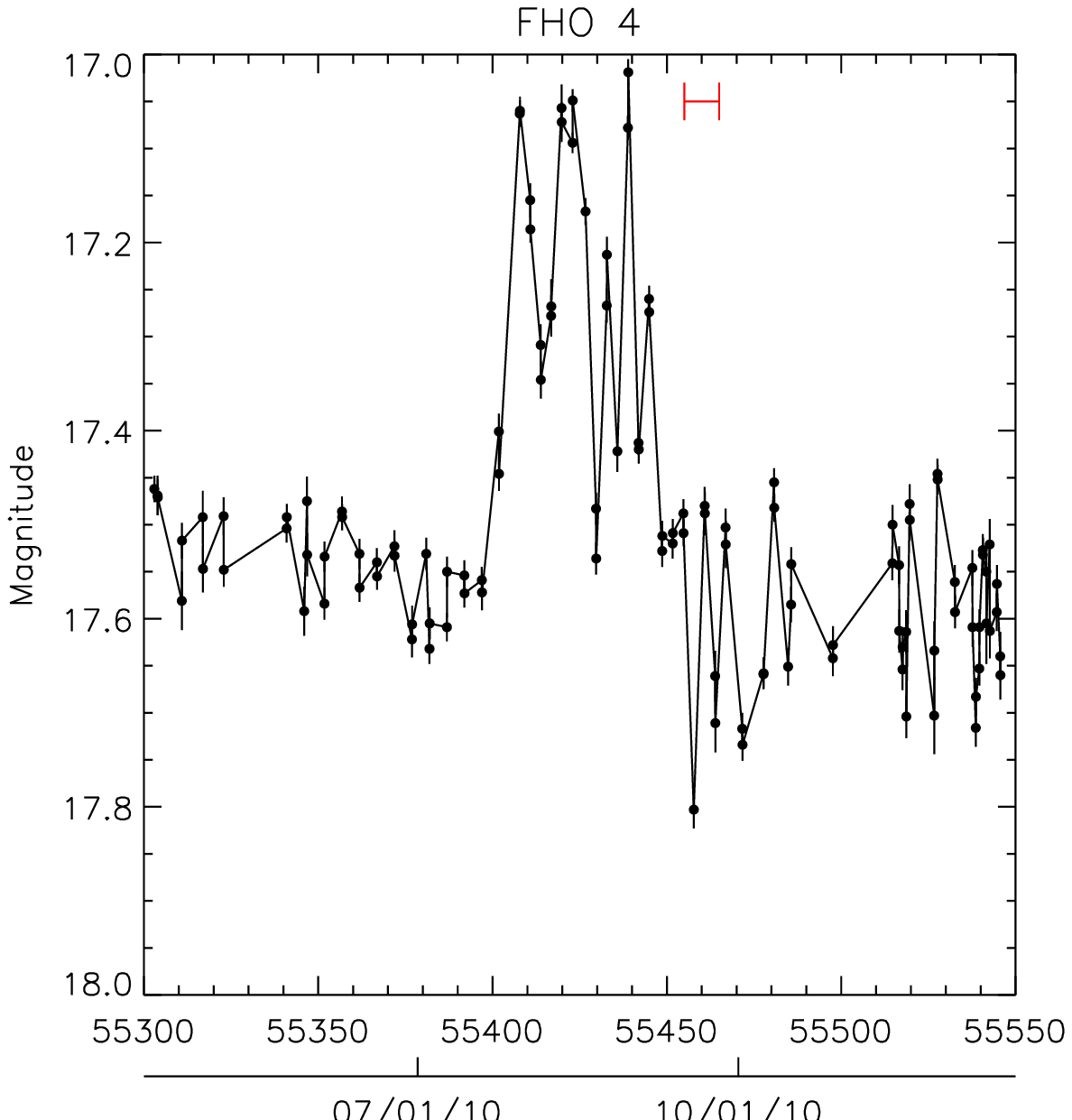}
\qquad\quad\includegraphics[width=0.47\textwidth]{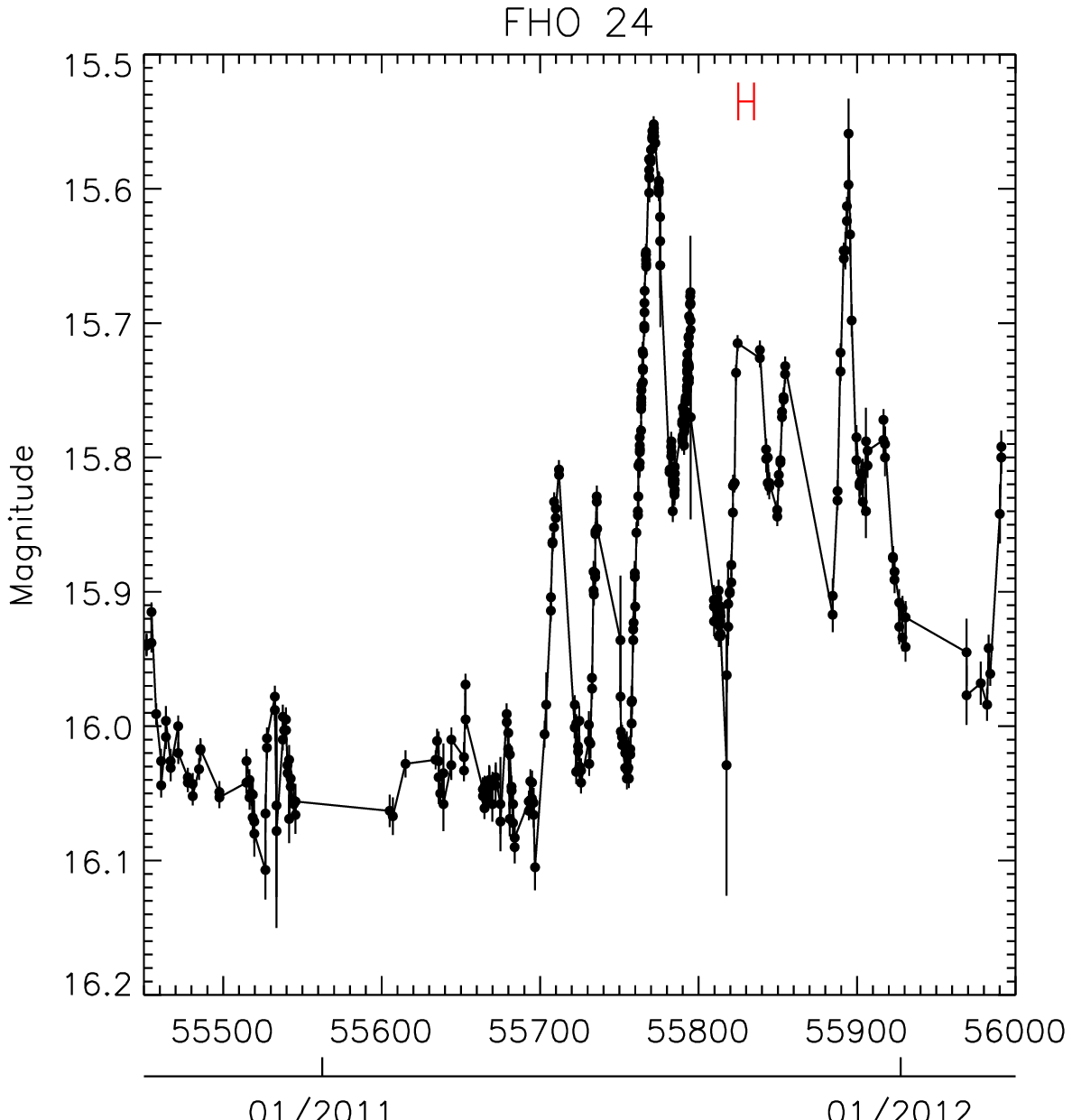}
\caption{Burst profiles for three bursters with durations of tens of days.  No points having any of the flags listed in Section~\ref{flaglist} are plotted. In this and subsequent plots, the points are connected by line segments to clarify the order of closely spaced observations.
FHO~2 has the simplest event profile, showing a smooth rise and fall over a 20-30~day interval. The bursts of FHO~4 are longer and show a more complex profile. The light curve for FHO~24 shows a large number of contiguous bursts rather than a few isolated events like the other two.}\label{outbursts}
\end{figure}

Of the bursting sources in Table~\ref{candidates}, FHO~2, FHO~4, and FHO~24 show multiple bursts lasting tens of days each. The separations between bursts vary: tens of days in the case of FHO~24, 100-300 days in the case of FHO~2, and $\sim 500$~days for FHO~4. We show all three sources in Figure~\ref{outbursts}. The timescales and shapes of these events, particularly FHO~2 and FHO~4, resemble the simulated variations in $\dot{M}$ shown in Figure~4 of \citet{RUK2004b}.
Although they do not stand out in the context of our sample, where burst durations vary continuously from $<1$-150~days, FHO~2, FHO~4, and FHO~24 are noteworthy as the first bursts reported in young stars having timescales of tens of days. 
To our knowledge, these light curves represent the first observations consistent with the predicted inner-disk instabilities. 

Models predict that short-term accretion outbursts should have amplitudes of a few tenths of a magnitude. For example, scaling to a fiducial star with $0.8~\Msun$ and $2~\Rsun$, the simulations of \citet{RUK2004b} predict an accretion rate of $2 \times 10^{-8}~\Msun \textrm{yr}^{-1}$ in quiescence and 6-$8 \times 10^{-8}~\Msun \textrm{yr}^{-1}$ in outburst. The fiducial star would have a luminosity of $1.14~\Lsun$ \citep{SiessModels}, with quiescent and outburst accretion luminosities of $0.25~\Lsun$ and $0.88~\Lsun$, respectively, implying a brightening of $\sim 0.4$ magnitudes between quiescence and outburst. The three candidate stars have amplitudes between 0.2 and 0.5~mag, consistent values given star-to-star variations in star and disk parameters.

The behavior of the light curves is inconsistent with white-light flares analogous to those seen on the Sun. 
White-light flares tend to have a steep rise followed by an exponential decay. None of the three bursts show an exponential profile, and the timescales of tens of days are much longer than the minutes to hours durations observed in the Sun or in low-mass stars. 
A superposition of many short flares is also unlikely: FHO~2 and FHO~24 show little variability on timescales of a single day, as might be expected from a stochastic sum of shorter events. In addition, a 0.4~mag burst lasting 30 days corresponds to an energy release, depending on the (unknown) spectrum of the transient emission, of $\sim 6 \times 10^{39}$~erg for a $1.4 \Lsun$ star. The entire stellar magnetic field, integrating a dipole field from an assumed $R_\star \sim 2 \Rsun$ to infinite radius, contains only $\sim 5 \times 10^{38} \left(B_\mathrm{surf}/\textrm{1~kG}\right)^2$~erg. Even a 3~kG field, near the largest values observed in T~Tauri stars \citep{Bouvier_PPV}, cannot provide enough energy to power the bursts.

The brightness enhancements for these three objects are also unlikely to be dust-clearing episodes. Only FHO~4 is a Class~II source, while FHO~2 and FHO~24 are both Class~III sources with excess only in the MIPS 24~$\mu$m band. It is doubtful that these two stars have enough circumstellar dust in the inner disk to allow significant extinction-driven variability. 
We note that stars with infrared excess only in the MIPS bands can still show ongoing accretion on the order of $0.1$-$0.5 \times 10^{-8}~\Msun \textrm{yr}^{-1}$ \citep{2009ApJ...704L..15M,2012ApJ...747..103E}, so the absence of an IRAC excess does not rule out either low-level accretion or, plausibly, brief periods of accretion at a higher rate. 
Extinction is a possible origin for the variability of FHO~4; color data taken over the course of one of its bursts could test this hypothesis. 

\section{The Fader Phenomenon}\label{dippers}\label{dipperproperties}

The two prototypical faders are AA~Tau, which fades repeatedly by 1.4~mag over 30\% of an 8.2-day cycle \citep{AATau_first,AATau_main}, and UX~Ori, which fades by 3~mag for tens of days at irregular intervals \citep{UXOri_review}. Both AA~Tau and UX~Ori are well understood as the result of recurring extinction by circumstellar material, from a warped inner disk edge in the case of AA~Tau or from more irregular structures in the case of UX~Ori.

Of the 29 sources that show some kind of fading behavior, only two, LkH$\alpha$~174 and FHO~12, show the periodic modulation characteristic of AA~Tau. 
Four more sources, LkH$\alpha$~150, FHO~7, FHO~15, and FHO~27, show multiple fading events with durations of tens of days, as seen for UX~Ori stars. However, their typical amplitudes of 1~mag or less are much smaller than the 3-4~mag fades associated with UX~Ori stars.

The remaining 23 faders do not resemble either of the previously established categories. The natural assumption is that these sources also have their variability dominated by circumstellar extinction, with different spatial scales or different geometries causing the light curves to behave differently. 

All the sources except for the two AA~Tau analogs are aperiodic, and, as illustrated in Figure~\ref{dipgallery}, they often bear little resemblance to each other. For example, FHO~19 has narrow fades repeating every 8-10 days, but without enough coherence to be periodic. In contrast, NSV~25414 and FHO~3 both have frequent but irregular events, with the interval between adjacent fades varying by more than a factor of two. At the other extreme, FHO~21 and FHO~22 each show only one fading event per year, while LkH$\alpha$~150 and FHO~25 fade only once in the entire survey period. Most fading events are short, but those of LkH$\alpha$~150 and [OSP2002]~BRC~31~1 last for hundreds of days. Most stars have fading events of roughly constant depth, but FHO~15 and FHO~20 have significant amplitude variability. Most fading events are symmetric, but FHO~11 and FHO~27 show strongly lopsided events.

Like the bursters, the faders do not separate naturally into sources with distinct timescales. We show in Figure~\ref{dipwidths} the joint distributions of fade amplitudes, fade widths, and fade separations for all 29 faders. The absence of gaps in the plot suggests that, as with the bursters, short- and long-timescale faders have a common origin. 

As with the bursters, we tested for a correlation with the presence of circumstellar material. Using the Kendall's~$\tau$ statistic, we found no evidence for a correlation between the Spitzer IRAC $[3.6]-[8.0]$ or IRAC/MIPS $[3.6]-[24]$ colors and the presence of fading among our sample of 117 IR-excess variables, at 20\% and 18\% confidence respectively. However, we do not find a single example of a fader among Class~III sources (i.e., significant excess only at 24~$\mu$m), as would be expected if circumstellar material near the star is needed for fading events to occur. 

To test whether the fading events could instead be the result of variable foreground extinction, we searched for a correlation between stars' near-infrared color, where we can avoid variability-induced systematic errors through the use of coeval 2MASS photometry, and the presence or absence of fading behavior. Since unreddened M dwarfs have $J-K \lesssim 1$, stars with $1 \lesssim J-K \lesssim 3$ must have significant extinction, while stars with $J-K < 1$ may have only moderate extinction. If fading events are caused by foreground dust, we might expect fading to be more prevalent among the reddest stars.
Using the Kendall's~$\tau$ statistic, we found no evidence for a correlation between the $J-K$ color and the presence of fading among our sample of 117 IR-excess variables, at 26\% confidence. 

While we find that neither the degree of infrared excess nor proxies for near-infrared reddening are good predictors for the presence of fading behavior, the absence of faders among the Class~III sources is consistent with fading events requiring the presence of inner disk dust and therefore the possibility of occasionally enhanced extinction along the line of sight. We discuss how additional data could allow more conclusive tests in Section~\ref{future}.

\section{Individual Sources of Interest}\label{oddballs}

In Sections~\ref{bursters} and \ref{dippers} we have examined the 41 burster or fader candidates as an ensemble. However, many of the sources have a character of their own. While we present brief descriptions of all the sources in Table~\ref{candidates}, in this section we focus on a small number of stars whose behavior seems particularly difficult to explain. We present the available data on each and challenge interested readers to develop models for these sources.

\subsection{FHO~26}\label{fho26}

\begin{figure}
\includegraphics[width=0.47\textwidth]{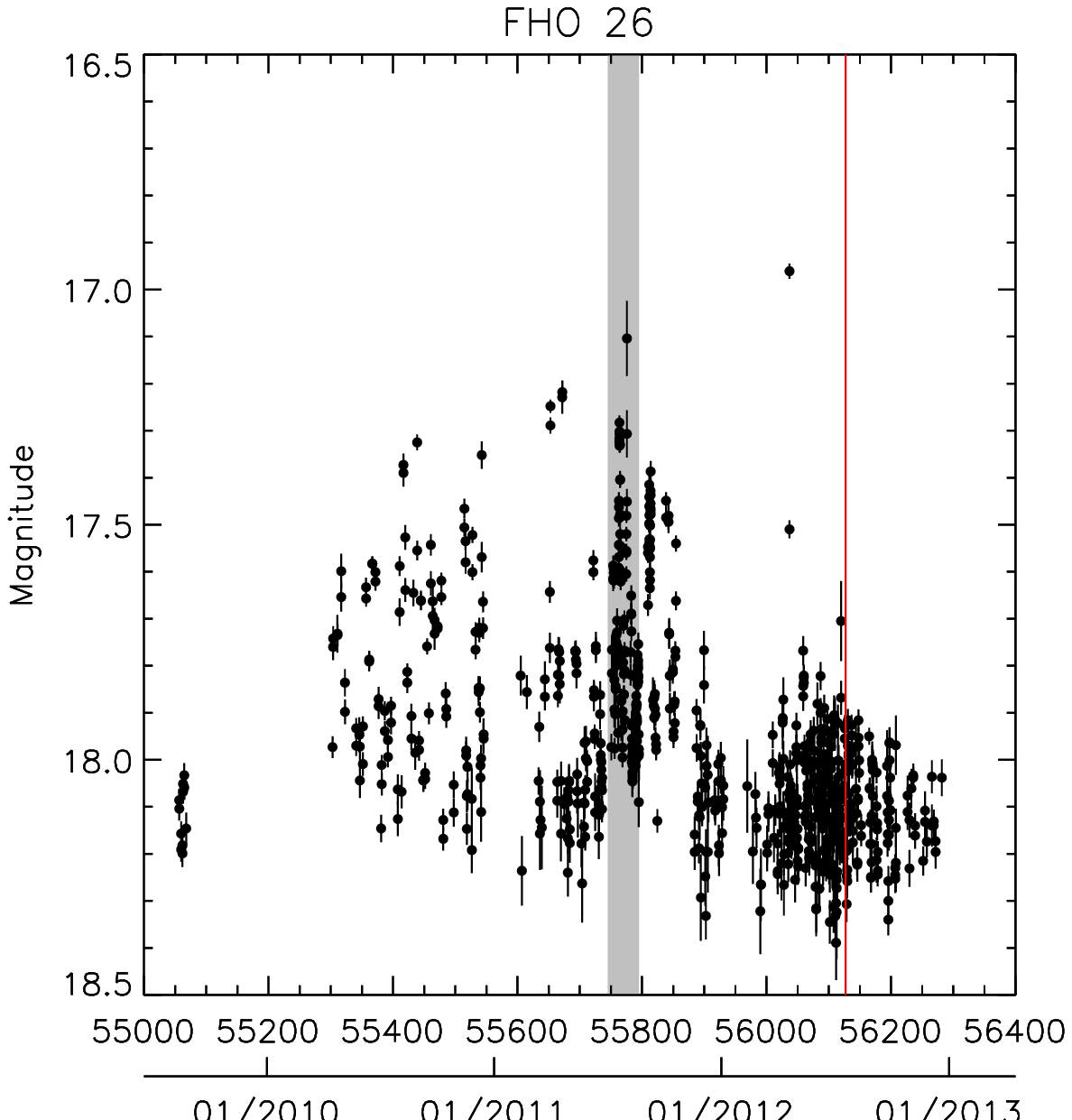}
\vspace{10pt}
\includegraphics[width=0.47\textwidth]{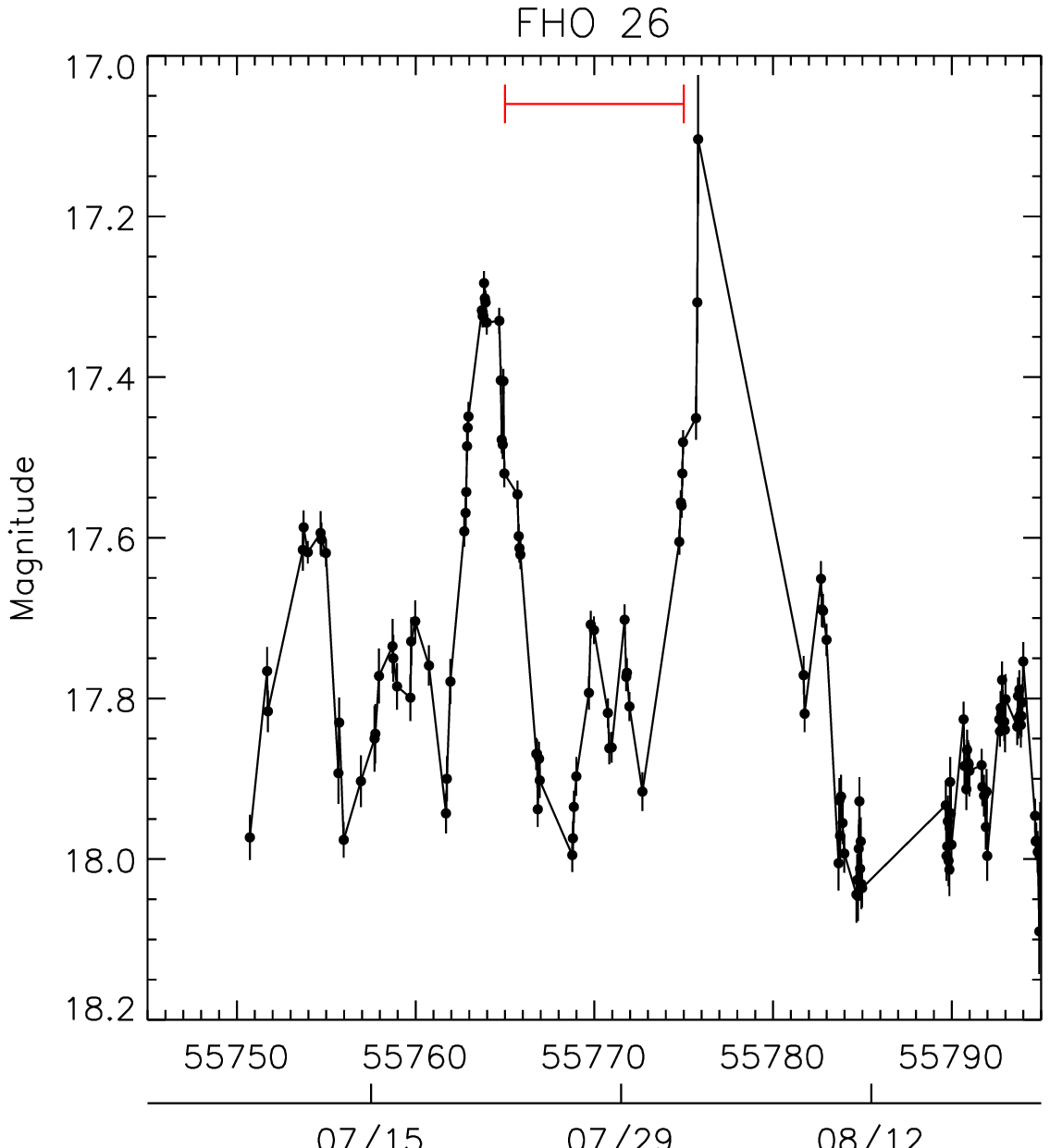}
\includegraphics[width=0.45\textwidth]{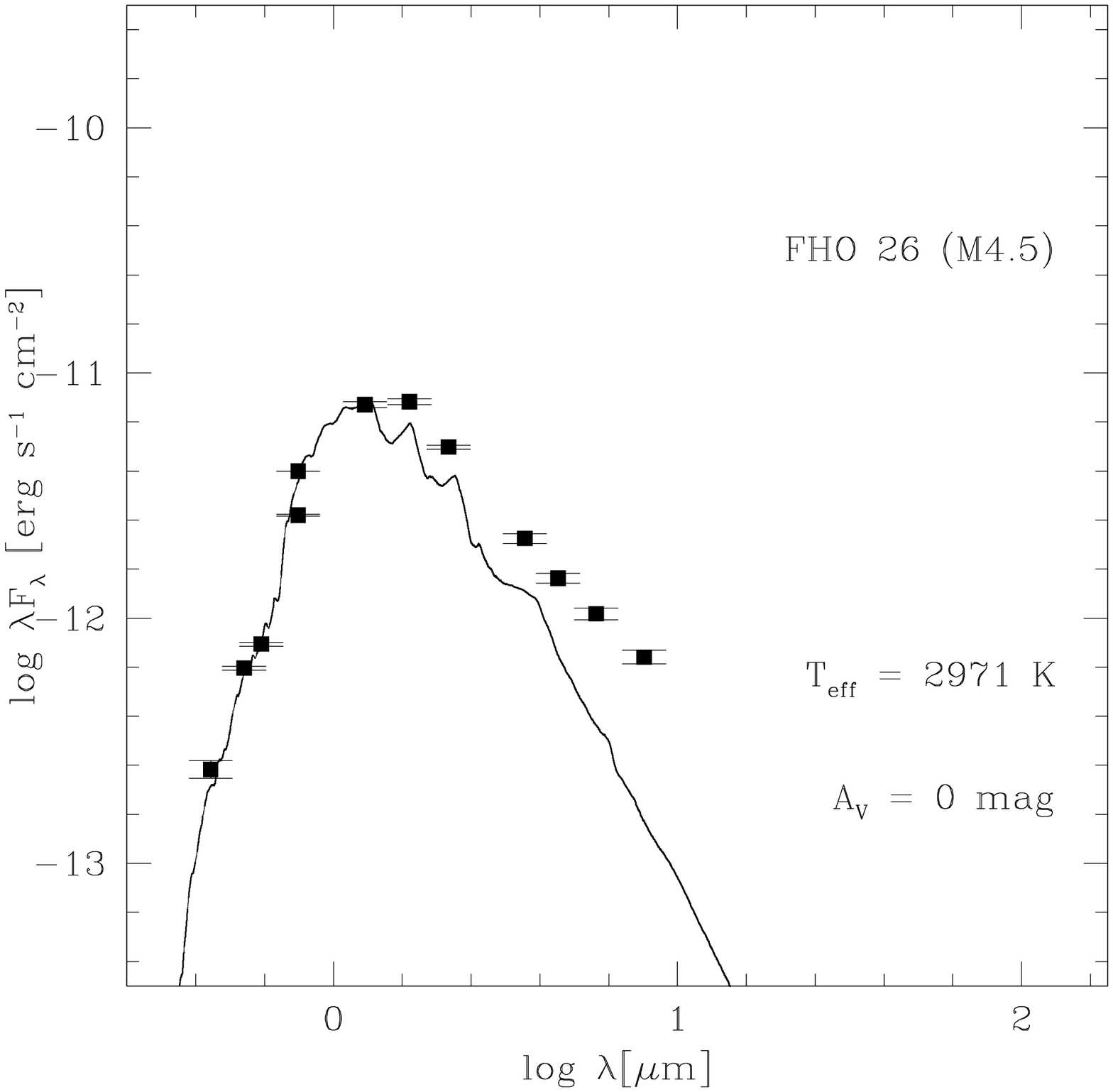}
\qquad\quad\ \includegraphics[width=0.47\textwidth]{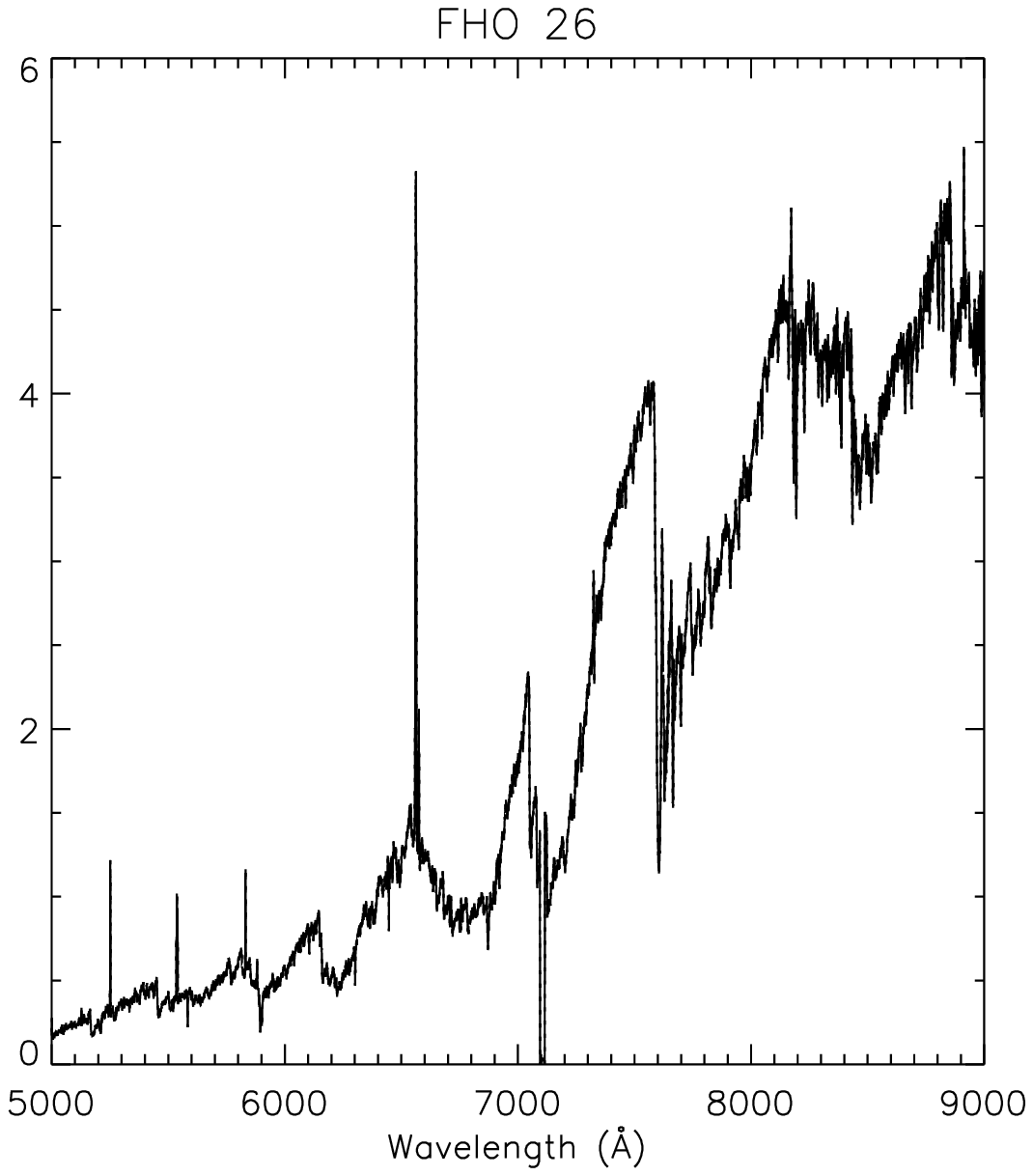}
\caption{A star whose regular bursting activity stopped at the end of 2011. The upper left panel shows the full 3-year light curve, with the shaded period expanded in the upper right panel to illustrate typical bursts for this source and the vertical red line marking the time at which the July 2012 spectrum in the lower right panel was taken. The red scale bar represents a 10-day interval. No points having any of the flags listed in Section~\ref{flaglist} are plotted. The lower left panel shows the spectral energy distribution for this source. The points are taken from non-simultaneous optical, near-infrared, and Spitzer photometry. The solid curve is a reddened NextGen model atmosphere \citep{NextGen_Dwarf} with temperature corresponding to the star's spectral type, matched to the optical through $J$-band fluxes.}\label{lc_1220bv}
\end{figure}

FHO~26 showed several-day-long, $\sim 0.7$~mag bursts in 2010 and 2011 (see upper right panel of Figure~\ref{lc_1220bv}) but became quiescent in late 2011. In 2012, except for two brief bursts, it has shown only a 0.2~mag, 5.6-day periodic modulation. The 2010-2011 bursts do not phase up under the 2012 period. FHO~26 has a modest infrared excess, as shown in the lower left panel of Figure~\ref{lc_1220bv}.

We show in the lower right panel a spectrum of the source taken in July 2012, well into the quiescent phase and near the peak of the periodic variability. The spectrum shows an M4.5 photosphere with emission from H$\alpha$ (-13~\AA\ equivalent width).

\subsection{[OSP2002]~BRC~31~1}\label{brc311}

\begin{figure}
\includegraphics[width=0.47\textwidth]{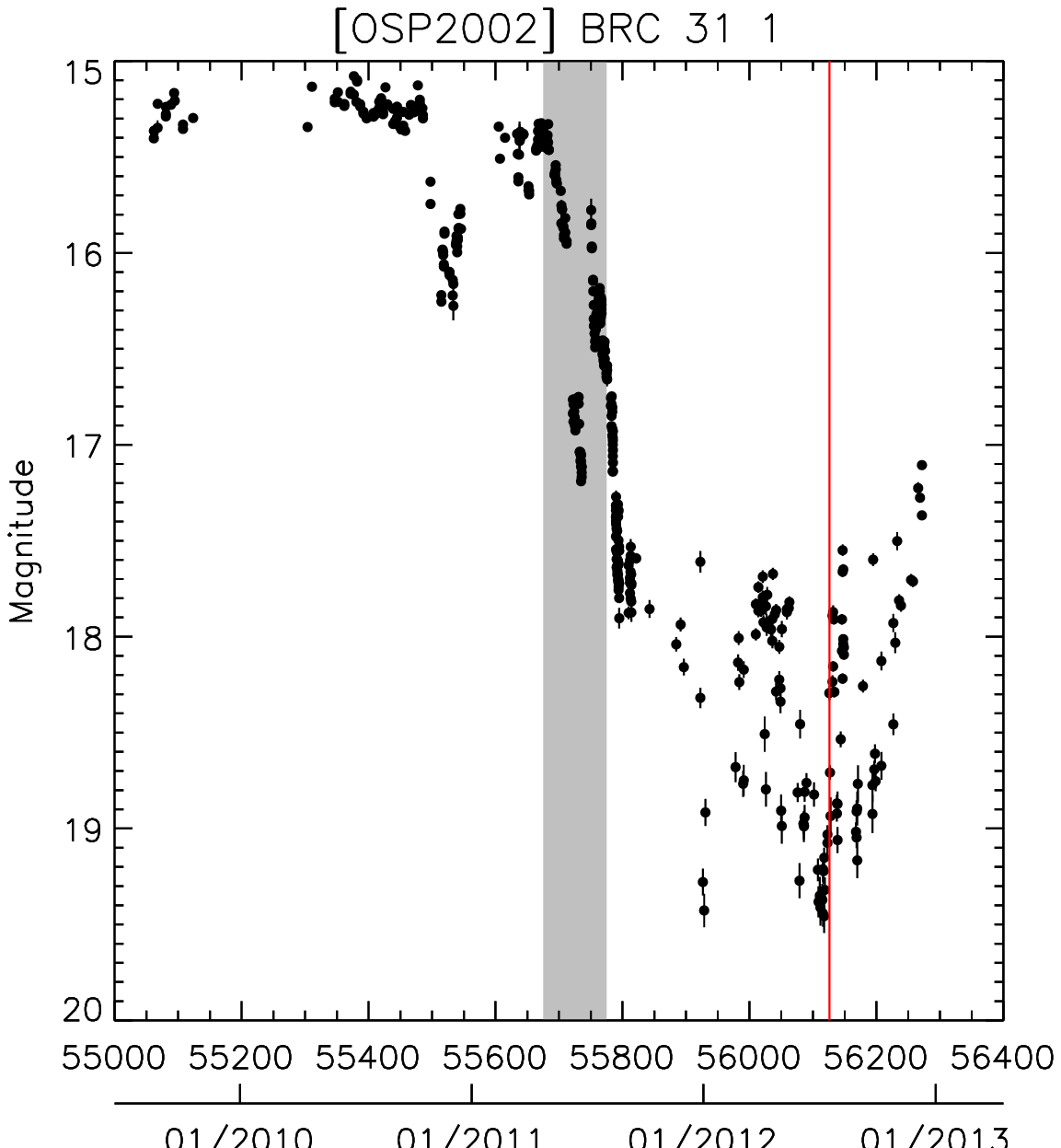}
\vspace{10pt}
\includegraphics[width=0.47\textwidth]{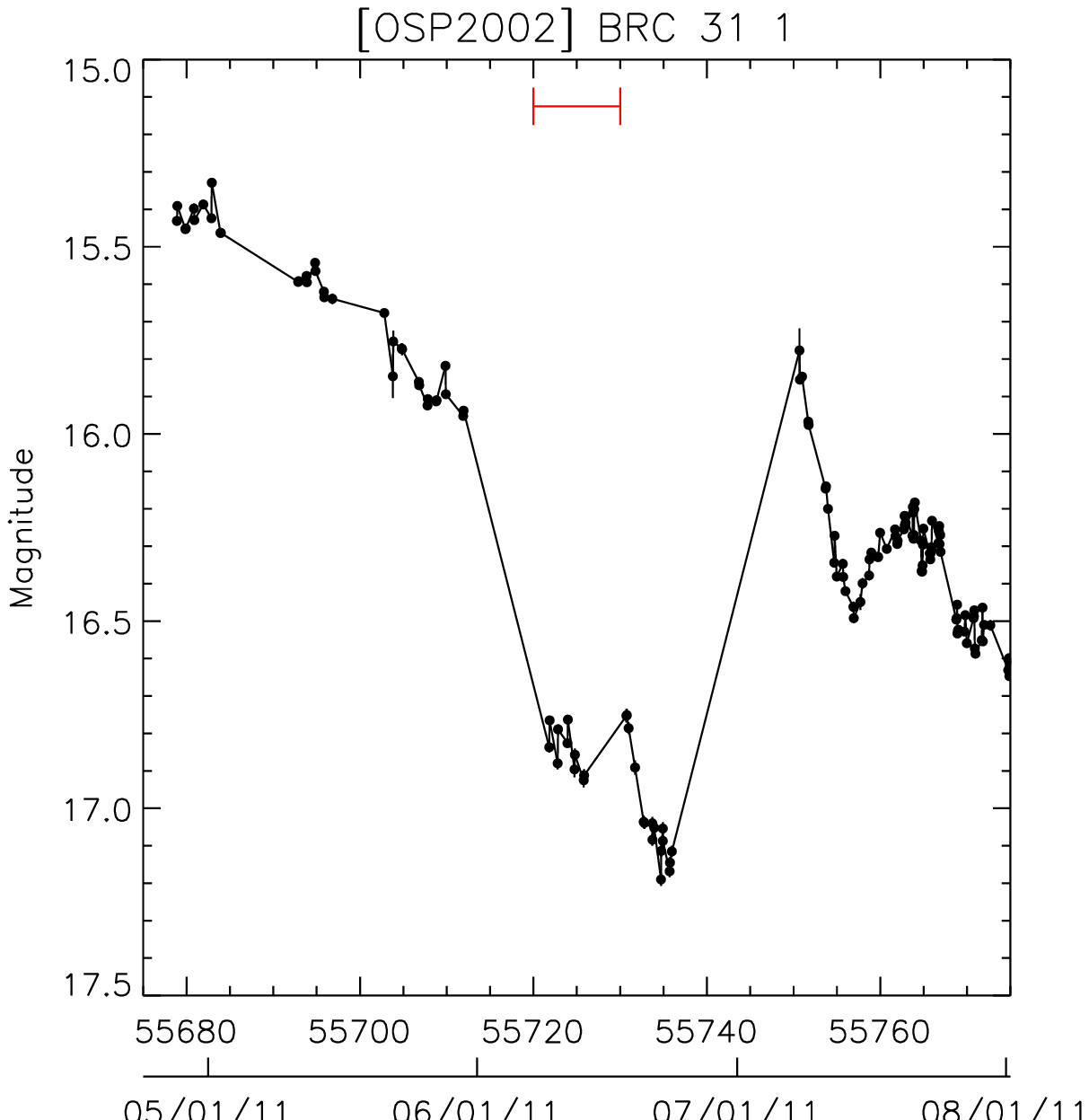}
\includegraphics[width=0.45\textwidth]{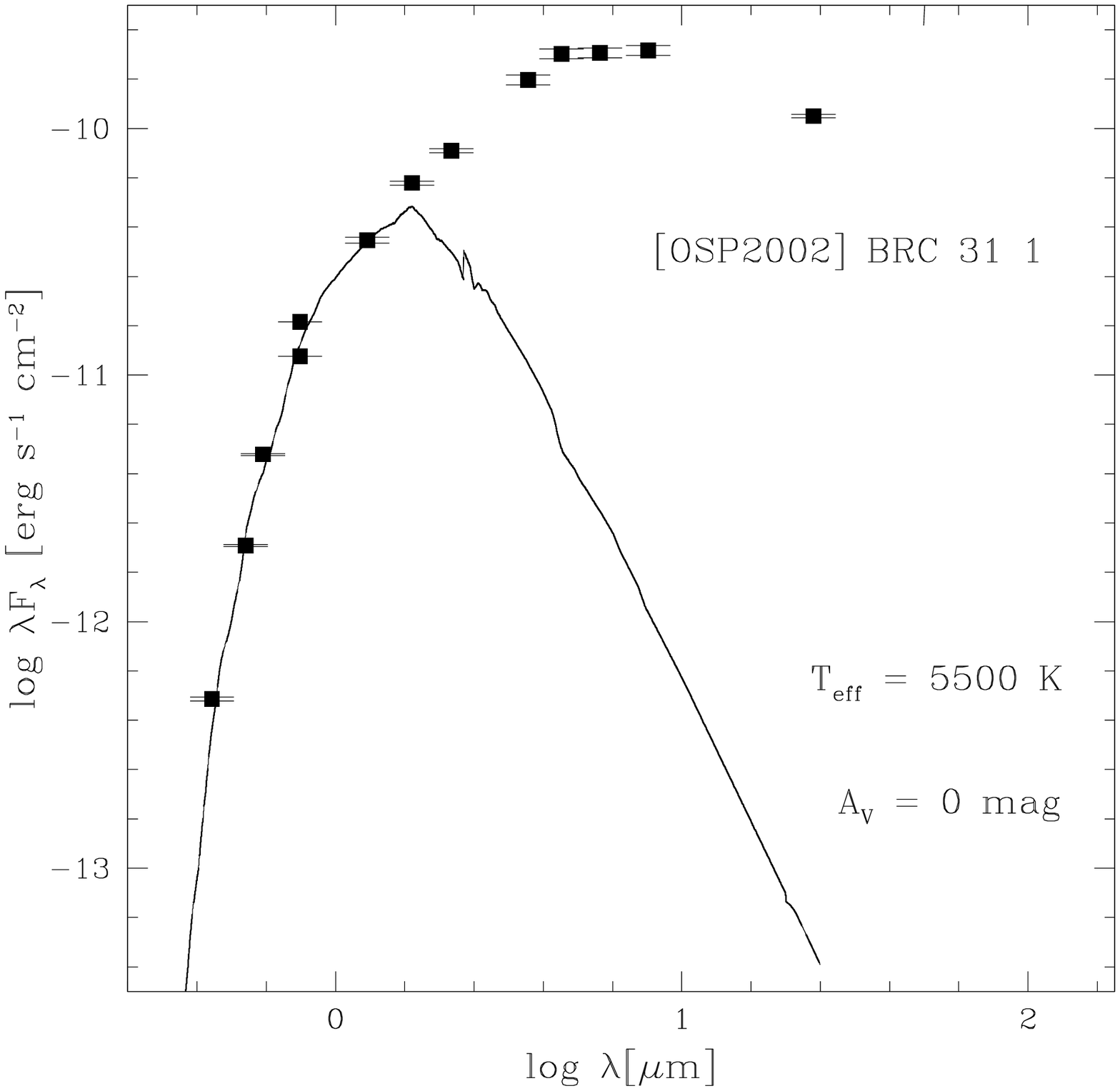}
\qquad\quad\ \includegraphics[width=0.47\textwidth]{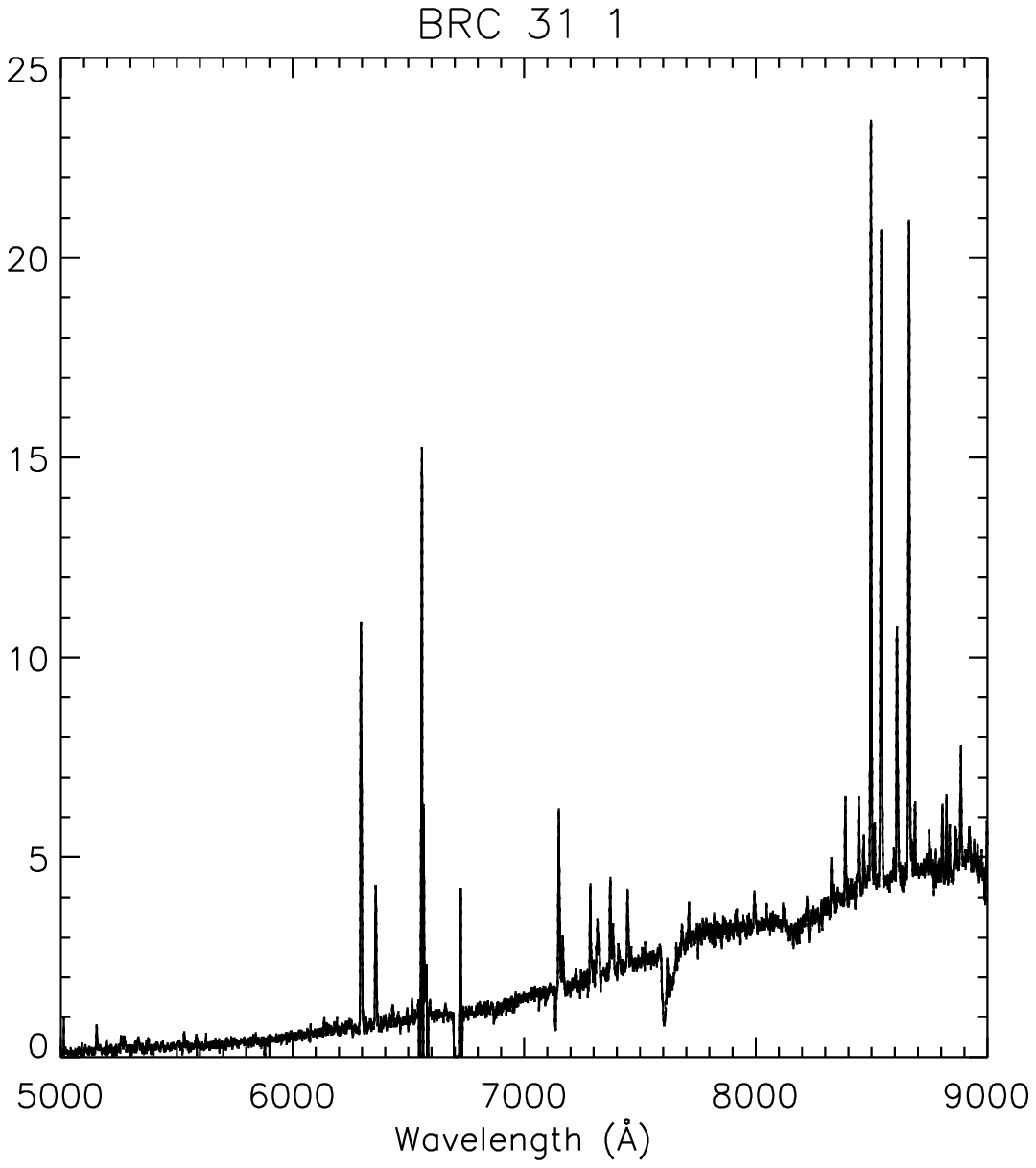}
\caption{Same as Figure~\ref{lc_1220bv}, but for a star showing a sudden decline in 2011. The star was not detected in roughly half the epochs in 2012; the non-detections are not shown. The upper right panel highlights the decline, including a temporary dip that interrupted it. Since we could not determine a spectral type for this source, the photosphere shown in the lower left panel is for an assumed effective temperature.}\label{lc_1220cb}
\end{figure}

[OSP2002]~BRC~31~1 grew fainter by nearly three magnitudes between April and August~2011 but showed relatively little variability before the fade, as shown in the upper left panel of Figure~\ref{lc_1220cb}. Our spectrum, taken during the star's faint state, shows a forest of emission lines including H$\alpha$, Ca~II, [O~I], [Fe~II], [S~II], [Ni~II], Fe~II, and many others.
A spectrum of [OSP2002]~BRC~31~1 from 1998 shows only H$\alpha$, Ca~II, and Fe~II at the same strength as in 2012, plus much weaker [O~I] and [Fe~II] lines. We see few absorption lines in the spectrum in either epoch.

The 1998 and 2012 spectra are similar to high- and low-state spectra, respectively, of the long-term variable PTF10nvg \citep{ptf10nvg_new}. Since BRC~31~1, like PTF10nvg, is a Class~I infrared excess source, it is possible that BRC~31~1 is a similar system: a high-inclination source with circumstellar material obscuring the inner disk, stellar photosphere, and accretion zone, but not obscuring a spatially extended jet. We note the light curve resembles that of V1184~Tau presented by \citet{2008Ap.....51....1G}.

\subsection{FHO~18}\label{fho18}

\begin{figure}
\includegraphics[width=0.47\textwidth]{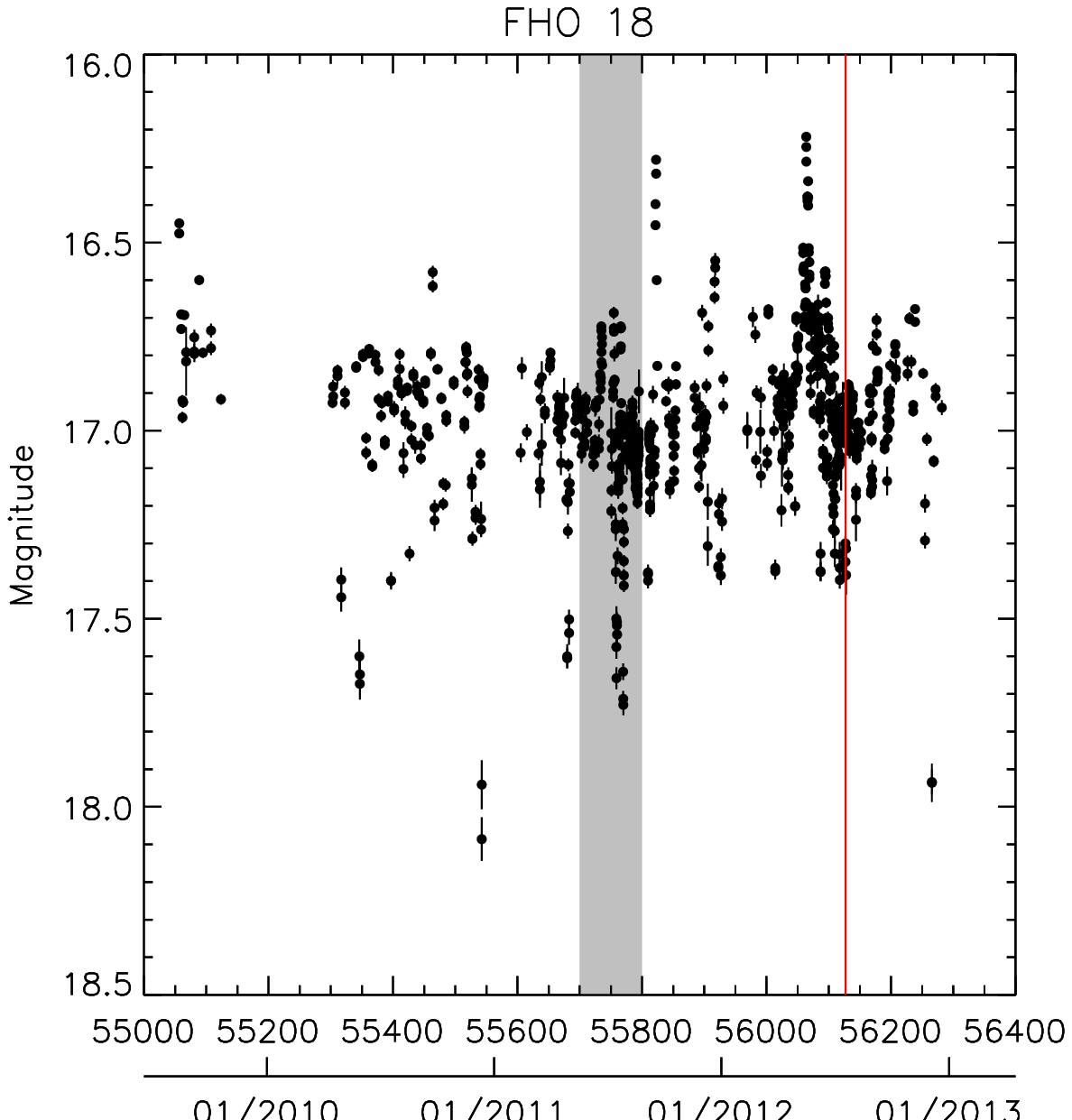}
\vspace{10pt}
\includegraphics[width=0.47\textwidth]{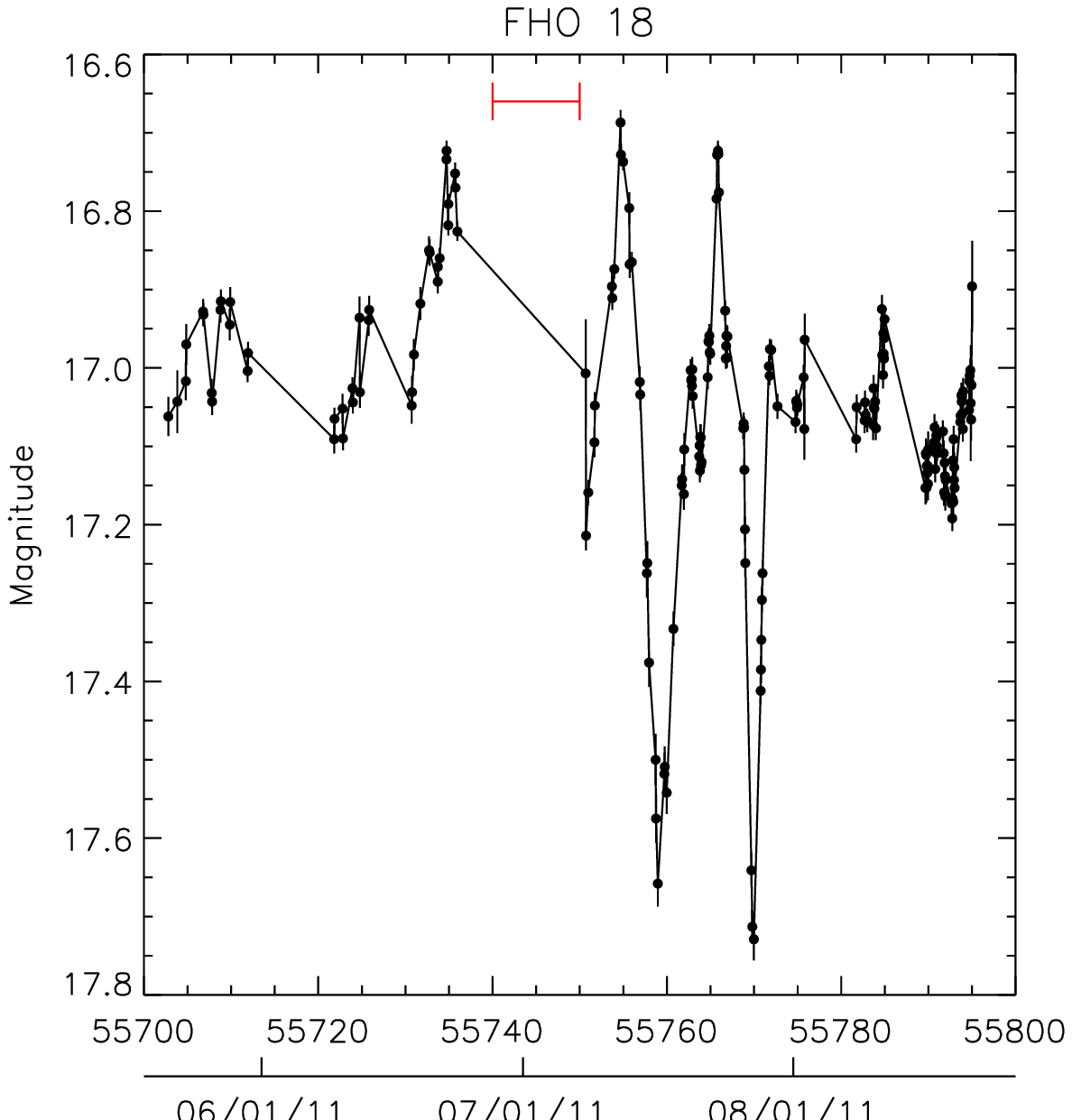}
\includegraphics[width=0.45\textwidth]{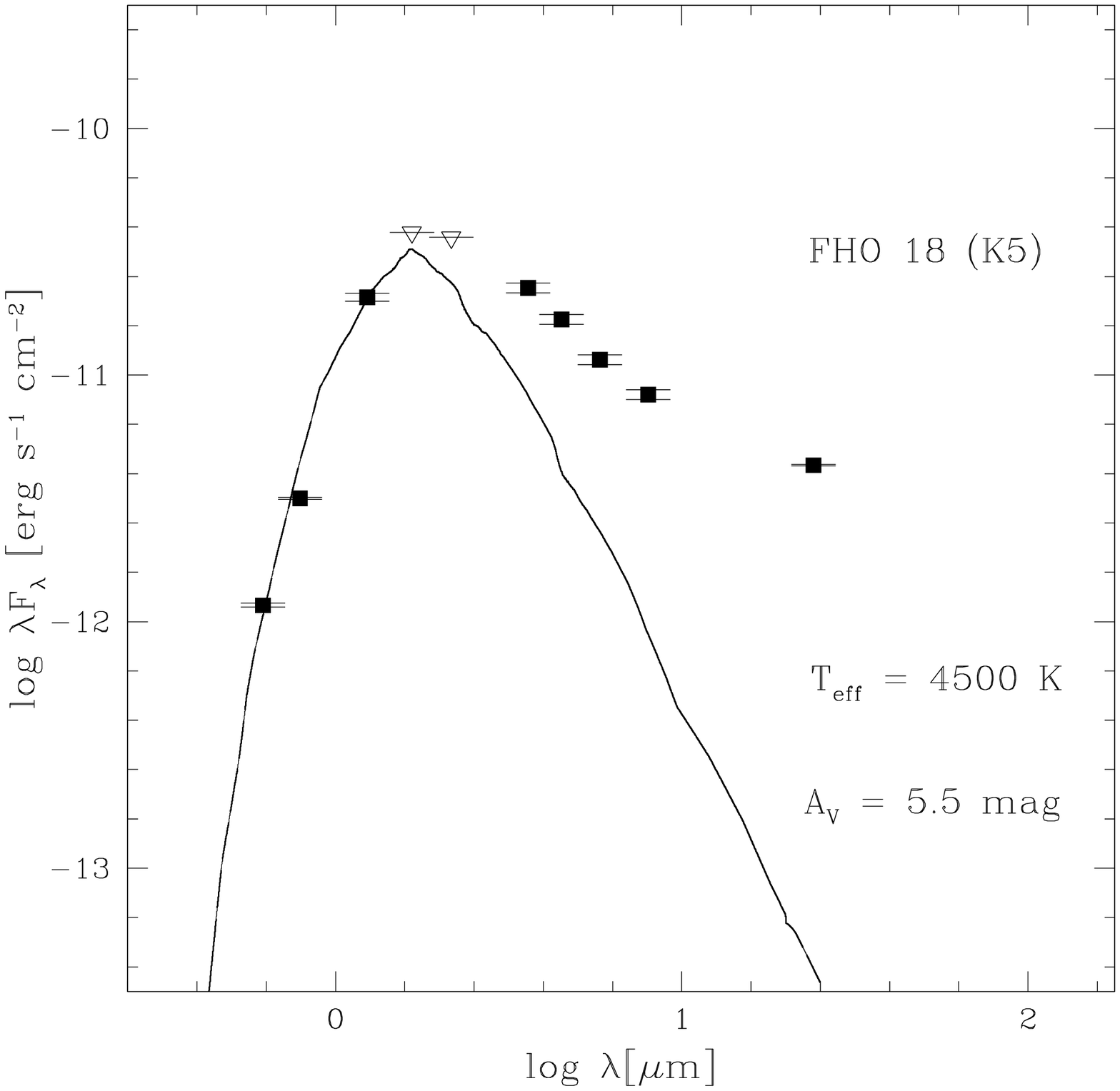}
\qquad\quad\ \includegraphics[width=0.47\textwidth]{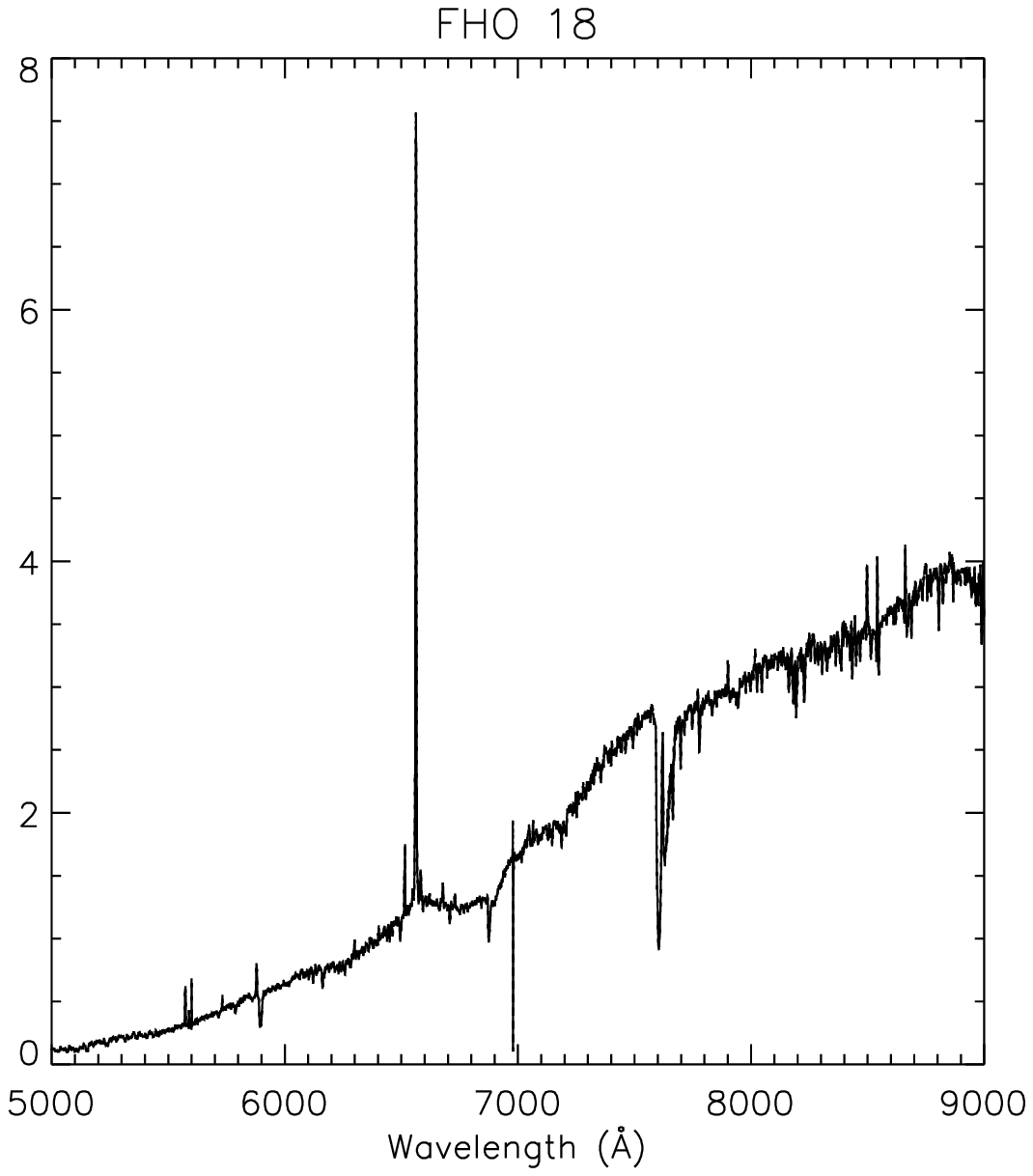}
\caption{Same as Figure~\ref{lc_1220bv}, but for a star showing an odd combination of bursting and fading. The upper right panel shows the fades and their precursor bursts.}\label{lc_1220cd}
\end{figure}

FHO~18, shown in the upper panels of Figure~\ref{lc_1220cd}, faded twice by 0.6~mag in quick succession in 2011. Immediately before each fading event, it brightened by 0.3~mag (upper right panel). This behavior was not repeated for other fading events during our monitoring period. Aside from these two fades and their precursor bursts, FHO~18 appears to be a typical Class~II young star.

Our DEIMOS spectrum of FHO~18 was taken during a 0.4~mag fade. The spectrum shows a K5 star with H$\alpha$ emission (-23~\AA\ equivalent width) as well as weaker Ca~II and He~I emission. However, as this fade was \emph{not} preceded by a burst the spectrum does not directly constrain the star's unusual behavior in mid-2011.

\subsection{FHO~27}\label{fho27}

FHO~27 had only 0.5-0.6~mag variability with a roughly constant or slightly rising mean magnitude throughout 2009-2010, but then began to show deep (up to 2~mag) fading events from late 2011 onward. At the same time, the upper envelope of the light curve began to gradually dim, leveling off in mid-2012 after a total decrease of roughly 0.8-0.9~mag. The minimum magnitude reached during each fading event rose from 17.9 in the first fade to 17.3 in late 2012, so that the most recent fades have only a few tenths of a magnitude depth. 

\begin{figure}
\includegraphics[width=0.47\textwidth]{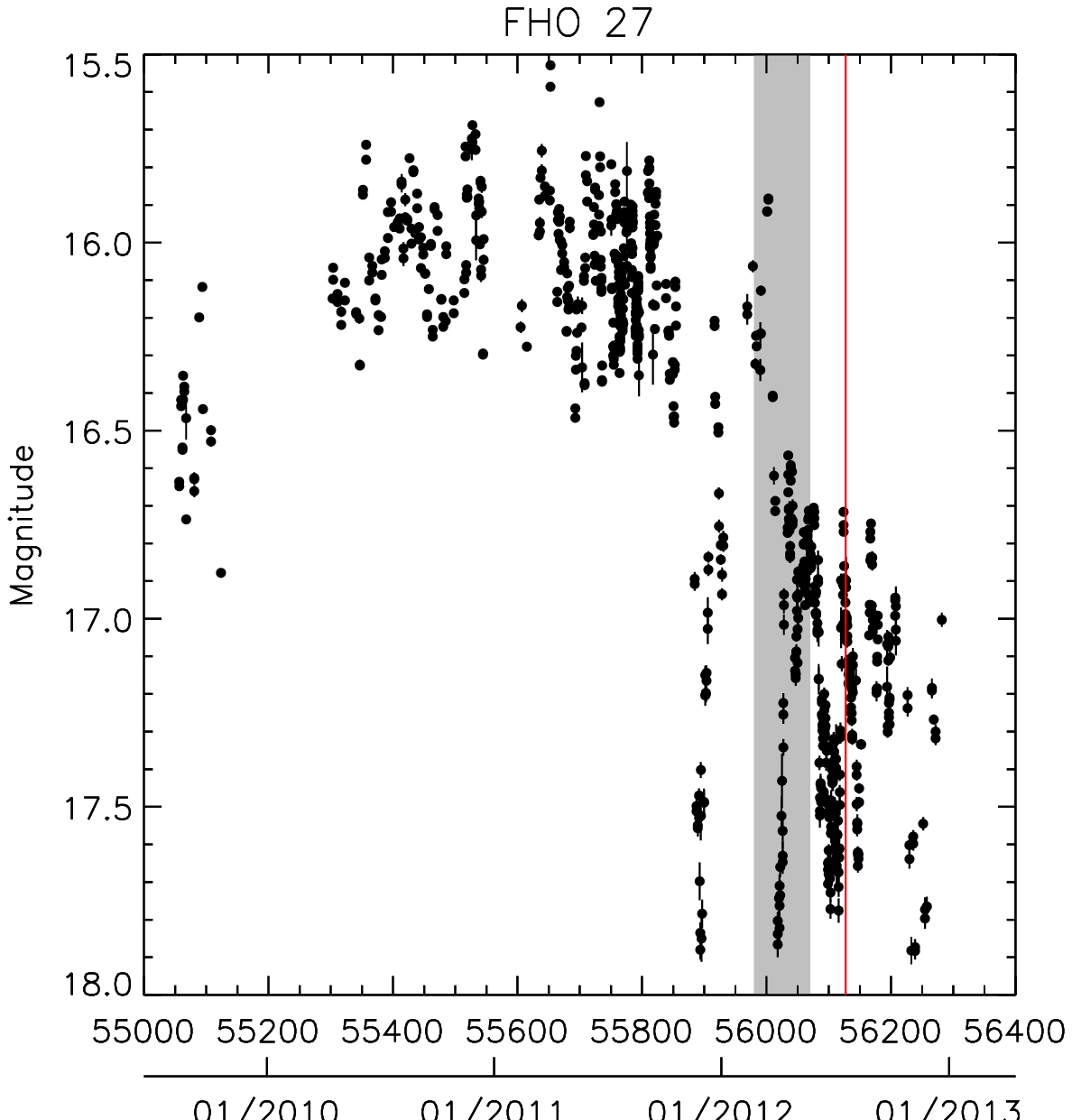}
\vspace{10pt}
\includegraphics[width=0.47\textwidth]{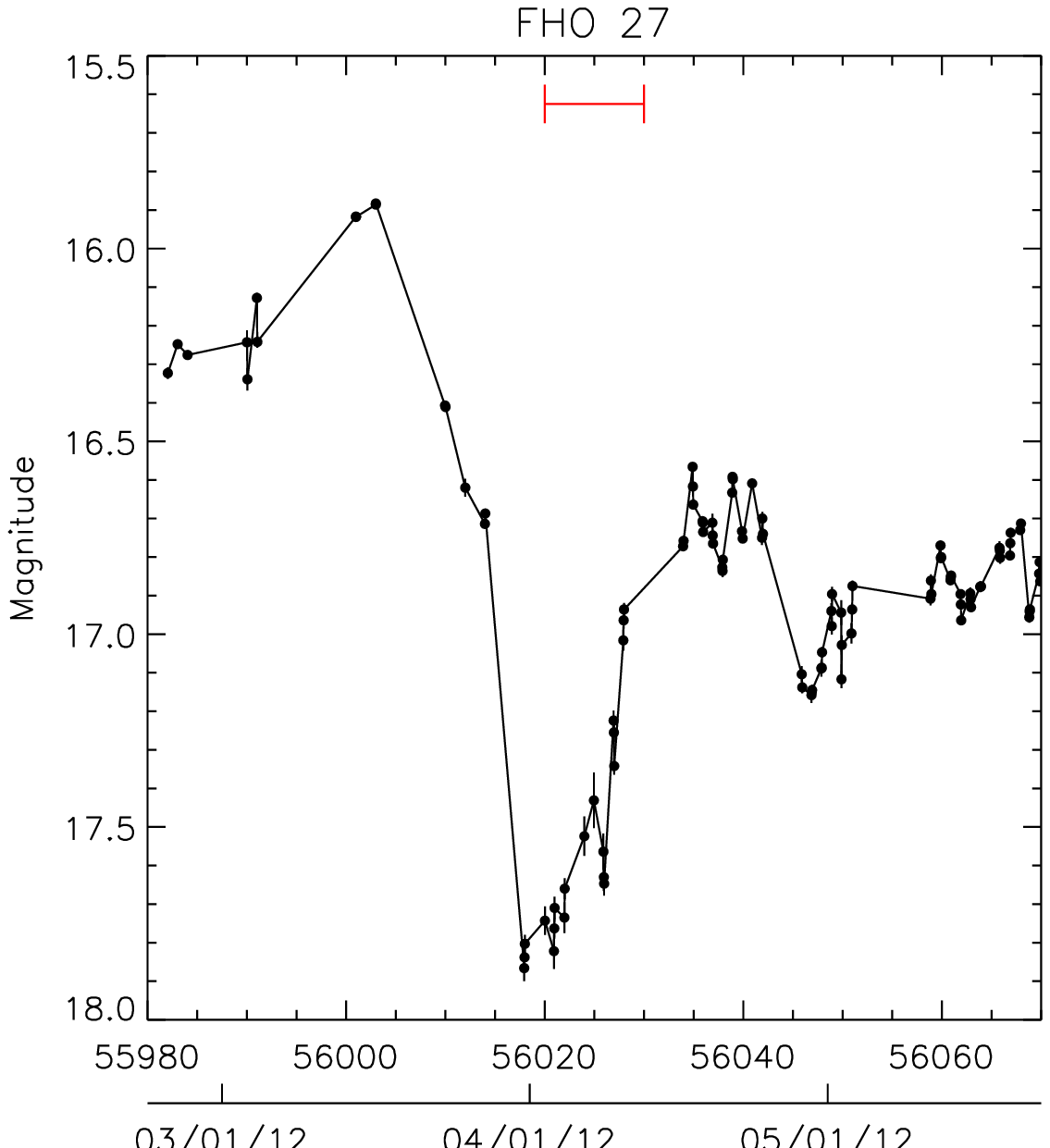}
\includegraphics[width=0.45\textwidth]{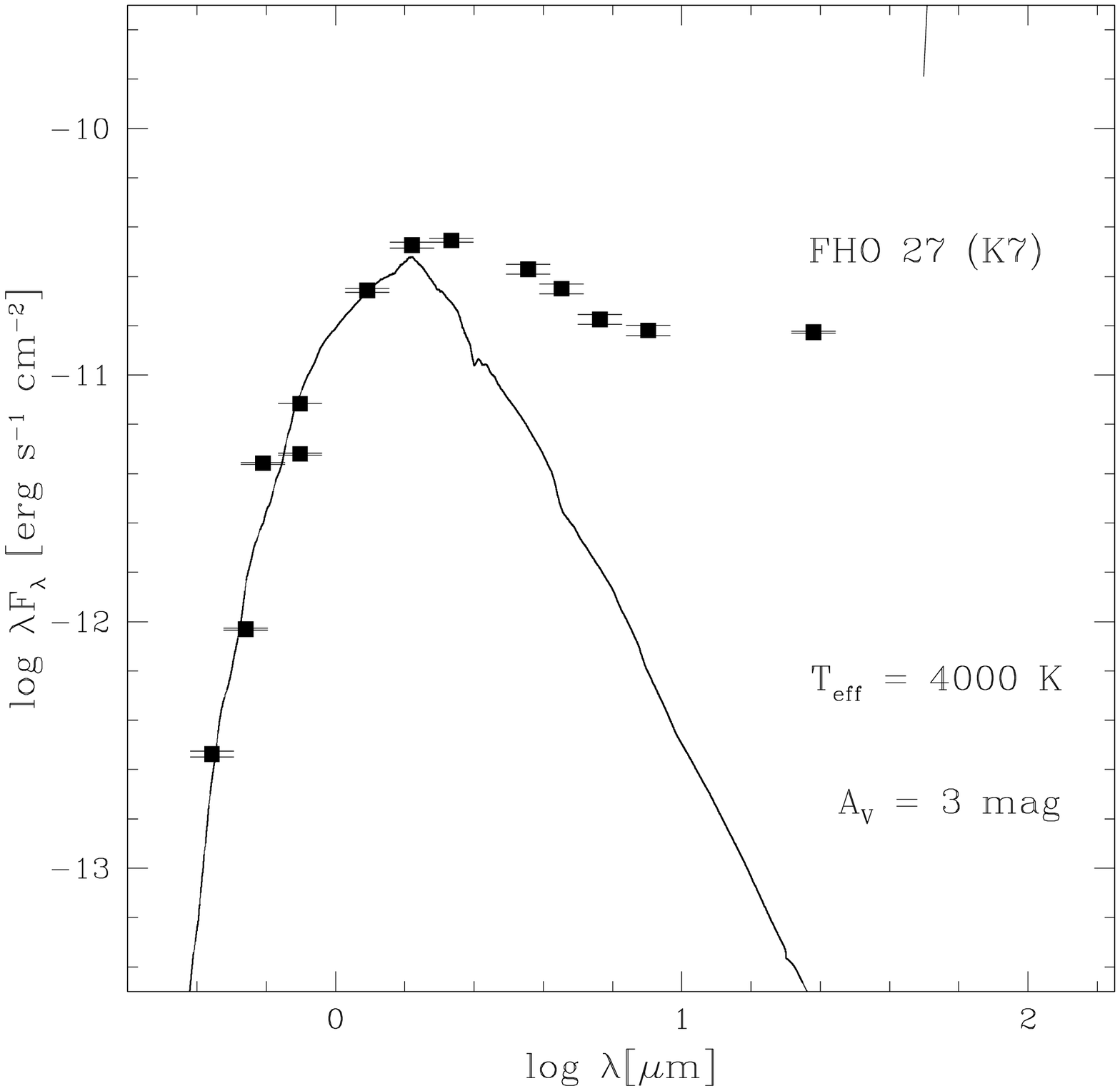}
\qquad\quad\ \includegraphics[width=0.47\textwidth]{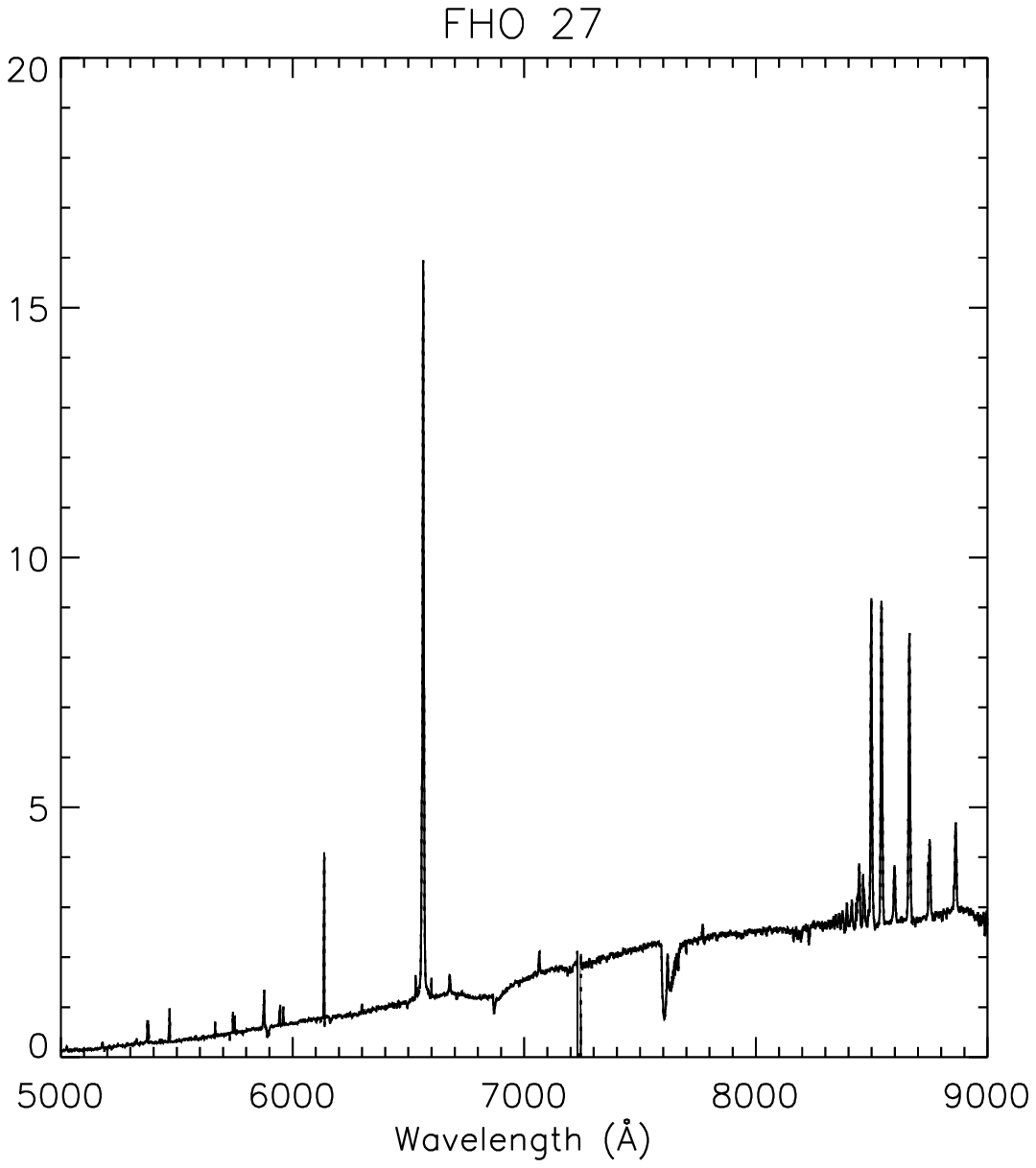}
\caption{Same as Figure~\ref{lc_1220bv}, but for a star showing a series of fades superimposed on a year-long decay. The upper right panel highlights the asymmetric profile of one of the fades.}\label{lc_1220bf}
\end{figure}

While the fading events repeat every 30-90 days, they are not periodic. 
In addition, each fade has a dramatically different profile from those before and after, with many of the profiles showing strong asymmetries (Figure~\ref{lc_1220bf}, upper right panel), and some fades being much shallower than the rest (e.g., a 0.3~mag fade in late April 2012 was sandwiched between a 1.1~mag and a 0.6~mag fade). 

The spectral energy distribution of the source, shown in the lower left panel of Figure~\ref{lc_1220bf}, has a strong infrared excess; \citet{RGS2011} classify FHO~27 as a flat-spectrum source. 

We acquired one spectrum of FHO~27 in July~2012, during the star's long-term low state but between the deeper fading events. The K7 spectrum in the lower right panel of Figure~\ref{lc_1220bf} shows strong H$\alpha$ (-80~\AA\ equivalent width), Paschen series, and Ca~II emission. Weaker lines in the spectrum include He~I, [O~I], and O~I.

\subsection{FHO~28}\label{fho28}

\begin{figure}
\includegraphics[width=0.47\textwidth]{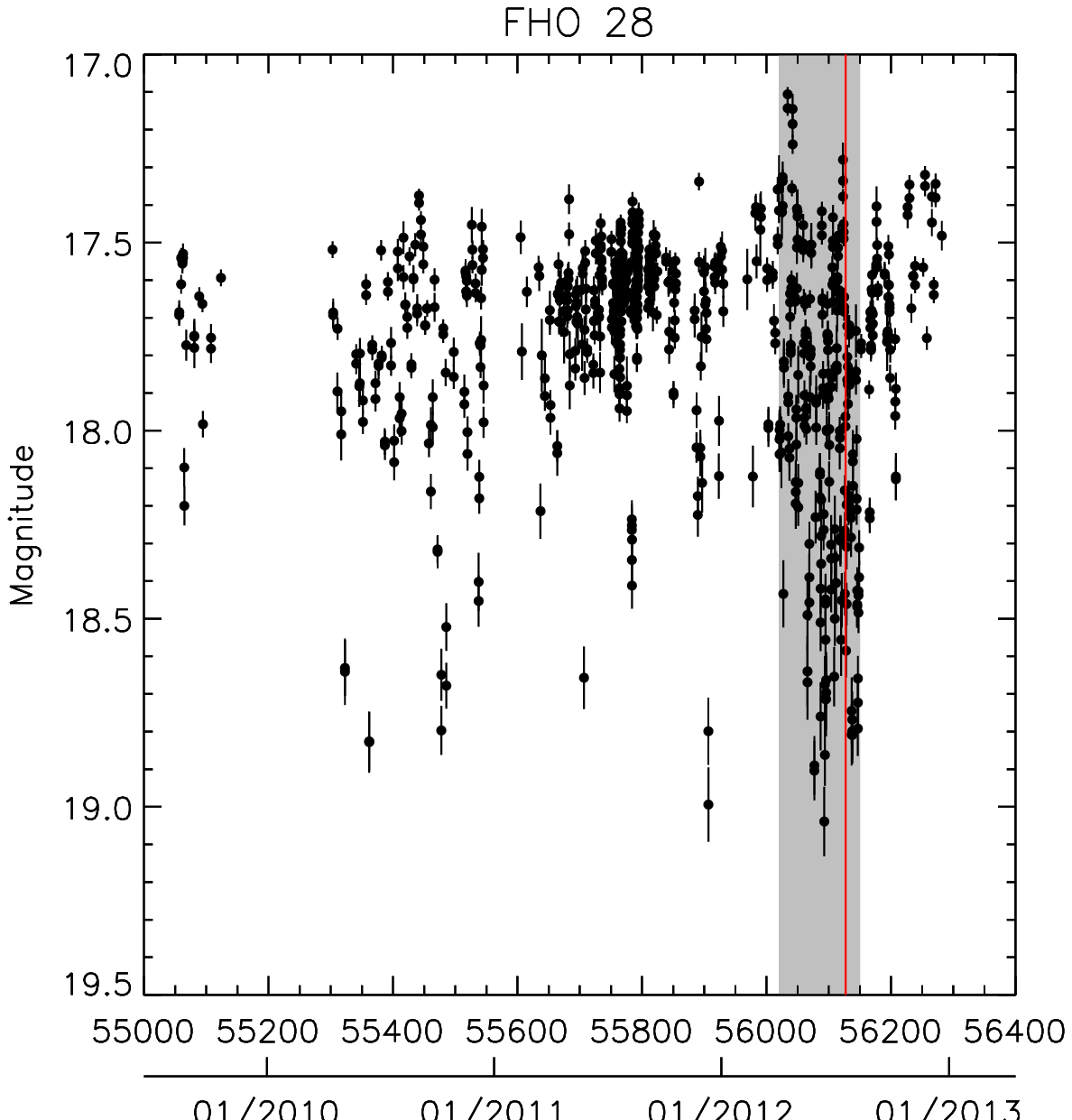}
\vspace{10pt}
\includegraphics[width=0.47\textwidth]{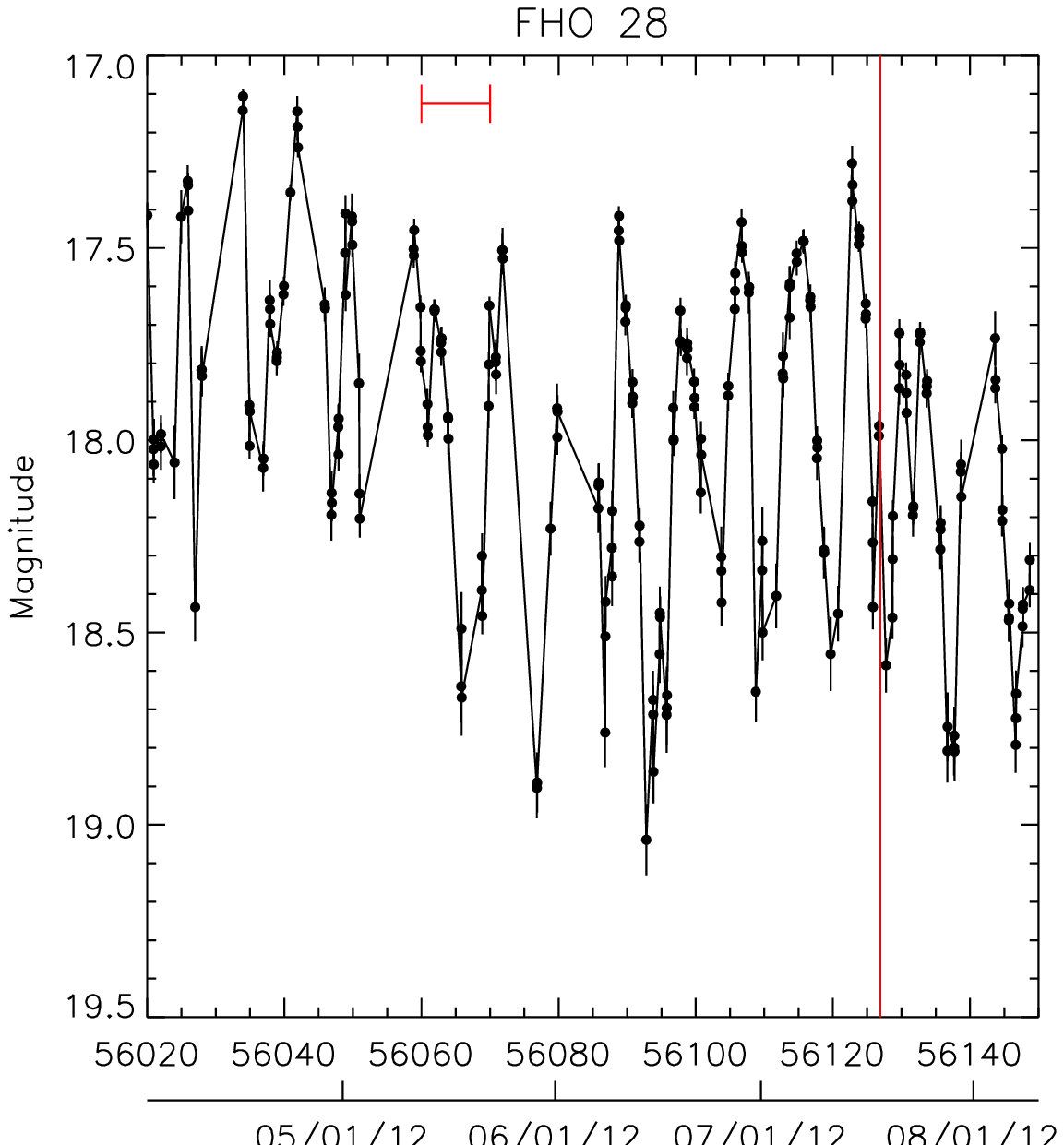}
\includegraphics[width=0.45\textwidth]{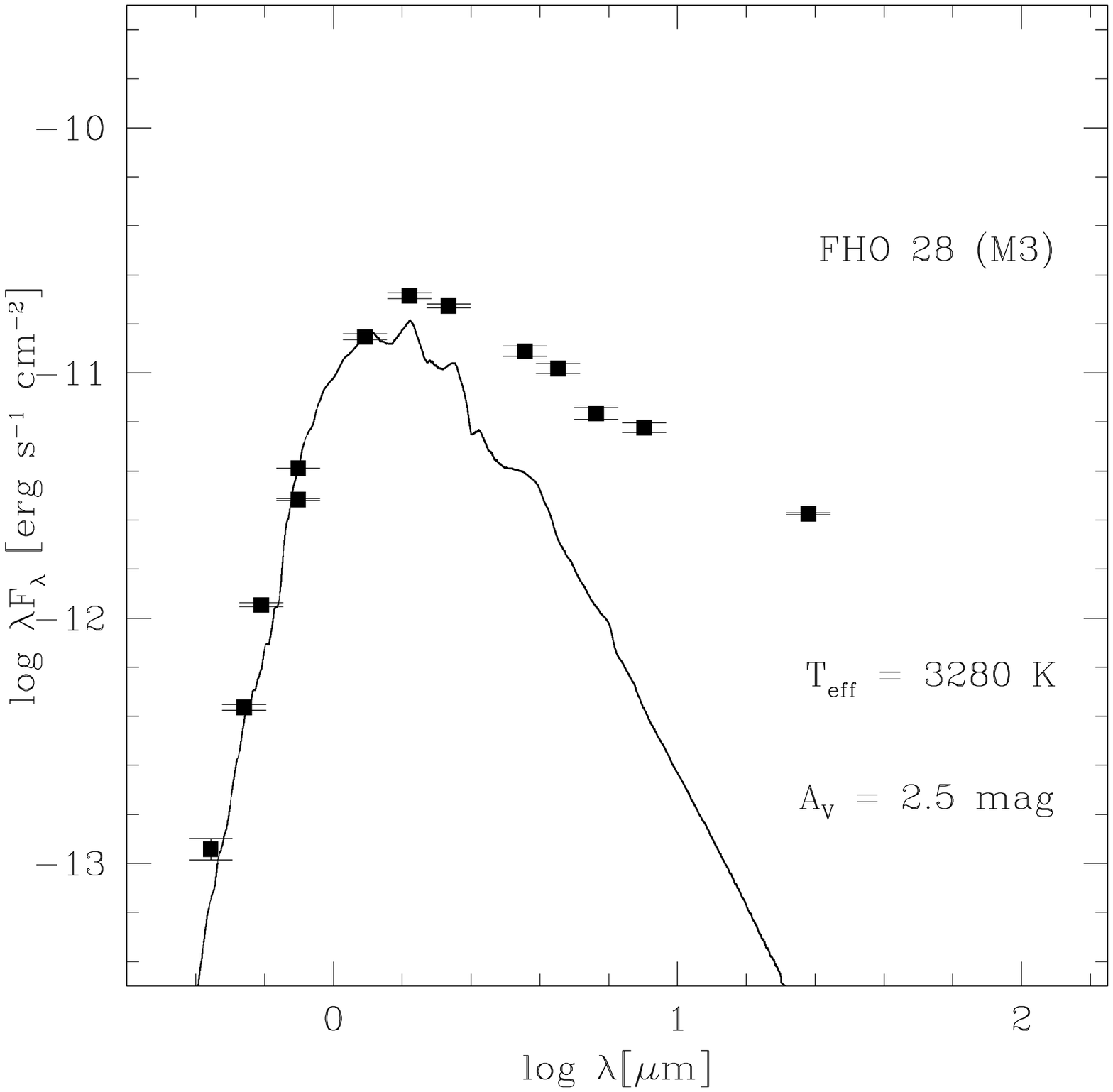}
\qquad\quad\ \includegraphics[width=0.47\textwidth]{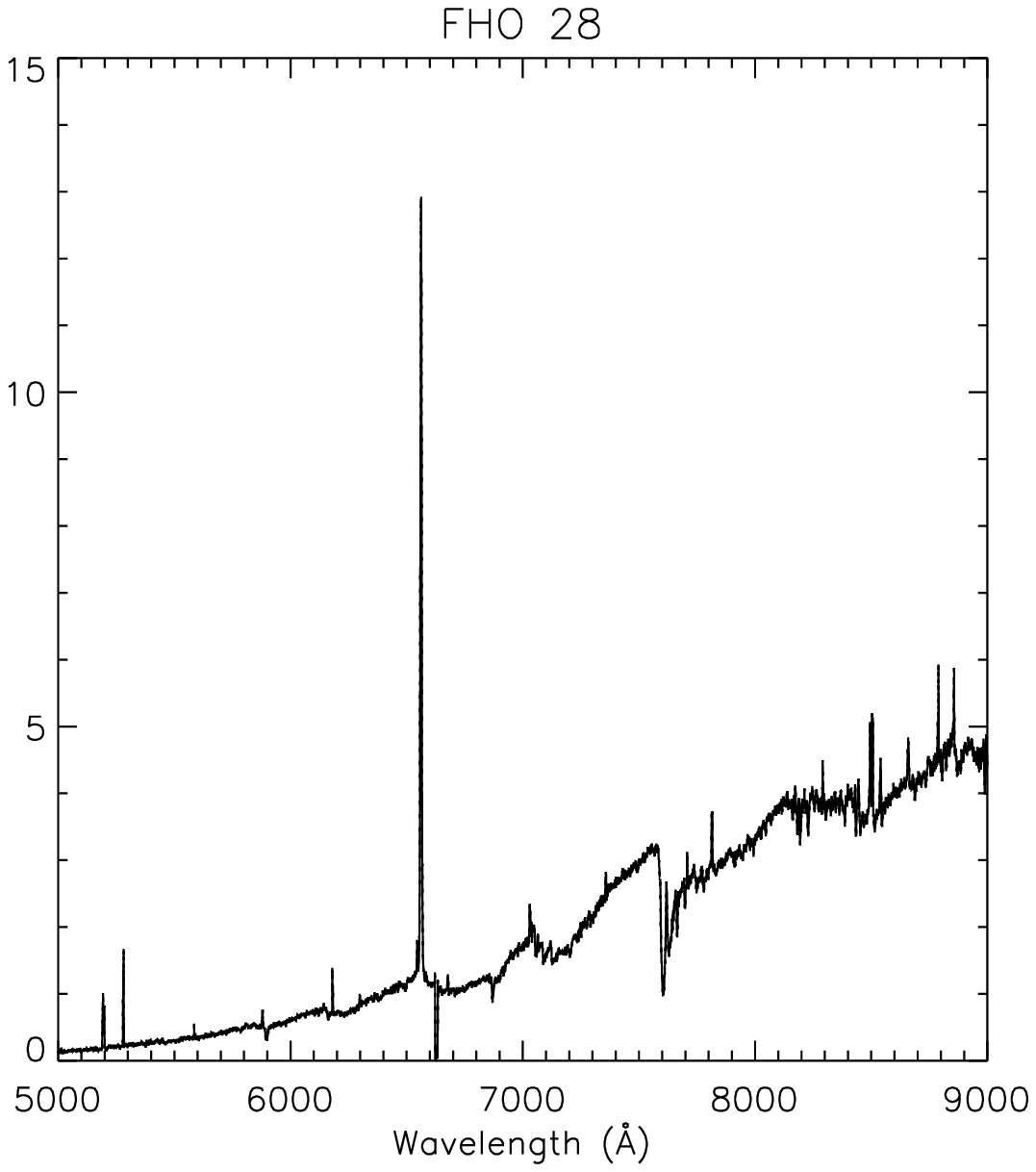}
\caption{Same as Figure~\ref{lc_1220bv}, but for a star showing an increased frequency of fades in 2012. The upper right panel illustrates that the fades in the more active phase were nearly superimposed.}\label{lc_1220br}
\end{figure}

Like FHO~27, FHO~28 was dominated by 0.6~mag irregular variability in the first few years of our survey, interrupted by occasional 1~mag fades. Then, in early 2012, it began showing rapid variability with the same maximum brightness, but with a much higher amplitude of 2~mag. The high-amplitude variability lasted 130~days before the star returned to its earlier behavior. Since the source was not strictly periodic during its high-amplitude phase, it is not clear whether the variability has been fully resolved at our daily cadence, in which case the fades are roughly 9 days apart, or whether we are seeing a strobing effect of a more rapid 23-hour variation. The light curve is shown on the top two panels of Figure~\ref{lc_1220br}.

The spectral energy distribution, shown in the lower left panel of Figure~\ref{lc_1220br}, shows a Class~II infrared excess. A spectrum of FHO~28 (Figure~\ref{lc_1220br}, lower right panel), taken during its strongly varying phase, shows an M3 star with strong H$\alpha$ emission (-60~\AA\ equivalent width) and weak Ca~II lines. An older spectrum shows that H$\alpha$ was much weaker (-20~\AA) in 1998, although since we don't know the photometric state of FHO~28 at the time, it is not clear whether the difference between the two spectra is related to the star's increased activity in 2012.

FHO~28 is yet another example of how the photometric behavior of young stars can change abruptly from one year to the next. This source would \emph{not} be classified as a fader if we had only data from its active phase, as the photometry shows no preference between high and low magnitudes (Figure~\ref{lc_1220br}, upper right panel). It is the comparison to previous years that allows us to establish that the brighter magnitudes represent the unperturbed state of the star.

\section{Summary and Discussion}\label{discussion}

\subsection{Key Results}

We have presented first results from a new variability survey of young stars that probes a large dynamic range of timescales, from roughly a day to roughly a year. We have used this new data set to uniformly identify stars with episodic variability, regardless of whether the episodes had day-scale, month-scale, or year-scale durations and regardless of whether the episodes were periodic. From a sample of 186 candidate members of the North America Nebula complex, we have identified 14 that showed episodes of brightening (``bursters'') and 29 that showed episodes of dimming (``faders''), sometimes mixed with more erratic variability. Two stars showed both bursting and fading at separate epochs. We have presented basic photometric and spectroscopic properties of both bursters and faders.

We have found that:
\begin{enumerate}
\item Most high-amplitude variables have a strong infrared excess, while low-amplitude variables may or may not have a strong excess. While similar correlations have been noted before, i.e., that classical T Tauri stars tend to have higher amplitudes than weak-lined T Tauri stars, we show here that a correlation between degree of infrared excess and variability amplitude also holds among stars with infrared excess.
\item Even within the individual burster and fader classes, we see a wide variety of timescales, amplitudes, and burst or fade profiles. This includes events that occur only once or twice in our three-year monitoring period, and would be missed in a shorter survey. It is not clear whether these varied behaviors imply varied underlying mechanisms. We find no gap separating groups of bursters or faders with different amplitudes or timescales (Figure~\ref{dipwidths}), suggesting that they are all members of a single population, but in-depth study of representative objects will be needed to settle the issue.
\item We identify three bursters whose photometric and spectroscopic characteristics are consistent with published models of accretion driven by instabilities at the boundary between the stellar magnetosphere and the circumstellar disk. To our knowledge, this is the first time candidate objects corresponding to these models have been identified.
\item A substantial number of sources show variability over long timescales. Among other examples, FHO~14 and FHO~28 showed enhanced fading activity in an interval 100-200 days long. [OSP2002]~BRC~31~1 changed from a 15th magnitude star in 2010 to a 18-19th magnitude star in 2012. [OSP2002]~BRC~31~8 and FHO~29 both showed bursting modulated by a timescale of roughly 300 days. Except for the sudden decay of [OSP2002]~BRC~31~1, these are behaviors that have not been associated with bursting or fading activity before, for lack of sufficient sampling.
\end{enumerate}

\subsection{Comparison to Previous Work}\label{litcompare}

While much previous time domain work on young stars has focused on finding and characterizing periodic variables, there have been some studies of more general variability. Here we discuss whether our population statistics are consistent with the existing literature.

We see bursting behavior in 14~sources, or $12\pm3\%$ of the $R < 18$~mag variables with an infrared excess. To the best of our knowledge, no one has reported a short-term optical burst fraction for pre-main-sequence stars, so we have no published results to which to compare this figure.

We see fading behavior in 29~sources, or $25\pm4\%$ of the variables with an infrared excess and $16\pm3\%$ of all sources with a good PTF light curve and an infrared excess. For comparison, \citet{AlencarDips} found 28\% of stars selected from X-ray or H$\alpha$ emission, all variables, showed periodic fading behavior in unfiltered optical light. \citet{YSOvar_first} found fades (periodic or not) in the mid-infrared in 5\% of variables and 3\% of their total sample, selected by proper motion, X-ray or H$\alpha$ emission, or infrared excess. Finally, \citet{SigOri} found I-band fading behavior in 6\% of their variables and 5\% of their total sample, selected by kinematics, H$\alpha$ emission, forbidden line emission, lithium absorption, or infrared excess. Each of these surveys was a few weeks in duration, shorter than our survey, but had higher cadence by factors of 10-200.

To test whether our results are consistent with previous work after accounting for differences in our observing strategies, we clipped our light curves to a 30-day period of high-cadence observations, up to eight per night, between JD~2455765.5 and 2455795.5. This allowed us to compare our data to \citet{YSOvar_first}, who observed their field for a month at roughly a 2-hour cadence. We found that 12 of our faders (LkH$\alpha$~174, V1701~Cyg, and FHO 3, 5, 14, 16, 18, 19, 20, 21, 22, and 28) were recognizable as such during the 30-day period, indicating that with only a month of high-cadence data we would have reported a $10\pm3\%$ fader fraction out of the variables in our sample or $6\pm2\%$ of the infrared-selected sample. 
This is slightly higher than, though consistent with, the \citeauthor{YSOvar_first} results. Since our ground-based survey had more data gaps than the Spitzer observations of \citeauthor{YSOvar_first}, however, our fader rate had we observed with their exact cadence may have been higher.
On the other hand, it is possible that we are overestimating our recovery rate, since we had already identified these stars as faders using the full data set and were aware of their nature while examining the clipped light curves, introducing hindsight bias.

\subsection{Limitations of the Present Work}

\label{biases}
We were careful to identify bursting and fading events using only the light curves themselves, and not any ancillary data such as SEDs or spectra, to avoid psychological biases in interpreting ambiguous cases. However, we could not eliminate all ambiguity: the qualitative nature of event identification inevitably made some kinds of events easier to detect than others. 
The easiest events to identify were either those where the event lasted for several days, so that the light curve resolved the event profile, or those where the event repeated many times over our observing baseline, so that we could be confident that a high or low point represented real variability rather than a statistical fluke or an isolated error in the data reduction. We tried to confirm, using thumbnail images, whether isolated high or low points represented brief but real changes in the stellar flux, but image inspection allowed us to verify only high-amplitude events. 
We therefore may be incomplete to variability on timescales of a few hours.

We also had difficulty identifying bursts or fades lasting longer than several months, particularly if they were superimposed on other variability. Some stars in our sample showed erratic variability on timescales of months, and it is not always clear from only three years of data whether a star that spent several months in a high (or low) state had undergone an anomalous change in brightness, or whether we were merely seeing an extreme in a continuous series of brightness fluctuations. We chose to err on the side of caution, and only counted sources where the light curve apart from the candidate burst or fade had much lower-amplitude variability. However, this introduced a bias against mixed variability modes.

\label{missingdata}
There are at least two sources in the North America Nebula complex that, while they meet our definition of bursters, are absent from our sample.
PTF10qpf \citep{ptf10qpf} was an $R = 16.5$ star at the beginning of the survey that brightened to $R \sim 12.5$ in mid-2010 and has remained there since. The source was disqualified from this paper's sample because it failed three criteria in the photometry produced by the PTF Photometric Pipeline: it was flagged as blended with nearby stars at nearly all epochs, it was flagged as saturated in nearly all epochs after the outburst, and its median magnitude of 12.9 was well above our flux limits. 
PTF10nvg \citep{ptf10nvg} did not rise past PTF's saturation limits; however, as noted in Section~\ref{completeness}, the PTF Photometric Pipeline had difficulty identifying sources around nebulosity. PTF10nvg is located just off a bright nebula filament, and neither it nor any other nearby sources were extracted.
These two omissions illustrate key sources of incompleteness in this work: crowding, nebula contamination, and a limited magnitude range. Fortunately, these problems do not apply to the majority of sources in the Spitzer-selected sample, which are well-separated, in low-background regions, and of less than one magnitude amplitude.

%\subsection{Future Work}
\label{future}

This work is based primarily on a long-term, single-band photometric survey, which has allowed us to identify and characterize new types of bursters and faders. However, the data we have presented here cannot identify without ambiguity the physics behind each kind of bursting or fading variability, or even whether all bursters or all faders represent different cases of a common variability mechanism. 
The following additional data would provide more insight into the nature of bursters or faders:
\begin{itemize}
\item Time series color information would help test whether all of the faders are caused by variable extinction along the line of sight to the star. Color data would also help us interpret bursters by providing color constraints to better estimate the energy released in the burst. We plan to present a color analysis of our bursters and faders in future work.
\item Spectroscopic monitoring, especially at high dispersion, would allow us to compare accretion and wind indicators in a star's high and low states, allowing us to distinguish which events are accretion-powered, which represent partial or total obscuration of the photosphere, and which are driven by something else entirely. 
\item Polarimetry would help identify which bursts or fades are associated with changes in the obscuration of the star, as it probes what fraction of the measured flux comes from the photosphere and what fraction is scattered from the disk \citep[e.g.,][]{1992A&AT....3...17G,AATau_first}. In particular, it could be used to directly test the hypothesis that all fades are obscuration events -- if they are, then they should all show stronger polarization at minimum light.
\end{itemize}

We have shown that the class of faders is far broader than previously appreciated, and that bursters, while fewer in number, show a comparable diversity. We have identified new phenomenology within both classes. These objects can serve as prototypes for future study of particular forms of bursting or fading activity.

\acknowledgments

We would like to thank 
Asaaf Horesh, Kunal Mooley, and Yi Cao for acquiring a spectrum of FHO~7 for us, 
Gregory Herczeg for reducing the HYDRA and Norris spectroscopic data, 
Susan Tokarz for reducing the Hectospec data, 
and 
Evan Kirby for helping us with the DEIMOS pipeline. 
We thank the anonymous referee for several comments that improved our presentation. 
We also thank John Carpenter and Nairn Baliber for much helpful discussion and suggestions for tests to run, 
and 
Luisa Rebull for providing us with a copy of the Spitzer data. 

%This paper makes use of a spectral energy distribution script written by Sylvain Guieu and available at \url{http://www.pinpinatla.com/NaNeb\_yso/selec2.php}.
%
This paper uses data products produced by the OIR Telescope Data Center, supported by the Smithsonian Astrophysical Observatory.
Some of the data presented herein were obtained at the W.M. Keck Observatory, which is operated as a scientific partnership among the California Institute of Technology, the University of California and the National Aeronautics and Space Administration. The Observatory was made possible by the generous financial support of the W.M. Keck Foundation.

%% See the AASTeX Web site at http://aastex.aas.org/facilities/

%% for the paper. Each keyword will be checked against the master list during
%% copy editing. Individual instruments or configurations can be provided 
%% in parentheses, after the keyword, but they will not be verified.

\textit{Facilities:} \facility{PO:1.2m (P48 Survey Camera)}, \facility{Keck:II (DEIMOS)}, \facility{MMT (Hectospec)}, \facility{WIYN (HYDRA)}, \facility{Hale (Norris)}

\bibliographystyle{apj}
\bibliography{flaredip,../../ysovar,../../ptf,../../references}

\clearpage

%DIPPERS
%Min amplitude: 0.3~mag, [FHO~23]
%Max amplitude: 2.5~mag, BRC~31~1 OR 1.8~mag, [FHO~7], FHO~27
%Min dip length: <1~day, [FHO~5], FHO~30
%Max dip length: 330+~days, [BRC~31~1]
%Min repeat time: 9~days, [FHO~19], FHO~28
%Max repeat time: ???, LkH$\alpha$~150, FHO~8, FHO~25, [[CP2005]~17]
%Max repeat time: 1 year, [FHO~21], FHO~22
%
%FLARERS
%Min amplitude: 0.2~mag, [FHO~2]
%Max amplitude: 1.4~mag, [FHO~29]
%Min flare length: 2-3~days, LkH$\alpha$~139, [V1716~Cyg]
%Max flare length: 150~days, [FHO~17]
%Min repeat time: 10~days, [FHO~26]
%Max repeat time: ???, LkH$\alpha$~185
%Max repeat time: 1 year, LkH$\alpha$~139, [FHO~4], FHO~2, FHO~29

\clearpage

\begin{deluxetable}{l c c c c c}
%\tabletypesize{\scriptsize}
%\tabletypesize{\small}
\tablewidth{0pt}
\tablecaption{Fits to RMS as a Function of Magnitude by Chip}
\tablehead{\colhead{Chip} & \colhead{$a$} & \colhead{$b$} & \colhead{$p$} & \colhead{Total Sources} & \colhead{Variables}}
\startdata
0 & 0.033 & 0.0053 & 0.68 &  6090 & 491 (8.1\%) \\
1 & 0.018 & 0.0130 & 0.52 &  9330 & 490 (5.3\%) \\
2 & 0.023 & 0.0145 & 0.49 & 14057 & 533 (3.8\%) \\
6 & 0.026 & 0.0072 & 0.62 & 11036 & 457 (4.1\%) \\
7 & 0.025 & 0.0081 & 0.60 &  7720 & 337 (4.4\%) \\
8 & 0.022 & 0.0105 & 0.55 & 10393 & 397 (3.8\%) \\
\enddata
\tablecomments{Best-fit values for the parameters in Equation~\ref{rmsfit} for each of the six target chips. The parameter $a$ can be interpreted as the systematic noise floor at bright magnitudes, and $p$ is the power-law dependence of RMS on flux at faint magnitudes. The last two columns show the total number of PTF sources on each chip as well as the number selected by making an RMS cut at 1.75 times the value given by Equation~\ref{rmsfit}.}
\label{rmsthresholds}
\end{deluxetable}

\begin{deluxetable}{l c r r r r r}
%\tabletypesize{\scriptsize}
\tabletypesize{\footnotesize}
%\tabletypesize{\small}
\tablewidth{0pt}
\tablecaption{Selection Effects in the PTF Sample}
\tablehead{\colhead{IR Source Type} & \colhead{\# in RGS2011} & \colhead{in PTF Field} & \colhead{PTF Counterparts} & \colhead{With $R < 18$} & \colhead{Flags in $<50\%$ Epochs} & \colhead{High RMS}}
\startdata
MIPS-Only     &  25 &  25 (100\%) &   0 ( 0\%) &            &             &  \\
Class I       & 273 & 242 ( 89\%) &  11 ( 5\%) &   3 (27\%) &   3 (100\%) &  3 (100\%) \\
Flat-Spectrum & 272 & 242 ( 89\%) &  53 (22\%) &  20 (38\%) &  13 ( 65\%) & 10 ( 77\%) \\
Class II      & 604 & 542 ( 90\%) & 321 (59\%) & 160 (50\%) & 120 ( 75\%) & 79 ( 66\%) \\
Class III     & 112 &  82 ( 73\%) &  76 (93\%) &  43 (53\%) &  31 ( 72\%) & 16 ( 52\%) \\
IRAC-Only     & 796 & 613 ( 77\%) & 140 (23\%) &  27 (19\%) &  19 ( 70\%) &  9 ( 47\%) \\
\hline
Total         &2082 &1746 ( 84\%) & 601 (34\%) & 253 (42\%) & 186 ( 74\%) &117 ( 63\%) \\
\enddata
\tablecomments{\citet{RGS2011} gave SED classes only for sources that were detected in both IRAC and MIPS. Sources that were detected in only one or the other are listed for comparison, but were not used to estimate the incompleteness from source confusion and flux limits.}
\label{ptfcomplete}
\end{deluxetable}

\begin{deluxetable}{l l c c c c c c c}
%\tabletypesize{\scriptsize}
\tabletypesize{\footnotesize}
%\tabletypesize{\small}
%\tablewidth{0pt}
\tablecaption{Statistics of the Candidate Bursters and Faders}
\tablehead{\colhead{[RGS2011] ID} & \colhead{Short} & \colhead{SED} & \colhead{$\bar{R}$} & \colhead{$R_{med}$} & \colhead{RMS} & \colhead{$\Delta R$} & \colhead{Total} & \colhead{Unflagged} \\ 
\colhead{} & \colhead{Name} & \colhead{Class} & \colhead{(mag)} & \colhead{(mag)} & \colhead{(mag)} & \colhead{(mag)} & \colhead{Detections} & \colhead{Detections}}
\startdata
205032.32+442617.4 & FHO~1              & II   & 17.4 & 17.5 &  0.13 & 1.12 &  873 &  870 \\
205036.93+442140.8 & [OSP2002] BRC~31~1 & I    & 16.8 & 16.6 &  1.24 & 4.94 &  750 &  505 \\
205040.29+443049.0 & LkH$\alpha$~139    & II   & 14.6 & 14.6 &  0.13 & 0.92 &  883 &  778 \\
205042.78+442155.8 & [OSP2002] BRC~31~8 &      & 16.9 & 16.9 &  0.18 & 1.97 &  877 &  872 \\
205100.90+443149.8 & V1701~Cyg          & II   & 15.5 & 15.4 &  0.36 & 1.89 &  769 &  629 \\
205114.80+424819.8 & FHO~2              & III  & 13.8 & 13.8 &  0.075& 0.91 &  857 &  750 \\
205115.14+441817.4 & LkH$\alpha$~150    & II   & 16.4 & 16.3 &  0.29 & 2.22 &  879 &  833 \\
205119.43+441930.5 & FHO~3              & II   & 17.1 & 17.0 &  0.48 & 2.74 &  873 &  865 \\
205120.99+442619.6 & LkH$\alpha$~153    & II   & 15.1 & 15.1 &  0.13 & 0.80 &  882 &  866 \\
205123.59+441542.5 & FHO~4              & II   & 17.6 & 17.6 &  0.13 & 1.13 &  875 &  875 \\
205124.70+441818.5 & FHO~5              & II   & 18.0 & 17.9 &  0.24 & 2.06 &  872 &  839 \\
205139.26+442428.0 & FHO~6              & II   & 14.8 & 14.7 &  0.26 & 1.49 &  884 &  872 \\
205139.93+443314.1 & FHO~7              & II   & 16.7 & 16.5 &  0.53 & 2.66 &  879 &  877 \\
205145.99+442835.1 & FHO~8              & II   & 17.8 & 17.7 &  0.27 & 2.71 &  874 &  870 \\
205155.70+443352.6 & FHO~9              & II   & 15.9 & 15.8 &  0.26 & 1.76 &  881 &  865 \\
205158.63+441456.7 & FHO~10             & Flat & 16.7 & 16.7 &  0.12 & 0.90 &  879 &  825 \\
205203.65+442838.1 & FHO~11             & II   & 18.0 & 17.9 &  0.13 & 0.98 &  870 &  836 \\
205228.33+442114.7 & FHO~12             & II   & 16.5 & 16.4 &  0.26 & 1.71 &  878 &  863 \\
205230.89+442011.3 & LkH$\alpha$~174    & II   & 16.7 & 16.7 &  0.28 & 1.51 &  878 &  878 \\
205252.48+441424.9 & FHO~13             & II   & 18.1 & 18.1\tablenotemark{a} &  0.42 & 2.63 &  874 &  795 \\
205253.43+441936.3 & FHO~14             & II   & 18.0 & 17.9 &  0.14 & 1.11 &  873 &  853 \\
205254.30+435216.3 & FHO~15             & II   & 17.1 & 17.0 &  0.34 & 2.01 &  877 &  785 \\
205314.00+441257.8 & FHO~16             & II   & 17.1 & 17.0 &  0.18 & 1.21 &  877 &  877 \\
205315.62+434422.8 & FHO~17             & II   & 17.3 & 17.5 &  0.49 & 2.21 &  848 &  720 \\
205340.13+441045.6 & FHO~18             & II   & 17.0 & 17.0 &  0.21 & 1.87 &  875 &  875 \\
205410.15+443103.0 & FHO~19             &      & 18.0 & 17.9 &  0.24 & 1.88 &  869 &  867 \\
205413.74+442432.4 & FHO~20             & II   & 16.3 & 16.2 &  0.21 & 2.36 &  876 &  876 \\
205424.41+444817.3 & FHO~21             & II   & 16.7 & 16.6 &  0.24 & 1.59 &  876 &  876 \\
205445.66+444341.8 & FHO~22             &      & 17.4 & 17.3 &  0.31 & 2.99 &  874 &  872 \\
205446.61+441205.7 & FHO~23             & II   & 17.4 & 17.3 &  0.16 & 0.87 &  763 &  665 \\
205451.27+430622.6 & FHO~24             & III  & 15.9 & 15.8 &  0.13 & 0.70 &  860 &  860 \\
205503.01+441051.9 & FHO~25             & Flat & 16.1 & 16.1 &  0.12 & 1.31 &  876 &  872 \\
205534.30+432637.1 & $[$CP2005$]$~17    & II   & 17.2 & 17.2 &  0.10 & 0.88 &  850 &  850 \\
205659.32+434752.9 & FHO~26             &      & 18.0 & 18.0\tablenotemark{a} &  0.23 & 1.72 &  861 &  827 \\
205759.84+435326.5 & LkH$\alpha$~185    & II   & 14.6 & 14.6 &  0.074& 0.70 &  884 &  884 \\
205801.36+434520.5 & FHO~27             & Flat & 16.6 & 16.3 &  0.61 & 2.35 &  878 &  866 \\
205806.10+435301.4 & V1716~Cyg          & II   & 16.5 & 16.5 &  0.16 & 1.10 &  879 &  879 \\
205825.55+435328.6 & FHO~28             & II   & 17.8 & 17.7 &  0.40 & 2.40 &  874 &  871 \\
205839.73+440132.8 & FHO~29             & Flat & 16.7 & 16.8 &  0.59 & 3.17 &  879 &  877 \\
205905.98+442655.9 & NSV~25414          & II   & 14.7 & 14.6 &  0.45 & 2.21 &  884 &  823 \\
205906.69+441823.7 & FHO~30             & II   & 17.2 & 17.2 &  0.14 & 1.23 &  872 &  869 \\
\enddata
\tablecomments{$\bar{R}$ denotes the mean PTF magnitude, $R_{med}$ the median PTF magnitude, and $\Delta R$ the peak-to-peak amplitude.}
\tablenotetext{a}{While this star is fainter than $R_{med} = 18$ in the latest data release, the target selection was done using an earlier release, at which time the source had $R_{med} < 18$.}
\label{lcstats}
\end{deluxetable}

\begin{deluxetable}{l p{1.5cm} c l p{4cm} p{4cm}}
\tabletypesize{\footnotesize}
\tablecaption{Phenomenology of Candidate Bursters and Faders}
\tablehead{\colhead{[RGS2011] ID} & \colhead{Short} & \colhead{$R_{med}$} & \colhead{Event} & 
\colhead{Lightcurve} & \colhead{Spectrum} \\ 
\colhead{} & \colhead{Name} & \colhead{(mag)} & \colhead{} & 
\colhead{Notes} & \colhead{Notes}}
\startdata
205032.32+442617.4 & FHO~1 & 17.5 & Burster & 
Several bursts, each lasting only 1-2 hours. &  \\
205036.93+442140.8 & [OSP2002] BRC~31~1 & 16.3 & Fader & 
Faded in mid-2011, still at minimum. See section \ref{brc311}. & Spectrum dominated by emission lines in both 1998 and 2012. Both epochs show H$\alpha$, Ca~II, Paschen series, and Fe~II emission; 2012 also has [O~I], [Fe~II], [S~II], and [Ni~II]. \\
205040.29+443049.0 & LkH$\alpha$~139 & 14.6 & Burster & 
One burst lasting 3~days in 2011. &  \\
205042.78+442155.8 & [OSP2002] BRC~31~8 & 16.9 & Burster & 
300-day modulation, with daily 1-2 hour bursts near maximum of the modulation. &  \\
205100.90+443149.8 & V1701~Cyg & 15.3 & Fader & 
Fades lasting several days, roughly once a month. &  \\
205114.80+424819.8 & FHO~2 & 13.8 & Burster & 
Bursts lasting $\sim 50$ days, every 100-300 days. See Section~\ref{shortoutbursts}. &  \\
205115.14+441817.4 & LkH$\alpha$~150 & 16.3 & Fader & 
Faded by 1~mag in early 2012 for 3-4 months. Long rise with $\pm 0.4$~mag variations during recovery. &  \\
205119.43+441930.5 & FHO~3 & 16.9 & Fader & 
2-day fades at intervals from 4 to 7 days. &  \\
205120.99+442619.6 & LkH$\alpha$~153 & 15.0 & Burster & 
One burst lasting 2-15 days, and several lasting less than 1 day each  &  \\
205123.59+441542.5 & FHO~4 & 17.6 & Burster & 
Two bursts lasting $\sim 60$ days, separated by 350 days. More complex profile than FHO~2. See Section~\ref{shortoutbursts}. & M2 star with H$\alpha$, He~I, [N~II], Ca~II emission. H$\alpha$ and Ca~II half as strong in 2012 as in 1998. \\
205124.70+441818.5 & FHO~5 & 17.9 & Fader & 
Many short 1~mag fades lasting $\sim 1$~day, mixed with some longer ($\sim 3$~day) but shallower ($\sim 0.6$~mag) fades. &  \\
205139.26+442428.0 & FHO~6	& 14.8 & Fader & 
Many short fades lasting $\sim 4$~days, separated by 20-50 days, superimposed on lower-amplitude erratic variability. &  \\
205139.93+443314.1 & FHO~7 & 16.3 & Fader & 
Three fading events, 50, 150, and $>180$ days long, each with complex bursts in their cores; one additional event, lasting $<70$~days, started at the end of the 2010 season. Decline seen over 2012, though not as pronounced as in FHO~27. & K5 star with H$\alpha$ and Ca~II emission. \\
\tablebreak
205145.99+442835.1 & FHO~8 & 17.7 & Fader & 
One 0.5~mag fading event lasting 100 days. &  \\
205155.70+443352.6 & FHO~9 & 15.9 & Fader & 
One 0.6~mag fading event lasting $\sim 6$ days in July 2011. 4-day, 0.6~mag, fading events separated by 10-20 days throughout rest of light curve. &  \\
\multirow{2}{*}{205158.63+441456.7} & \multirow{2}{*}{FHO~10} & 16.7 & Fader & 
Two fades $\sim 0.3$~mag deep lasting 4-5 days, and one lasting $<10$ days. Events separated by several months. Mixed with erratic variability of $\sim 0.2$~mag. & \multirow{2}{4cm}{} \\
 & 	&  & Burster & 
Two bursts $\sim 0.4$~mag high lasting $<3$ days each, separated by 7 days. Mixed with erratic variability of $\sim 0.2$~mag. &  \\
205203.65+442838.1 & FHO~11	& 17.9 & Fader & 
Slow decay over $\sim 100$~days followed by a rapid rise in $\sim 30$~days. Weaker, shorter fade 2 years before had a fast decay followed by a slow rise. &  \\
205228.33+442114.7 & FHO~12	& 16.5 & Fader & 
1.5-day fading events repeating every 5.8 days. & K7 star with strong H$\alpha$ emission, as well as He~I and [O~I] emission. \\
205230.89+442011.3 & LkH$\alpha$~174 & 16.7 & Fader & 
Fading events lasting 3 days, repeating every 7.7 days. Roughly 1/3 of the cycles do not have a fade. & K5 star with H$\alpha$, Ca~II, and He~I emission. \\
205252.48+441424.9 & FHO~13 & 18.0 & Fader & 
Fades lasting several days, every 10-20 days. Most fades have depths of $\sim 1$~mag; roughly every $\sim 200$~days a fade is deeper, $\sim 1.4$~mag. &  \\
205253.43+441936.3 & FHO~14 & 18.0 & Fader & 
0.4-0.7~mag fading events lasting 6-12 days every 20-30 days. One 0.2~mag, 150-day fading event with several of the shorter fades within it. &  \\
205254.30+435216.3 & FHO~15 & 17.1 & Fader & 
Three low states lasting 100, 30, and 70~days, in order, separated by one year; all three show high variability at minimum. First two fades 1.3~mag deep, third only 0.8~mag. & K8 with H$\alpha$, He~I, and O~I emission. \\
205314.00+441257.8 & FHO~16 & 17.1 & Fader & 
Combination of 0.6~mag fades, lasting 2-4 days, and 0.3~mag fades, lasting 60-80 days. &  \\
\tablebreak
205315.62+434422.8 & FHO~17 & 17.6 & Burster & 
Several 0.4~mag bursts lasting 1-3 days, followed by a quiescent period, followed by a 1.5~mag burst lasting 150~days. &  \\
\multirow{2}{*}{205340.13+441045.6} & \multirow{2}{*}{FHO~18} & 17.0 & Fader & 
Two 0.4~mag fades lasting 5 and 3 days, 11 days apart. Both fades immediately preceded by 0.3~mag bursts. See section \ref{fho18}. & \multirow{2}{4cm}{K5 star with H$\alpha$, He~I, Ca~II, and [N~II] emission.}  \\
  &  &  & Burster & 
Two 0.8~mag bursts lasting 10 and 7 days, 240 days apart. Several 0.3~mag bursts separated by tens of days. &  \\
205410.15+443103.0 & FHO~19  & 18.0 & Fader & 
Several fades lasting 3 days each, repeating every 8-10 days. Fade depth varies between 0.5 and 0.9~mag. &  \\
205413.74+442432.4 & FHO~20 & 16.2 & Fader & 
2-5~day fading events; longer events tend to be deeper.  &  \\
205424.41+444817.3 & FHO~21 & 16.6 & Fader & 
Three fades, lasting $\sim 10$~days (first part not observed), 5 days, and 11 days, separated by 250 and 330 days. &  \\
205445.66+444341.8 & FHO~22 & 17.3 & Fader & 
Complex fades lasting 6-20 days, separated by 230 and 300 days. Hints of a double profile for each event. One additional 3-day fade 50~days after the third main fade. &  \\
205446.61+441205.7 & FHO~23	& 17.3 & Fader & 
Several fades lasting 2-6 days, separated by a few weeks. Fades range from 0.6~mag to 0.3~mag, the level of the underlying erratic variability.  &  \\
205451.27+430622.6 & FHO~24	& 15.9 & Burster & 
0.2~mag burst lasting $\sim 15$ days in 2010, followed by a series of 0.5~mag bursts in 2012 lasting 15-40~days each. See Section~\ref{shortoutbursts}. &  \\
205503.01+441051.9 & FHO~25 & 16.0 & Fader & 
One $\sim$ 5-10-day fade in late 2010 &  \\
205534.30+432637.1 & $[$CP2005$]$~17	& 17.1 & Fader & 
One 65-day fade in 2010. &  \\
205659.32+434752.9 & FHO~26 & 17.9 & Burster & 
Several bursts in 2010-2011, lasting 4-5 days each and separated by 10-30 days. No activity in 2012. See section \ref{fho26}. & M4.5 star with H$\alpha$ emission. \\
205759.84+435326.5 & LkH$\alpha$~185 & 14.6 & Burster & 
First half of a 0.3~mag burst before a data gap in mid-2011. Rise time 2 days. &  \\
205801.36+434520.5 & FHO~27	& 16.1 & Fader & 
Multiple fading events lasting 15-40~days and separated by intervals ranging from 30-60 days. Events superimposed on a steep decline over the course of 2012, more extreme than in FHO~7. Fading events get shallower over the course of the decline. See section \ref{fho27}. & K7 star with strong H$\alpha$, Ca~II, Paschen series, and He~I emission, and weaker lines of [O~I] and O~I. \\
\tablebreak
205806.10+435301.4 & V1716~Cyg & 16.5 & Burster & 
Two bursts, the first lasting 5-20 days and the second 3 days, separated by 35 days. Complex profiles.  &  \\
205825.55+435328.6 & FHO~28 & 17.7 & Fader & 
130-day interval of repeated 8-day fading events in 2012; only 5, well-separated events each in 2010 and 2011. 2011 fades were typically only 2~days long, while 2010 events were too sparsely sampled to constrain their length. See section \ref{fho28}. & M3 star with H$\alpha$ emission in both 1998 and 2012, though the line is stronger in 2012. The 2012 spectrum also has weak emission of Ca~II, [N~II], He~I. \\
205839.73+440132.8 & FHO~29	& 16.8 & Burster & 
High states in early 2010, early 2011, late 2011, and entire first half of 2012. 2010-2011 bursts repeat roughly every 270-300 days, but 2012 behavior does not fit the period. &  \\
205905.98+442655.9 & NSV~25414 & 14.6 & Fader & 
1~mag fading events lasting 10-15 days, with $\pm 0.5$~mag variability at minimum. Fades repeat every $\sim 30$~days. \\
205906.69+441823.7 & FHO~30 & 17.2 & Fader & 
Short 0.6~mag fades, typically 2~days or less, separated by between 10 and 60 days. 
Two 0.15~mag fades lasting 30 days each in mid-2011 and late 2012. All fades are superimposed on 0.4~mag erratic variability. &  \\
\enddata
\tablecomments{$R_{med}$ denotes the median PTF magnitude. Light curves for all these sources are available online from the PTF website.}
\label{candidates}
\end{deluxetable}

\end{document}